\documentclass[aps,prd,twocolumn,floatfix,preprintnumbers,nofootinbib, superscriptaddress]{revtex4-1}

\usepackage{blindtext}
\usepackage{epsf,graphicx,xcolor}
\usepackage{amsmath,amssymb,amstext}
\usepackage{slashed}
\allowdisplaybreaks
\usepackage{verbatim}

\usepackage{mathtools}
\newcommand{\sla}[1]{\ensuremath{\mathrlap{\!\not{\phantom{#1}}}#1}}
\usepackage{epstopdf}
\usepackage{url,hyperref}
\usepackage[utf8]{inputenc}
\usepackage[english]{babel}
\usepackage{float}
\usepackage[capitalize]{cleveref}

\newcommand{\pvec}[1]{\vec{#1}\mkern2mu\vphantom{#1}}

\begin{document}

\title{Soft-photon radiative corrections to the $e^- p \to e^- p l^- l^+$ process}
\author{Matthias Heller}
\affiliation{Institut f\"ur Kernphysik and $\text{PRISMA}^+$ Cluster of Excellence, Johannes Gutenberg Universit\"at, D-55099 Mainz, Germany}
\author{Niklas Keil}
\affiliation{Institut f\"ur Kernphysik and $\text{PRISMA}^+$ Cluster of Excellence, Johannes Gutenberg Universit\"at, D-55099 Mainz, Germany}
\author{Marc Vanderhaeghen}
\affiliation{Institut f\"ur Kernphysik and $\text{PRISMA}^+$ Cluster of Excellence, Johannes Gutenberg Universit\"at, D-55099 Mainz, Germany}

\date{\today}

\begin{abstract}
We calculate the leading-order QED radiative corrections to the process $e^- p\rightarrow e^- p l^- l^+ $ in the soft-photon approximation, in two different energy regimes which are of relevance to extract nucleon structure information. In the low-energy region, this process is studied to better constrain the hadronic corrections to precision muonic Hydrogen spectroscopy. In the high-energy region, the beam-spin asymmetry for double virtual Compton scattering allows to directly access the Generalized Parton Distributions. We find that the soft-photon radiative corrections have a large impact on the cross sections and are therefore of paramount importance to extract the nucleon structure information from this process. 
For the forward-backward asymmetry the radiative corrections are found to affect the asymmetry only around or below the 1\% level, whereas the beam-spin asymmetry is not affected at all in the soft-photon approximation, which makes them gold-plated observables to extract nucleon structure information in both the low- and high-energy regimes.
\end{abstract}

\maketitle
\section{Introduction}
Double-virtual Compton scattering (dVCS) on a proton, the process $\gamma^\ast p \to \gamma^\ast p$ with initial and final virtual photons ($\gamma^\ast$), is a prime process to study and test models describing the electromagnetic structure of the nucleon beyond the information contained in the elastic form factors. 

At low-energies, it allows to extract nucleon structure constants, which enter 
the expansion of the nucleon Compton amplitude. The real Compton scattering (RCS) limit, the process $\gamma p \to \gamma p$ with both photons real, has been used over many years as an experimental tool to access the nucleon electromagnetic polarizabilities, see e.g.  Ref.~\cite{Drechsel:2002ar,Griesshammer:2012we,Hagelstein:2015egb,Pasquini:2018wbl} for reviews. The virtual Compton scattering (VCS) process, $\gamma^\ast p \to \gamma p$ with initial space-like virtual photon, which can be accessed as a subprocess of the $e^- p \to e^- p \gamma$ reaction,   has also been studied extensively over the past three decades to access the generalized nucleon polarizabilities~\cite{Guichon:1995pu,Guichon:1998xv,Drechsel:2002ar,Fonvieille:2019eyf}. These structure quantities allow to obtain, through a Fourier transform, a spatial representation of the deformation of the charge and magnetization distributions of the nucleon under the influence of an external static electromagnetic field~\cite{Gorchtein:2009qq}. 

The most general case of a double-virtual Compton process, with both initial and final virtual photons, has until now been studied only in special limits. The most useful extension is given by the forward double-virtual Compton scattering (VVCS) process, where the initial and final photons have the same non-zero space-like virtuality. In contrast to the processes discussed above, the forward VVCS process is not directly measurable. It enters however in the leading hadronic corrections
to the muonic Hydrogen Lamb shift and hyperfine splitting. The interest in its improved estimate was spurred in 2010 by the  ultra-precise determination of the proton charge radius from muonic Hydrogen Lamb shift measurements~\cite{Pohl:2010zza}, which reported a radius value which was 4\% smaller than the 2010 recommended value by the Committee on Data for Science and Technology (CODATA)\cite{RevModPhys.84.1527} based on results from electron-proton scattering and ordinary Hydrogen spectroscopy measurements, and represents a $7 \sigma$ difference. Over the past decade, major progress has been made in resolving this puzzle, see Refs.~\cite{Pohl:2013yb,Carlson:2015jba,Gao:2021sml} for some recent reviews. The dominant theoretical model error in the extraction of the proton charge radius from muonic Hydrogen Lamb shift measurements 
to date results from the subtraction function entering the VVCS process~\cite{Carlson:2011zd,Birse:2012eb,Antognini:2013rsa}. At second order in the photon virtuality, this function is constrained by the magnetic polarizability, which is determined experimentally~\cite{ParticleDataGroup:2020ssz}.     
To fourth order in the photon virtuality, one low-energy constant in this subtraction function is at present empirically unconstrained~\cite{Lensky:2017bwi}, and one relies on chiral effective field theory calculations~\cite{Birse:2012eb, Alarcon:2013cba} or phenomenological estimates~\cite{Tomalak:2015hva}.  
In Ref.~\cite{Pauk:2020gjv} it was proposed to access this low-energy nucleon structure constant empirically through the forward-backward asymmetry in the $e^- p \to e^- p \, l^- l^+$ process, with $l=e$ or $l=\mu$. The dVCS amplitude contributing to  that process, $\gamma^* p \rightarrow \gamma^* p$, has an incoming photon with space-like virtuality and an outgoing photon with time-like virtuality. 

A second kinematical region in which the virtual Compton processes are being used as a prime tool to study the partonic structure of the nucleon is at high energies, for near-forward kinematics, either through 
the $e^- p \to e^- p \gamma$ process with initial space-like photon with large virtuality, the deeply-virtual Compton scattering process (DVCS), or through the di-lepton photoproduction process $\gamma p \to l^-l^+ p$ process with outgoing time-like photon with large virtuality, the time-like Compton scattering (TCS).  
In such kinematical regime, pertubative Quantum Chromo Dynamics (QCD) allows to express the proton structure entering the DVCS and TCS processes through Generalized Parton Distributions (GPDs), which access the correlation between the longitudinal momentum distribution of partons in a proton and their two-dimensional transverse spatial distributions. We refer the reader to Refs.~\cite{Ji:1996ek,Mueller:1998fv,Radyushkin:1996nd,Ji:1996nm}
for the original articles on GPDs and to
Refs.~\cite{Goeke:2001tz,Diehl:2003ny,Belitsky:2005qn,Boffi:2007yc,Guidal:2013rya,Kumericki:2016ehc} for reviews of the field.   
Accessing the resulting three-dimensional momentum-spatial  distributions of valence quarks in a nucleon through exclusive processes has been one of the driving motivations for the JLab 12 GeV upgrade~\cite{osti_1345054}. Furthermore, accessing the 
sea-quark and gluonic structure of nucleons and nuclei through such processes is one of the main science questions that will be addressed at the future Electron-Ion Collider (EIC) machine~\cite{Accardi:2012qut}. 

A further extension of either the DVCS or TCS process in the high-energy near-forward region has been proposed through the $e^- p \to e^- p l^-l^+$ reaction (with $l^-$ either an $e^-$ or $\mu^-$), which accesses the double deeply virtual Compton scattering (DDVCS) process with incoming space-like photon and outgoing time-like photon. The DDVCS process is of particular interest as it allows to extend the DVCS beam spin asymmetry measurements, which directly access GPDs, into the so-called ERBL domain~\cite{Guidal:2002kt,Belitsky:2002tf}.
A feasibility study of the DDVCS experiment has shown that the SoLID@JLab project with its high-luminosity and large acceptance is very promising to perform such measurements~\cite{Accardi:2020swt}. 

In order to use the $e^- p \to e^- p l^-l^+$ reaction as a tool of proton structure, it is imperative to quantitatively estimate the QED radiative corrections to this process, which is the main objective of the present work. 
Our work extends previous studies of radiative corrections for the VCS process~\cite{Vanderhaeghen:2000ws}, as well as more recently for the TCS process~\cite{Heller:2018ypa,Heller:2019dyv,Heller:2020lnm}. In Ref.~\cite{Heller:2020lnm} it was found for the $\gamma p \rightarrow l^+ l^- p$ process that the relevant asymmetries to extract the real and imaginary parts of the TCS amplitudes, the forward-backward and beam-helicity asymmetries, are nearly unaffected by the radiative corrections. In contrast the TCS cross sections receive sizeable corrections: in the low-energy region up to $10\%$, and in the high-energy kinematical region up to $20\%$. As for the single space-like or single time-like Compton scattering cases, it is crucial to have a good quantitative  understanding of radiative corrections also in the double-virtual case in order to be able to extract relevant information about the proton structure from future experimental data. As a first estimate of the size of radiative corrections we will use the soft-photon approximation in this work. We distinguish between three different, gauge-invariant types of corrections, from which one contributes to the VCS case, a second one contributes to the TCS case, both of which are obtained as limits of our work, and a third type of correction which is new for the double-virtual case. We study the size of these corrections on the level of unpolarized cross sections as well as on the forward-backward and beam-spin asymmetries.

The outline of the present paper is as follows. In Sec.~\ref{sec:tree} we introduce the relevant Feynman diagrams at tree level. We distinguish between two different contributions: the Bethe-Heitler and the Compton scattering processes. In Sec.~\ref{sec:models} we introduce the two different nucleon structure models which we use to describe the dVCS amplitude. In  the low-energy regime we calculate the contribution from the Born process in terms of the protons form factors as well as the $\Delta(1232)$ resonance excitation in combination with a low-energy expansion of the dVCS amplitude. In the high-energy regime we use the QCD factorization theorem to express the dVCS amplitude in terms of GPDs. In Sec.~\ref{sec:virt_corr} we calculate the virtual radiative corrections in the soft-photon approximation from the three gauge invariant types of contributions. We give analytic expressions for the finite and infrared divergent parts of all three contributions in terms of a factorizing contribution on the level of cross section. In Sec.~\ref{sec:bremsstrahlung} we calculate the contribution to the cross sections due to soft-photon bremsstrahlung. Taking real radiation into account, we cross-check analytically the  cancellation with the infrared divergences from the virtual corrections. In Sec.~\ref{sec:results} we show our results for the observables in both the low-energy and high-energy kinematical regimes. We conclude in Sec.~\ref{sec:conclusion}. Technical details are discussed in two appendices.

\section{Di-lepton electroproduction at tree level}
\label{sec:tree}

\begin{figure}[h]
\includegraphics[scale=0.5]{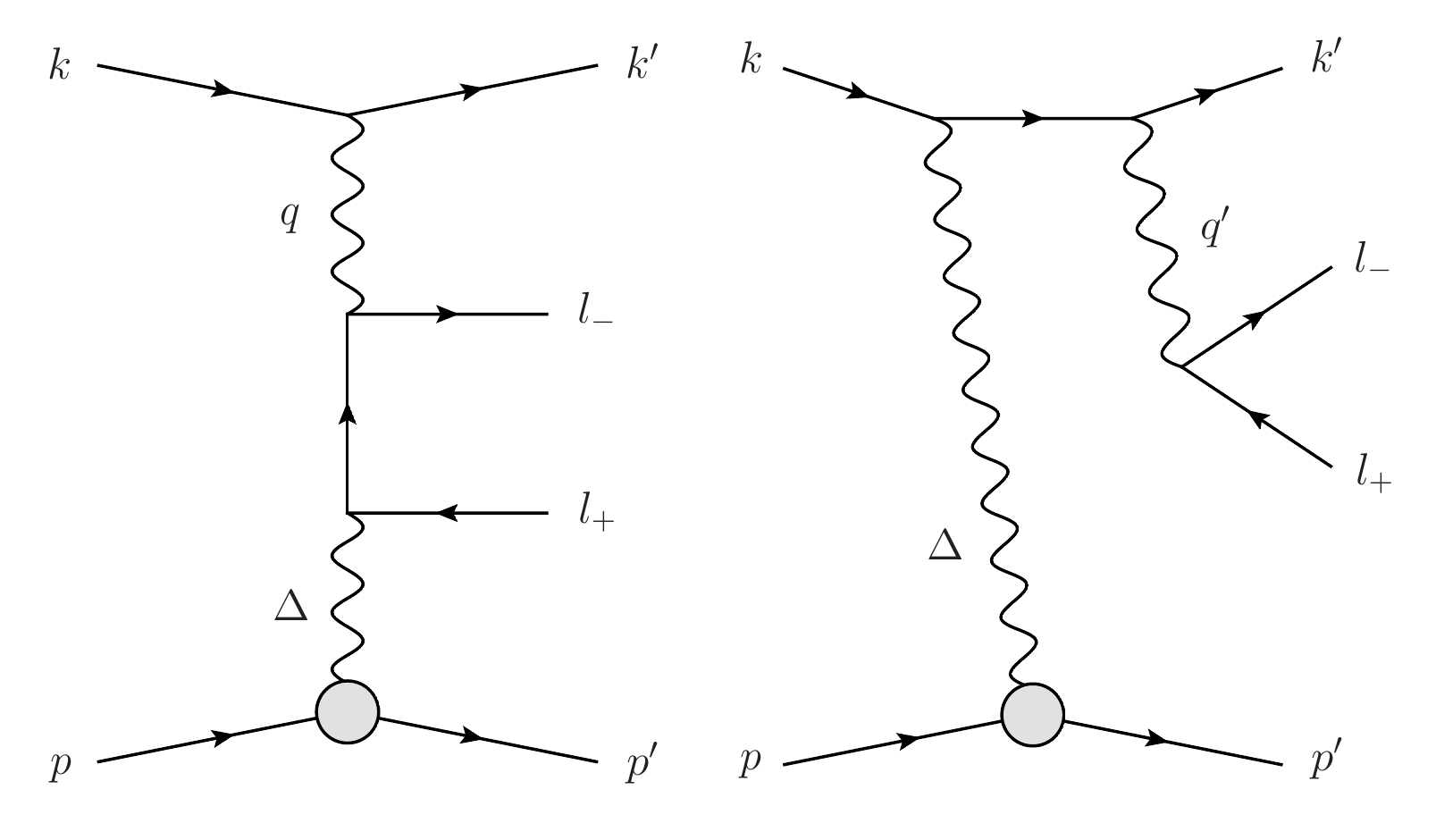}
\caption{Tree level QED diagrams contributing to the $e^- p \to e^- p l^- l^+$ process. We distinguish between the spacelike (left) and the timelike (right) Bethe-Heitler processes. The crossed diagrams, for which in the spacelike process the order of the vertices on the produced di-lepton line are interchanged, and for which in the timelike process the order of the vertices on the electron beam line are interchanged, are not shown.  
\label{fig:tree_diagrams}}
\end{figure}

\begin{figure}[h]
\centering
\includegraphics[scale=0.5]{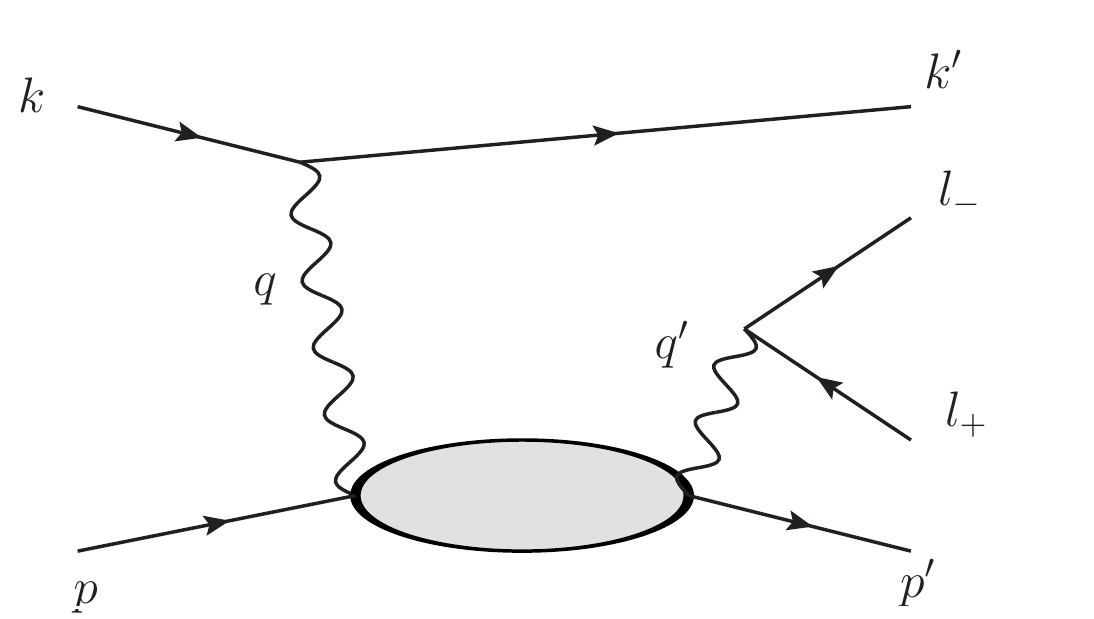}
\caption{Tree level diagrams for the Compton scattering. The blob represents the (elastic and inelastic) interaction of the virtual photon with the nucleon.}
\label{fig:TreeDVCS}
\end{figure}

In this work we study the di-lepton electroproduction process 
\begin{equation}
    e^-(k)+N(p)\rightarrow e^-(k^\prime)+N(p^\prime)+l^-(l_-)+l^+(l_+),
    \label{eq:reaction}
\end{equation}
as a probe of proton ($N$) structure, 
with $l^-$ either an $e^-$ or a $\mu^-$, 
where the quantities in brackets denote the particle four-momenta.
At tree level, we distinguish between three different contributions, which we denote as the spacelike (SL) and timelike (TL) Bethe-Heitler (BH) processes, see Fig.~\ref{fig:tree_diagrams}, as well as the double virtual Compton process (dVCS), see Fig.~\ref{fig:TreeDVCS}.

\begin{figure}
    \centering
    \includegraphics[trim=80 80 0 70, scale=0.75]{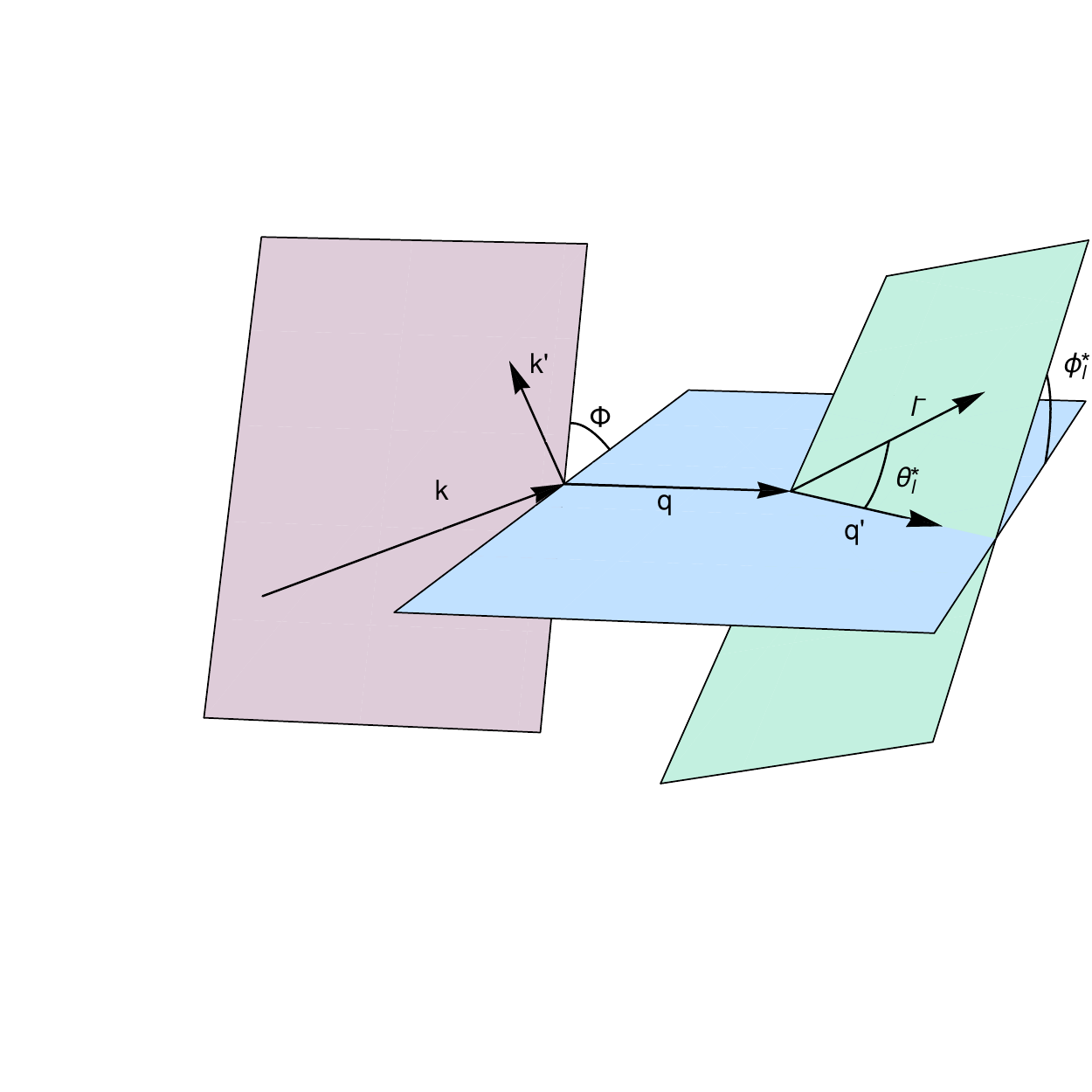}
    \caption{Planes defining the scattering angles which characterize the $e^- p \to e^- p l^-l^+$ process. The angles $\Phi$ and $\phi^*_l$ are defined with respect to the blue plane, which is the scattering plane of the virtual photons with four-momenta $q$ and $q'$.}
    \label{fig:kin_plane}
\end{figure}

To specify the kinematics, it is useful to introduce the following four-momenta:
\begin{align}
    q=&k-k^\prime,\quad q^\prime=l_++l_-,\quad \Delta=p^\prime-p.
\end{align}
The process~(\ref{eq:reaction}) is defined by eight kinematical invariants, which we choose as:
\begin{align}
    s=&(k+p)^2,\qquad
    Q^2=-(k-k^\prime)^2,\nonumber\\
    W^2=&(q+p)^2,\qquad
    t=\Delta^2,\nonumber\\
    s_{ll}=&q^{\prime 2},\qquad
    \Phi,\;\theta^*_l,\;\phi^*_l,
\end{align}
where $\Phi$ denotes the angle of the initial electron plane relative to the production plane. Furthermore, the angle $\theta^*_l$ ($\phi^*_l$) denotes the polar (azimuthal) angle respectively of the negative lepton in the rest frame of the $l^-l^+$ lepton pair. In Fig. \ref{fig:kin_plane} we show the three different scattering planes defined by these angles.

We denote by $m$ the mass of the electron, by $m_l$ the mass of the produced leptons, and by $M$ the mass of the proton. The on-shell relations of the external particles are therefore:
\begin{align}
    k^2=k^{\prime2}=m^2,\quad
    l_-^2=l_+^2=m_l^2,\quad
    p^2=p^{\prime2}=M^2,
\end{align}
and the invariant $s$ is obtained from the Lab electron beam energy $E_e$ as $s = M^2 + m^2 + 2 M E_e$. 

The matrix element for the spacelike Bethe-Heitler (BH,SL)  process (left diagram in Fig.~\ref{fig:tree_diagrams}) is given by:
\begin{align}
    \mathcal{M}_{0;dir}^\text{BH,SL}=&\frac{-ie^4}{Q^2 t} 
    \bar{u}\left(k', h'\right)\gamma^\mu u\left( k, h\right)
    \nonumber \\ 
    \times& 
    \bar{u}\left(l_-, h_-\right)\left[\gamma_\mu \frac{\slashed{l}_- - \slashed{q} + m_l}{(l_- -q)^2-m_l^2} \gamma_\alpha\right. \nonumber\\
    &\left. \hspace{1.5cm}
    +\gamma_{\alpha}\frac{\sla{q}-\sla{l_+}+m_l}{(q-l_+)^2-m_l^2}\gamma_{\mu}  \right]
    v\left(l_+, h_+\right) \nonumber \\
    \times&\bar{N}\left(p', s'\right)\Gamma^\alpha \left(p',p\right)N\left(p, s\right),
    \label{eq:BHSL}
\end{align}
while the timelike Bethe-Heitler (BH,TL) process (right diagram in Fig.~\ref{fig:tree_diagrams}) is given by:
\begin{align}
     \mathcal{M}_{0;dir}^\text{BH,TL}=&\frac{ie^4}{s_{ll} t} \bar{u}\left( l_-, h_-\right) \gamma^\mu v\left( l_+, h_+ \right)  \nonumber \\
     \times&\bar{u}\left( k', h' \right)\left[  \gamma_\mu \frac{\slashed{k}' + \slashed{q}' +m}{\left(k'+q' \right)^2-m^2}\gamma_\alpha\right.\nonumber\\
     &\left. \hspace{1.5cm}
     +\gamma_\alpha \frac{\slashed{k} - \slashed{q}' +m}{\left(k-q' \right)^2-m^2}\gamma_\mu\right] u (k, h)
     \nonumber \\
     \times&
     \bar{N}\left(p', s'\right)\Gamma^\alpha \left(p',p\right)N\left(p, s\right).
    \label{eq:BHTL}
\end{align}
In Eqs.~(\ref{eq:BHSL},\ref{eq:BHTL}), $h(h')$ denote the helicities of the intial (scattered) electrons, 
$h_-$ and $h_+$ are the helicities of the produced lepton pair, and $s (s')$ are the helicities of the intial (final) proton respectively. 
Furthermore, $\Gamma^\alpha$ is the electromagnetic nucleon vertex given by:
\begin{eqnarray}
\Gamma^\alpha(p^\prime, p) = F_1(t) \gamma^\alpha + F_2(t)   \frac{i \sigma^{\alpha \alpha^\prime} \Delta_{\alpha^\prime}}{2 M},
\label{eq:nucvertex}
\end{eqnarray}
 where $F_1$ ($F_2$) are the Dirac (Pauli) form factors of the nucleon respectively.  
 
The matrix element for the double virtual Compton scattering (dVCS) process (Fig.~\ref{fig:TreeDVCS}) is expressed as:
\begin{align}
    \mathcal{M}_{0;dir}^\text{dVCS}=&\frac{ie^4}{Q^2 s_{ll}} \bar{u}\left( k', h' \right) \gamma_\mu u \left(k, h\right) 
    \nonumber \\
    \times& \bar{u}\left(l_-, h_- \right)\gamma_\nu v\left(l_+, h_+ \right) \nonumber\\
    \times&\bar{N}\left(p', s'\right)M^{\mu \nu} N\left(p, s \right),
    \label{eq:Mdvcs}
\end{align}
where $M^{\mu\nu}$ denotes the Compton tensor which depends on the model to describe the interaction of photons with the nucleon, and which will be specified below.

In the case of $e^-e^+$ production we have to take into account that the electrons with momenta $k'$ and $l_-$ are indistinguishable. Thus for $e^-e^+$ production, we have to consider, besides the direct ({\it dir}) contribution of Eqs.~(\ref{eq:BHSL},\ref{eq:BHTL},\ref{eq:Mdvcs}), also the contribution of all exchange ({\it ex}) diagrams where both electrons in the final state are interchanged. The Bethe-Heitler matrix elements corresponding with these exchange terms are given by (note that this only contributes in the case $m_l=m$):
\begin{align}
    \mathcal{M}_{0;ex}^\text{BH,SL}=&\frac{ie^4}{(k-l_-)^2 t} 
    \bar{u}\left(l_-, h_-\right)\gamma^\mu u\left( k, h\right)
    \nonumber \\
    \times& 
    \bar{u}\left(k', h'\right)\left[\gamma_\mu \frac{\slashed{l}_- - \slashed{q} + m}{(l_- -q)^2-m^2} \gamma_\alpha\right. \nonumber\\
    &\left. \hspace{1.5cm}
    + \gamma_{\alpha}\frac{\sla{k}-\sla{q'}+m}{(k-q')^2-m^2}\gamma_{\mu}  \right]
    v\left(l_+, h_+\right) \nonumber \\
        \times&\bar{N}\left(p', s'\right)\Gamma^\alpha \left(p',p\right)N\left(p, s\right),
        \label{eq:BHSLex} \\
     \mathcal{M}_{0;ex}^\text{BH,TL}=&\frac{ie^4}{(l_++k')^2 t} \bar{u}\left(k', h'\right) \gamma^\mu v\left( l_+, h_+ \right) \nonumber \\
     \times&\bar{u}\left( l_-, h_- \right)\left[  \gamma_\mu \frac{\slashed{k}' + \slashed{q}' +m}{\left(k'+q' \right)^2-m^2}\gamma_\alpha\right.\nonumber\\
     &\left. \hspace{1.5cm} 
     +\gamma_\alpha \frac{\slashed{q} - \slashed{l}_+ +m}{\left(q-l_+ \right)^2-m^2}\gamma_\mu\right] 
     u (k, h)
     \nonumber \\
         \times&\bar{N}\left(p', s'\right)\Gamma^\alpha \left(p',p\right)N\left(p, s\right),
             \label{eq:BHTLex}
\end{align}
and the exchange term corresponding with the dVCS matrix element is given by:
\begin{align}
    \nonumber
    \mathcal{M}_{0;ex}^\text{dVCS}=&\frac{-ie^4}{(k-l_-)^2 (l_++k')^2} 
    \bar{u}\left( l_-, h_- \right) \gamma_\mu u \left(k, h\right) \nonumber \\
    \times& 
    \bar{u}\left(k', h'\right)\gamma_\nu v\left(l_+,h_+\right)
    \nonumber \\
    \times& \bar{N}\left(p', s'\right)M^{\mu \nu} N\left(p, s\right).
    \label{eq:Mdvcsex}
\end{align}
To ensure the Pauli principle, one has to anti-symmetrize the amplitude under exchange of both electrons in the final state.  Therefore, the full matrix elements for the $e^- p \to e^- p e^-e^+$ process is obtained as difference between the amplitudes for direct ({\it dir}) and exchange ({\it ex}) diagrams:
\begin{align}
    &\mathcal{M}_{0}^\text{BH,SL}=\mathcal{M}_{0;dir}^\text{BH,SL}-\mathcal{M}_{0;ex}^\text{BH,SL},
    \nonumber \\
    &\mathcal{M}_{0}^\text{BH,TL}=\mathcal{M}_{0;dir}^\text{BH,TL}-\mathcal{M}_{0;ex}^\text{BH,TL}\nonumber\\
    &\mathcal{M}_{0}^\text{dVCS}=\mathcal{M}_{0;dir}^\text{dVCS}-\mathcal{M}_{0;ex}^\text{dVCS},
    \label{eq:antisymm}
\end{align}
while for $\mu^-\mu^+$ production only the direct diagrams contribute.

The fully differential cross section for the 
$e^- p \to e^- p l^-l^+$ process is given by
\begin{eqnarray}
    &&\left(\frac{\text{d}\sigma}{\text{d}Q^2\text{d}W^2\text{d}\Phi\text{d}t\text{d}s_{ll}\text{d}\Omega_l^\ast} \right)_0=\frac{1}{(4\pi)^7}\frac{1}{2(s-M^2)^2}\nonumber \\
    &&\hspace{1.cm}\times\frac{\beta_{s_{ll}}}{[((W+M)^2+Q^2)((W-M)^2+Q^2)]^{\frac{1}{2}}}\nonumber\\
    &&\hspace{1.cm}\times\overline{\sum_i}\sum_f\left|\mathcal{M}^\text{BH,SL}_0+\mathcal{M}^\text{BH,TL}_0+\mathcal{M}^\text{dVCS}_0\right|^2, \quad
\end{eqnarray}
where d$\Omega_l^\ast$ refers to the phase space of the produced lepton of the dilepton pair in the $l^- l^+$ rest frame, and where $\beta_{s_{ll}}$ is the lepton velocity in the $l^-l^+$ rest frame:   
\begin{equation}
    \beta_{s_{ll}}=\sqrt{1-\frac{4m_l^2}{s_{ll}}}.
\end{equation}

\section{Models for the double virtual Compton amplitude}
\label{sec:models}
\begin{figure}[h]
	\centering
  \includegraphics[scale=0.5]{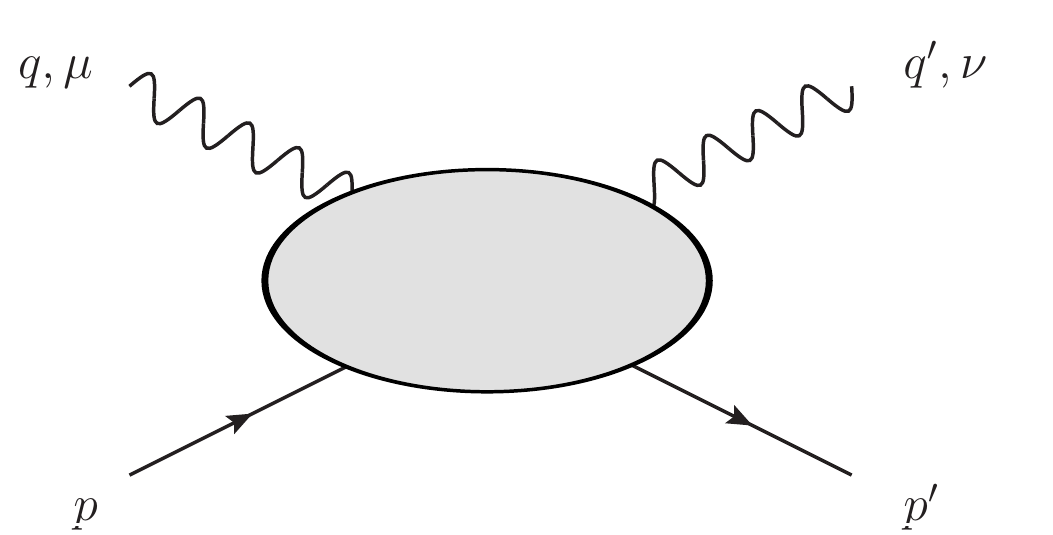}
	\caption{Diagram representing the double virtual Compton process. \label{fig:compton}}
\end{figure}

The double virtual Compton tensor $M^{\mu\nu}$ entering Eq.~(\ref{eq:Mdvcs}) is calculated from the process
\begin{equation}
    \gamma^*(q)+N(p)\rightarrow \gamma^*(q^\prime) + N(p^\prime).
\end{equation}
We show the Feynman diagram for this process in Fig. \ref{fig:compton}. The blob in this diagram represents the interaction of the incoming and outgoing virtual photons with the proton. 
In the following we use the average photon ($\bar{q}$) and proton ($P$) momenta,
\begin{equation}
    \bar{q}\equiv\frac{1}{2}(q+q^\prime),\qquad P\equiv\frac{1}{2}(p+p^\prime).
\end{equation}

The general double virtual Compton tensor $ M^{\mu\nu}$ can be constructed using $q^\mu$, $q^{\prime \mu}$, $p^\mu$, $g^{\mu\nu}$ and $\gamma^\mu$ as building blocks. From these blocks, one finds $34$ independent tensors with two indices~\cite{Tarrach:1975tu}. 
Using gauge invariance it was shown that the number of independent amplitudes can be reduced from $34$ to $18$~\cite{Tarrach:1975tu}. 
However it was realized in Ref.~\cite{Tarrach:1975tu}, that there is in general a problem in such representation. For specific kinematical points the $18$ tensors become linearly dependent and therefore do not form a basis at these specific points anymore. As a result the corresponding Compton amplitudes display kinematic singularities at these points. To bypass this problem, Tarrach introduced an overcomplete basis by introducing three additional tensors. Such an overcomplete basis does not have any kinematical constraints and is valid in the whole phase space.
It was realized in Ref.~\cite{Drechsel:1997xv} that the kinematic singularities and constraints of the Compton amplitude in a minimal basis are due to the Born terms, in which the intermediate state in the Compton process in Fig.~\ref{fig:compton} is a nucleon, and that for the non-Born contributions a minimal tensor basis consisting of 18 structures  free of kinematical singularities and constraints exists.  

In this work,  we will only need the  
helicity-averaged amplitude,   
which is described by 5 independent  tensors, and can be expressed as, following the notations of~\cite{Drechsel:1997xv}:
\begin{eqnarray}
M^{\mu \nu} &=& 
\sum_{i =1,2,3,4,19} B_i(\nu, q^2, q^{\prime 2}, q \cdot q^\prime) \, T_i^{\mu \nu} , 
\label{eq:dvcsunpol}
\end{eqnarray}
where $T_i^{\mu \nu}$ are the spin-independent and gauge invariant tensors, symmetric under exchange of the two virtual photons, and are given by:
\begin{eqnarray}
T_1^{\mu \nu} &=& - q \cdot q^\prime g^{\mu\nu}+ q^{\prime \mu} q^{\nu} \, , \nonumber \\
T_2^{\mu \nu} &=&  (2 M \nu)^2 \left( - g^{\mu\nu} +  \frac{q^{\prime \mu} q^{\nu}}{q \cdot q^\prime} \right) 
\nonumber \\
&-& \, 4 q \cdot q^\prime \left( P^{\mu} - \frac{q \cdot P}{q \cdot q^\prime} q^{\prime \mu} \right)   
\left( P^{\nu} - \frac{q \cdot P}{q \cdot q^\prime} q^{\nu} \right)  \, ,  \nonumber \\
T_3^{\mu \nu} &=& q^2 q^{\prime \, 2} g^{\mu\nu} + q \cdot q^\prime  q^{\mu} q^{\prime \nu}  
- q^2 \, q^{\prime \mu} q^{\prime \nu}  - q^{\prime \, 2} \, q^{\mu} q^{\nu}  \, , \nonumber \\
T_4^{\mu \nu} &=&  (2 M \nu) (q^2 + q^{\prime \, 2}) 
\left( g^{\mu\nu} -  \frac{q^{\prime \mu} q^{\nu}}{q \cdot q^\prime} \right) \nonumber \\ 
&+& \,2 \left( P^{\mu} - \frac{q \cdot P}{q \cdot q^\prime} q^{\prime \mu} \right)   
\left( - q^{\prime 2} q^{\nu} + q \cdot q^\prime  q^{\prime \nu} \right)  
\nonumber \\
&+& \, 2 \left( P^{\nu} - \frac{q \cdot P}{q \cdot q^\prime} q^{\nu} \right) 
\left( - q^2 q^{\prime \mu} +  q \cdot q^\prime  q^{\mu} \right)   
 \, ,  \nonumber \\
T_{19}^{\mu \nu} &=&   4 q^2  q^{\prime \, 2} 
\left( P^{\mu} - \frac{q \cdot P}{q^2} q^{\mu} \right)   
\left( P^{\nu} - \frac{q \cdot P}{q^{\prime \, 2}} q^{\prime \, \nu} \right)  \, .  
\label{eq:dvcstensors}
\end{eqnarray}
Furthermore, in Eq.~(\ref{eq:dvcsunpol}), the invariant amplitudes $B_i$ are functions of four Lorentz invariants, 
with $\nu \equiv q \cdot P / M$.    

In order to specify the double virtual Compton amplitude, we need to model the internal structure of the nucleon. In this work, we will consider two different models, which are tailored for applications in two different energy regimes. In a low-energy model, which is motivated for applications to describe the hadronic structure in precision atomic physics measurements such as the Lamb shift or hyperfine splitting in muonic Hydrogen, we consider the photons to interact with the nucleon and its lowest excitation, the $\Delta(1232)$ resonance. In a high-energy model, in which at least one of the photons is highly virtual, we use perturbative QCD which allows to factorize the Compton process on the nucleon in terms of a Compton amplitude on the quark convoluted with the amplitude 
to find the quarks inside the nucleon. The latter is parameterized through generalized parton distributions (GPDs).

\subsection{Low-energy double virtual Compton amplitude}
\label{sec:dvcs_low_energy}

\subsubsection{Born diagrams}

\begin{figure}[h]
	\centering
	\includegraphics[scale=0.6]{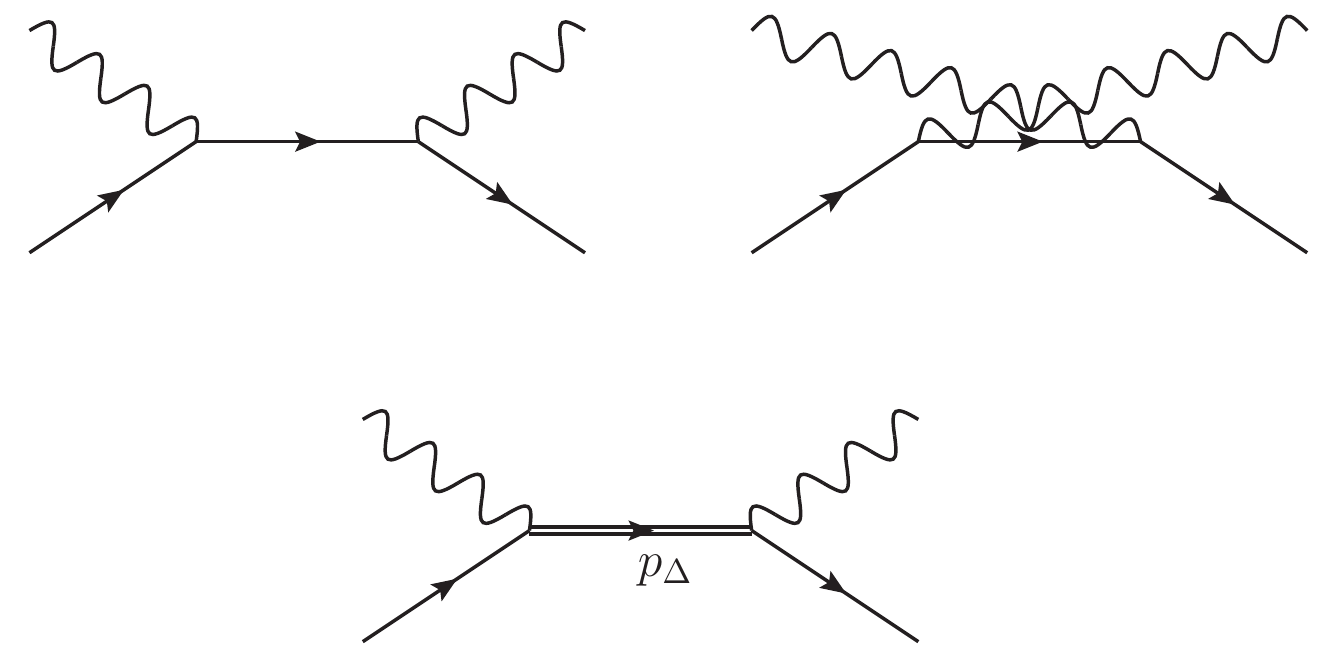}

	\caption{Born contribution (upper panel) and  $s$-channel $\Delta$-pole contribution (lower panel) to the Compton amplitude. While for the Born contribution only the sum of $s$- and $u$- channel diagrams is gauge invariant, the $s$-channel $\Delta$-pole contribution is gauge invariant by itself.}
	\label{fig:dvcs_low}
\end{figure}

In the low-energy regime, we describe the Compton tensor in terms of the leading Born (B) amplitude, given in terms of the proton form factors. The amplitude can be calculated from two Feynman diagrams shown in Fig.~\ref{fig:dvcs_low} (upper panel). Its contribution is given by
\begin{equation}
    M_{\rm B}^{\mu\nu}=\Gamma_f^\nu\frac{\sla{p}+\sla{q}+M}{(p+q)^2-M^2}\Gamma_i^\mu+\Gamma_i^\mu\frac{\sla{p}'-\sla{q}+M}{(p'-q)^2-M^2}\Gamma_f^\nu,
\end{equation}
where $\Gamma^\mu_i$ ($\Gamma^\nu_f$) are the initial (final) state proton vertices. Note that the FFs entering $\Gamma^\nu_f$ correspond with a timelike virtuality. For the numerical evaluation of these FFs we use the paramaterization of Ref.~\cite{Lomon:2012pn}, which allows the analytical continuation based on dispersion relations into the unphysical part of the timelike region, $0 < q^{\prime 2} < 4 M^2$. Note that in this region no direct experimental extraction exists.

\subsubsection{$\Delta$-pole model}

In addition to the Born amplitude, we need a model for the non-Born contribution at low energies. The covariant baryon chiral perturbation theory (BChPT) provides a systematic framework for the calculation of the double virtual Compton scattering process, see Ref.~\cite{Lensky:2017bwi}. 
The latter work has shown that BChPT is fully predictive at orders ${\mathcal O}(p^3)$ and ${\mathcal O}(p^4/\varDelta)$, in which $p$ stands for a small momentum and with $\varDelta \equiv M_\Delta - M$ the excitation energy of the $\Delta(1232)$ resonance.  
The ${\mathcal O}(p^3)$ contribution comes from the pion-nucleon ($\pi N$) loops, and the ${\mathcal O}(p^4/\varDelta)$ contribution comes from the Delta-exchange ($\Delta$-pole) graph, which is shown in Fig.~\ref{fig:dvcs_low} (lower panel),  and the pion-Delta ($\pi \Delta$) loops. 

For the near-forward real Compton cross section (i.e. integrated over dilepton phase space), it was found that around $W = 1.25$~GeV the Born + $\Delta(1232)$-pole contribution reproduces a full dispersive calculation based on empirical structure functions within an accuracy of 5\% or better~\cite{Pauk:2020gjv}. As we consider in this work kinematics around the $\Delta(1232)$ resonance, we will study as a first step the effect due to radiative corrections on the $\Delta(1232)$-pole contribution.  
 
The amplitude for the $\Delta$-pole contribution to the double virtual Compton tensor can be expressed as:
\begin{align}
  M_{s\Delta}^{\mu\nu} =&  \tilde{\Gamma}_{\gamma N \Delta}^{\alpha\nu}(p^\prime,p+q)\frac{(\sla{p}+\sla{q}+M_\Delta)(-g_{\alpha\beta}+\frac{1}{3}\gamma_\alpha\gamma_\beta)}{W^2 - M_\Delta^2+i M_\Delta\Gamma_\Delta(W^2)}\nonumber\\
&\times\Gamma_{\gamma N \Delta}^{\beta\mu}(p+q,p).\label{mdelta}
\end{align}

\begin{figure}[h]
\vspace{-.5cm}
	\centering
  \includegraphics[scale=0.6]{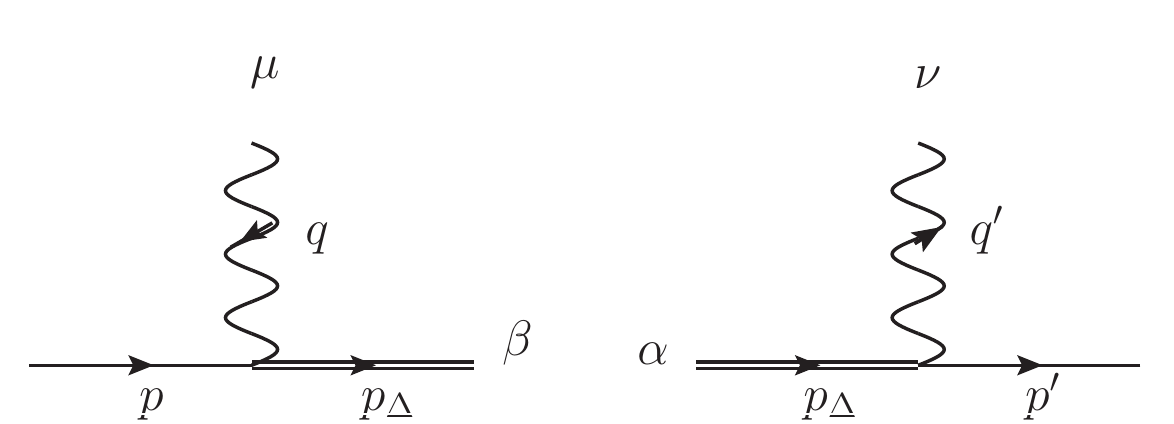}
	\caption{The $\gamma^* N \Delta$ vertex $\Gamma_{\gamma N \Delta}^{\beta\mu}$ (left diagram) and its adjoint, $\tilde{\Gamma}_{\gamma N \Delta}^{\alpha\nu}$ (right diagram). }
	\label{d_vertex}
\end{figure}
In Eq. \eqref{mdelta}, the $\gamma^* N \Delta$ vertex is denoted by $\Gamma^{\beta\mu}_{\gamma N \Delta}(p_\Delta,p)$ and its adjoint by $\tilde{\Gamma}^{\alpha\nu}_{\gamma N \Delta}(p^\prime,p_\Delta)$. Both vertices are shown in Fig.~\ref{d_vertex}. They are given for the $p \to \Delta^+$ transition by:
\begin{eqnarray}
\Gamma^{\beta\mu}_{\gamma N \Delta}(p_\Delta,p)
&=& \sqrt{\frac{3}{2}}\frac{(M_\Delta+M)}{M Q_+^2}\bigg\{g_M(q^2)i\epsilon^{\beta\mu\kappa\lambda}(p_\Delta)_\kappa q_\lambda\nonumber\\
&&- g_E(q^2)(q^\beta p_\Delta^\mu-q\cdot p_\Delta g^{\beta \mu})\gamma_5\nonumber\\
&&- g_C(q^2) \frac{1}{M_\Delta} \left[\sla{p}_\Delta(q^\beta q^\mu-q^2g^{\beta\mu}) \right. \nonumber \\
&&\left. \hspace{1cm}-\gamma^\beta(q\cdot p_\Delta q^\mu-q^2p^\mu_\Delta) \right]\gamma_5\bigg\},
\label{eq:ndelspace}
\end{eqnarray}
and
\begin{eqnarray}
\tilde{\Gamma}^{\alpha\nu}_{\gamma N \Delta}(p^\prime,p_\Delta) &=& -\sqrt{\frac{3}{2}}\frac{(M_\Delta+M)}{M Q_+^{\prime 2}}\bigg\{g_M(q^{\prime 2}) i\epsilon^{\alpha\nu\kappa\lambda}(p_\Delta)_\kappa q^\prime_\lambda\nonumber\\
&&-g_E(q^{\prime 2})(q^{\prime \alpha} p_\Delta^\nu-q^\prime \cdot p_\Delta g^{\alpha\nu})\gamma_5
\nonumber \\
&&-g_C(q^{\prime 2}) \frac{1}{M_\Delta} \gamma_5 \left[\sla{p}_\Delta(q^{\prime \alpha} q^{\prime \nu}-q^{\prime 2} g^{\alpha\nu}) \right. \nonumber \\
&&\hspace{1cm}\left. -\gamma^\alpha(q^\prime \cdot p_\Delta q^{\prime \nu} - q^{\prime 2} p^\nu_\Delta) \right]\bigg\},
\label{eq:ndeltime}
\end{eqnarray}
where we defined $Q_\pm=\sqrt{(M_\Delta\pm M)^2 - q^2}$ and likewise 
$Q^\prime_\pm=\sqrt{(M_\Delta\pm M)^2 - q^{\prime 2}}$. The FFs $g_M$, $g_E$ and $g_C$ appearing in Eq.~\eqref{eq:ndelspace} have spacelike virtuality, whereas the FFs in the adjoint vertex defined in Eq. \eqref{eq:ndeltime} have to be evaluated for timelike virtuality. We reexpress these FFs in terms of the more conventional magnetic dipole ($G_M^*$), electric quadrupole ($G_E^*$) and Coulomb quadrupole ($G_C^*$) transition FFs as:
\begin{align}
    g_M&=\frac{Q_+}{M+M_\Delta}(G_M^*-G_E^*),\nonumber\\
    g_E&=-\frac{Q_+}{M+M_\Delta}\frac{2}{Q_-^2}\{(M_\Delta^2-M^2+q^2)G_E^*-q^2G_C^*\},\nonumber\\
    g_C&=\frac{Q_+}{M+M_\Delta}\frac{1}{Q_-^2}\{4M_\Delta^2G_E^*-(M_\Delta^2-M^2+q^2)G_C^*\},
\end{align}
with the so-called Ash FFs parameterized, for spacelike virtuality $Q^2 = -q^2$, through the MAID2007 analysis as~\cite{Drechsel:2007if,Tiator:2011pw}:
\begin{align}
    G^{*}_{M}(Q^2)&=3.00 (1+0.01 Q^2)e^{-0.23 Q^2}G_D(Q^2),\nonumber\\
    G^{*}_{E}(Q^2)&=0.064 (1-0.021 Q^2)e^{-0.16 Q^2}G_D(Q^2),\nonumber\\
    G^{*}_{C}(Q^2)&=0.124 \frac{1+0.12 Q^2}{1+4.9Q^2/(4M^2)}\frac{4M_\Delta^2e^{-0.23 Q^2}G_D(Q^2)}{M_\Delta^2-M^2} ,
\end{align}
with $Q$ in GeV and the dipole FF $G_D(Q^2)=1/(1+Q^2/0.71)^2$. For small timelike virtualities, $0 < q^{\prime 2} < (M_\Delta - M)^2$, we extrapolate in Eq.~\eqref{eq:ndeltime} the expressions for spacelike virtualities by the substitution $Q^2 \to - q^{\prime 2}$. 

The dominant contribution is coming from the magnetic dipole transition FF $G^*_M$. In the following we use only that dominant contribution, corresponding with the leading term in the so-called $\delta$-expansion~\cite{Pascalutsa:2006up}, to calculate observables, i.e. we set $G_E^*=G_C^*=0$.

\subsubsection{Low-energy expansion}

The non-Born part of the dVCS amplitudes, denoted as $\bar B_i$, can be expanded 
for small values of $\nu, q^2, q^{\prime \, 2}$, and 
$q \cdot q^\prime$, 
with coefficients given by polarizabilities. 
The relations between these low-energy coefficients and the polarizabilities measured through real Compton scattering ($\gamma p \to \gamma p$) and virtual Compton scattering ($\gamma^\ast p \to \gamma p$) have been given in \cite{Lensky:2017bwi}. 

A special limit of the double virtual Compton process is given by its forward limit, denoted by VVCS, which corresponds with $q^\prime = q$ and $p^\prime = p$. This limit is of particular importance as it enters the two-photon hadronic corrections to the electronic and muonic Hydrogen energy levels.  
The helicity averaged VVCS process is described by two invariant amplitudes, denoted by $T_1$ and $T_2$, which are functions of the two kinematic invariants, $Q^2$ and $\nu$, as: 
\begin{eqnarray}
M^{\mu \nu}_{\mathrm{VVCS}} 
\equiv \frac{1}{\alpha_\mathrm{em}} 
\left\{ \hat g^{\mu \nu} T_1(\nu, Q^2) 
- \frac{\hat p^{\mu}  \hat p^{\nu}}{M^2}  T_2 (\nu, Q^2) \right\}, 
\label{eq:vvcs}
\end{eqnarray}
with $\hat g^{\mu \nu} \equiv g^{\mu\nu}- q^{\mu}q^{\nu} / q^2$, $\hat p^\mu \equiv p^{\mu}- p\cdot q / q^2 \,q^{\mu}$, and 
where $\alpha_\mathrm{em} = e^2 / 4 \pi \simeq 1/137$.  
The optical theorem allows to express the imaginary parts of $T_1$ and $T_2$ as:
\begin{eqnarray}
{\rm{Im}}\ T_1(\nu, Q^2) = \frac{e^2}{4M} F_1  \, , \quad
{\rm{Im}}\ T_2(\nu, Q^2)  = \frac{e^2}{4 \nu} F_2  \, ,  
\label{eq:optical}
\end{eqnarray}
where $F_1, F_2$ are the conventionally defined structure functions parameterizing 
inclusive electron-nucleon scattering, depending 
on $Q^2$ and $x \equiv Q^2 / 2 M \nu$.  
The two-photon exchange correction to the $\mu$H Lamb shift can be expressed as a weighted double integral over $Q^2$ and $\nu$ 
of the forward amplitudes $T_1$ and $T_2$~\cite{Carlson:2011zd}. Using the empirical input of $F_1$ and $F_2$, the  $\nu$ dependence 
of $T_2$ can be fully reconstructed using an unsubtracted dispersion relation, whereas the dispersion relation for $T_1$ requires one subtraction, which can be chosen at $\nu = 0$ as $T_1(0, Q^2)$. 
The subtraction function is usually split in a Born part, corresponding with the nucleon intermediate state, and a remainder, so-called non-Born part, denoted by $\bar T_1(0, Q^2)$. The Born part can be expressed in terms of elastic form factors and is well known, see e.g. \cite{Pasquini:2018wbl} for the corresponding expressions. The non-Born part cannot be fixed empirically so far. In general, one can however write down a low $Q^2$ expansion of $\bar T_1(0, Q^2)$ as:
\begin{eqnarray}
\bar T_1(0, Q^2) = \beta_{M1} Q^2 + \frac{1}{2} T_1^{''}(0) Q^4 + \mathcal{O}(Q^6),
\label{eq:T1bar0}
\end{eqnarray}
where the term proportional to $Q^2$ is empirically determined by the magnetic dipole polarizability $\beta_{M1}$~\cite{ParticleDataGroup:2020ssz}.  
Theoretical estimates for the subtraction term were given at order $Q^4$ in heavy-baryon chiral perturbation theory (HBChPT)~\cite{Birse:2012eb}, in BChPT, both at leading order (LO) due to $\pi N$ loops,
and at next-to-leading order (NLO), including both $\Delta(1232)$-exchange and $\pi \Delta$ loops~\cite{Alarcon:2013cba,Lensky:2017bwi}, as well as extracted from superconvergence sum rule (SR) relations~\cite{Tomalak:2015hva}. The different estimates for $\bar{T}_1^{\prime\prime}(0)$  are compared in Table~\ref{tab:subtraction}. Even for these theoretically well motivated approaches, the spread among the different estimates is quite large. The resulting uncertainty due to this subtraction term constitutes at present the main uncertainty in the theoretical $\mu H$ Lamb shift estimate. 
To reduce such model dependence, the dilepton electroproduction process on a proton has been proposed in~\cite{Pauk:2020gjv} as an empirical way to determine $\bar{T}_1^{\prime\prime}(0)$. 

\begin{table}[h]
\begin{tabular}{cc|c|c}
\hline\hline
Source & Ref. & $\frac{1}{2}\bar{T}_1^{\prime\prime}(0)$ & $\alpha_{\rm{em}}  b_{3,0}$ 
\\
\hline
HBChPT & \cite{Birse:2012eb} & $[-1.01, -0.35]$ &  \\
\hline
$\pi N$ loops &
& $-0.06$  & $0.001$ \\
$\pi \Delta$ loops &
& $-0.10$ & $-0.005$ \\
$\Delta$ exchange &
& $-1.98$ & $0.11$ \\
Total BChPT & \cite{Lensky:2017bwi}
& $-2.14 \pm 0.98$ & $0.11 \pm 0.05$  \\
\hline
superconvergence SR & \cite{Tomalak:2015hva} 
& $-0.47$ & $3.96$  \\
\hline\hline
\end{tabular}
\caption{Estimates of the $Q^4$ term of the subtraction function $\bar T_1(0,Q^2)$ (second column) and of the dVCS low-energy constant $b_{3,0}$ (third column), both in units $10^{-4}$ fm$^5$, in different theoretical approaches~\cite{Lensky:2017bwi}. The indicated range for the HBChPT result corresponds with the range given by Eq.~(15) in Ref.~\cite{Birse:2012eb}.
\label{tab:subtraction}
}
\end{table}

As the forward VVCS process of Eq.~(\ref{eq:vvcs}) is a special case of Eq.~(\ref{eq:dvcsunpol}), one can express the subtraction function entering the hadronic corrections to the $\mu H$ energy levels as~\cite{Lensky:2017bwi}:
\begin{eqnarray}
\bar T_1(0, Q^2) = \alpha_\mathrm{em} Q^2 \left( \bar B_1 + Q^2  \bar B_3   \right),
\end{eqnarray}
where both non-Born amplitudes $\bar B_{1}, \bar B_3$ are understood in the forward limit ($q = q^\prime$), 
i.e. $\bar B_{i}(0, q^2, q^2, q^2)$ for $i = 1,3$. 
In order to specify $\bar T_1(0,Q^2)$ up to the $Q^4$ term,   
we use the low-energy expansion in $k \in \{q, q^\prime\}$ of the amplitudes $\bar B_1, \bar B_3$~\cite{Lensky:2017bwi}:
\begin{eqnarray}
\bar{B}_1(0, q^2, q^{\prime 2}, q \cdot q^\prime) &=& \frac{1}{\alpha_\mathrm{em}} \left\{ \beta_{M1} - \frac{1}{6} \beta_{M2} q \cdot q^\prime 
\right. \nonumber \\
&&\left. \hspace{-1.5cm}- \left(\beta^\prime_{M1}(0) + \frac{\beta_{M1}}{8 M^2}  \right) ( q^2 + q^{\prime \, 2}) \right\} +  {\cal  O}(k^4), \nonumber \\ 
\bar{B}_3(0, q^2, q^{\prime 2}, q \cdot q^\prime)  &=& b_{3,0} + {\cal  O}(k^2), 
\label{lexdvcs}
\end{eqnarray}
where $\beta_{M2}$ is the magnetic quadrupole polarizability determined from real Compton scattering~\cite{Holstein:1999uu}, 
and $\beta^\prime_{M1}(0)$ is the slope at $Q^2 = 0$ of the generalized magnetic dipole polarizability which is accessed through 
virtual Compton scattering, see Ref.~\cite{Fonvieille:2019eyf} for a recent review. While the terms of ${\cal  O}(k^0)$ and ${\cal  O}(k^2)$ in the  low-energy structure of the amplitude $\bar B_1$ at  $\nu = 0$ are empirically constrained from real or virtual Compton scattering, the low-energy constant $b_{3, 0}$ is not determined empirically so far because the tensor structure $T_3^{\mu \nu}$ in Eq.~(\ref{eq:dvcstensors}) decouples when either the initial or final photon is real. As such the low-energy constant $b_{3,0}$ is the main unknown to date in the empirical determination of $\bar{T}_1^{\prime\prime}(0)$. Below, we study the sensitivity of the $e^- p \to e^- p l^-l^+$ process, including the soft-photon radiative corrections, to this low-energy constant.

\subsection{High-energy double virtual Compton amplitude in terms of GPDs
\label{sec:dvcs_high_energy}}

\begin{figure}[h]
\centering
  \hspace{-0.9cm}\includegraphics[scale=0.55]{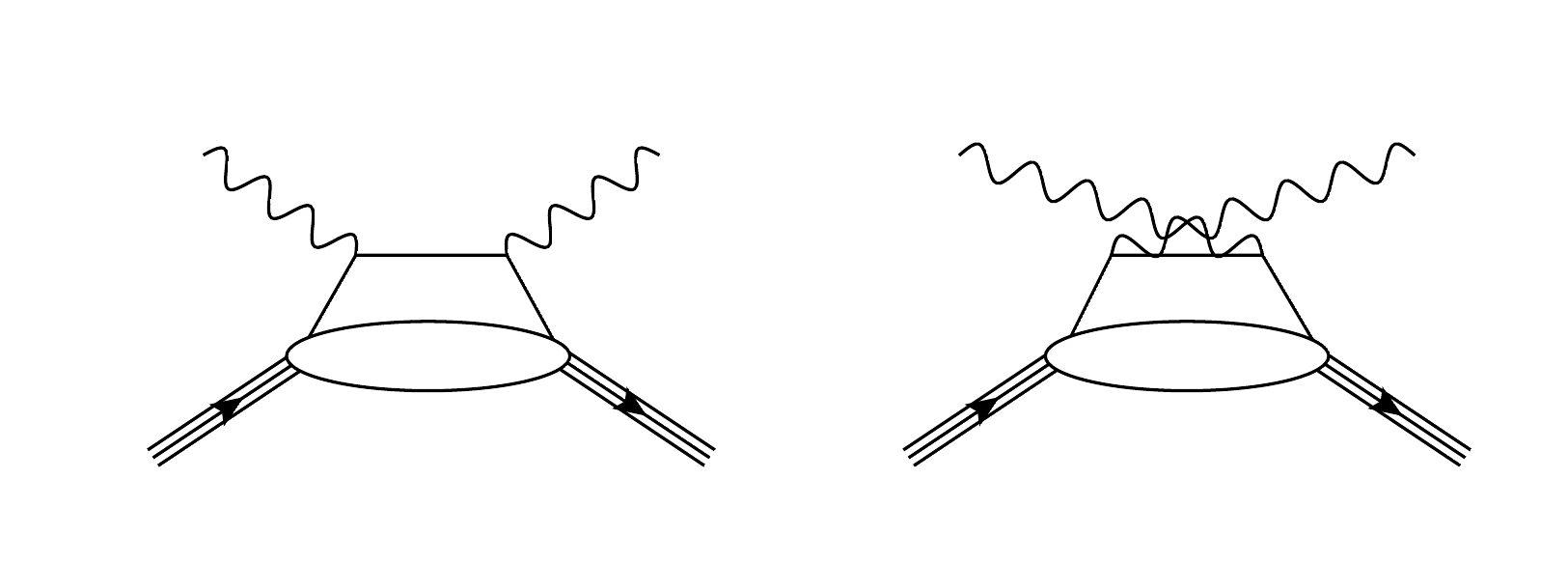}
  \caption{Handbag diagrams for the double deeply virtual Compton amplitude. The 
  single (composite) lines represent quarks (nucleons) respectively. The blobs represent the GPDs. 
  \label{dia:handbag}}
\end{figure}

For the high-energy Compton scattering we calculate the Compton tensor in terms GPDs using pertubative QCD. This can be done by calculating the leading-order handbag  diagrams shown in Fig. \ref{dia:handbag}. 

For the evaluation of these diagrams, we will need  the kinematic Lorentz invariants $\xi$ and $\xi^\prime$, defined as:
\begin{eqnarray}
    \xi&\equiv&-\frac{\Delta\cdot \bar{q}}{2P\cdot \bar{q}} 
    = \frac{-q^2 + q'^2}{2 (W^2 - M^2) -  q^2 - q'^2 + t} , 
    \label{eq:xi}\\
    \xi^\prime &\equiv& -\frac{\bar{q}^2}{2P\cdot \bar{q}} 
    = \frac{-q^2 - q'^2 + t/2}{2 (W^2 - M^2) -  q^2 - q'^2 + t}.
    \label{eq:xip}
\end{eqnarray}
Furthermore, we introduce the two light-like vectors $\tilde p^\mu$ and $n^\mu$ with $\tilde p \cdot n=1$, which are related to the four momenta $P^\mu$ and $\bar q^\mu$ as:
\begin{eqnarray}
P^\mu &=& \tilde p^\mu + \frac{\bar M^2}{2} n^\mu, \label{eq:P}\\
\bar q^\mu &=& - \tilde \xi^\prime \tilde p^\mu - \frac{\bar q^2}{2 \tilde \xi^\prime} n^\mu \label{eq:qbar}, 
\end{eqnarray}
where $\bar M^2 = M^2 - t / 4$. 
The variables $\tilde{\xi}$ and $\tilde{\xi}'$ are related to the invariants $\xi$ and $\xi^\prime$ introduced in Eqs.~(\ref{eq:xi},\ref{eq:xip}) as:
\begin{align}
\tilde{\xi}&=\xi\frac{1+\tilde{\xi}'^2 \bar{M}^2/ \bar{q}^2}{1-\tilde{\xi}'^2\bar{M}^2/\bar{q}^2}, 
\label{eq:tildexi} \\
    \tilde{\xi}'&=\xi'\frac{2}{1+\sqrt{1-4\xi'^2\bar{M}^2/ \bar{q}^2}}.
    \label{eq:tildexip}
\end{align}

Although at leading-twist, corresponding with the kinematical regime for which $\bar M^2 \ll \bar q^{2}$, one has
\begin{equation}
    \tilde{\xi} \rightarrow \xi,\qquad \tilde{\xi}' \rightarrow \xi',
\end{equation}
we will keep in the following analysis the 
(small) difference in the kinematical quantities. 

The leading twist-2 double deeply virtual Compton scattering (DDVCS) amplitude on a proton is given by:
\begin{eqnarray}
&&M^{\mu\nu}_{\rm{DDVCS, tw-2}} = \frac{1}{2}(-g_{\mu\nu})_{\perp}\int_{-1}^{1}dx
\, C_+(x,\tilde \xi^\prime)\nonumber\\
&&\hspace{1.5cm}\times \left\{H(x,\tilde \xi,t)\sla{n}+E(x,\tilde \xi,t)i\sigma^{\alpha\beta}n_\alpha \frac{\Delta_\beta}{2M}\right\}\nonumber\\
&&\hspace{1.5cm}+\frac{i}{2}(\varepsilon_{\nu\mu})_{\perp}\int_{-1}^{1}dx
\, C_-(x,\tilde \xi^\prime) \nonumber\\
&&\hspace{1.5cm}\times \left\{\tilde{H}(x,\tilde \xi,t)\sla{n}\gamma_5+\tilde{E}(x,\tilde \xi,t)\gamma_5\frac{\Delta\cdot n}{2M}\right\},
\label{eq:DDVCS}
\end{eqnarray}
where the coefficient functions $C_\pm(x,\tilde \xi^\prime)$ are defined as:
\begin{eqnarray}
C_\pm(x,\tilde \xi^\prime) \equiv \frac{1}{x-\tilde \xi^\prime+i\epsilon}
\pm\frac{1}{x+\tilde \xi^\prime-i\epsilon},    
\end{eqnarray}
and
\begin{align}
    (-g_{\mu\nu})_{\perp}&=-g_{\mu\nu}+\tilde{p}_\mu n_\nu+\tilde{p}_\nu n_\mu,\nonumber\\
    (\varepsilon_{\nu\mu})_\perp&=\varepsilon_{\nu\mu\alpha\beta}n^\alpha \tilde{p}^\beta,
\end{align}
where the lightlike four-vectors $\tilde p$ and $n$ are obtained from Eqs.~(\ref{eq:P},\ref{eq:qbar}) as
\begin{align}
    n^\mu&=\frac{1}{\tilde{\xi}^\prime \bar{M}^2/2-\bar{q}^2/(2\tilde{\xi}^\prime)}\left\{\tilde{\xi}^\prime P^\mu +\bar{q}^\mu\right\},\nonumber\\
    \tilde{p}^\mu&=\frac{-1}{\tilde{\xi}^\prime \bar{M}^2-\bar{q}^2/\tilde{\xi}^\prime}\left\{\bar{q}^2/\tilde{\xi}^\prime P^\mu +\bar{M}^2\bar{q}^\mu\right\}.
\end{align}

Furthermore, for the purpose of studying the influence of the radiative corrections on the DDVCS observables at small values of $-t$, we will only consider the contribution of the dominant GPD $H$ in our study below. For the numerical evaluation, we will use the GPD parametrizations from the
VGG model~\cite{Vanderhaeghen:1998uc, Vanderhaeghen:1999xj,Goeke:2001tz,Guidal:2004nd}, summarized in Ref.~\cite{Guidal:2013rya}, in terms of a double distribution (DD) and so-called D-term contribution (D), as:
\begin{eqnarray}
    H(x,\xi,t)&=&H_{\rm{DD}}(x,\xi,t)+ D(\frac{x}{\xi},t),
    \label{eq:gpdmodel}
\end{eqnarray}
with the double distribution part for the proton given by the weigthed sum of the light quark flavor distributions:
\begin{eqnarray}
    H_{\rm{DD}} = \frac{4}{9} H^u_{\rm{DD}}
    +\frac{1}{9}H^d_{\rm{DD}} +\frac{1}{9}H^s_{\rm{DD}}.
    \label{eq:gpdddmodel}
\end{eqnarray}
The isoscalar D-term contribution, is directly related to the subtraction function in a dispersive framework for the Compton amplitude. For its evaluation, we use the dispersive estimate of Ref.~\cite{Pasquini:2014vua}.

In order to satisfy exact electromagnetic gauge invariance for both incoming and outgoing virtual photons in the DDVCS process, we generalize the procedure introduced in Ref.~\cite{Vanderhaeghen:1999xj} to add transversal correction terms which are formally of higher-twist as follows:
\begin{eqnarray}
M^{\mu\nu}_{\rm{DDVCS}} &=&
M^{\mu\nu}_{\rm{DDVCS, tw-2}} 
- \frac{P^\mu}{2 P \cdot \bar q} 
(\Delta_\perp)_\kappa 
M^{\kappa\nu}_{\rm{DDVCS, tw-2}} \nonumber \\
&+& \frac{P^\nu}{2 P \cdot \bar q} (\Delta_\perp)_\lambda 
M^{\mu \lambda}_{\rm{DDVCS, tw-2}}  \nonumber \\
&-& \frac{P^\mu P^\nu}{4 (P \cdot \bar q)^2} (\Delta_\perp)_\kappa (\Delta_\perp)_\lambda 
M^{\kappa\lambda}_{\rm{DDVCS, tw-2}}, 
\end{eqnarray}
where the transverse part $\Delta_\perp$ of the four-momentum transfer to the nucleon is defined as:
\begin{eqnarray}
\left(\Delta_\perp \right)^\mu \equiv \Delta^\mu 
+ 2 \tilde \xi \, \tilde p^\mu 
- \tilde \xi \bar M^2 \, n^\mu.   
\end{eqnarray}

Using the identities
\begin{eqnarray}
q_\mu M^{\mu\nu}_{\rm{DDVCS, tw-2}} 
&=& \frac{1}{2} \left(\Delta_\perp \right)_\mu \, 
M^{\mu\nu}_{\rm{DDVCS, tw-2}}, \nonumber \\
q^\prime_\nu M^{\mu\nu}_{\rm{DDVCS, tw-2}} 
&=& - \frac{1}{2} \left(\Delta_\perp \right)_\nu \, 
M^{\mu\nu}_{\rm{DDVCS, tw-2}}, 
\end{eqnarray}
one immediately verifies that both 
$q_\mu M^{\mu\nu}_{\rm{DDVCS}} = 0$ and 
$q^\prime_\nu M^{\mu\nu}_{\rm{DDVCS}} = 0$. 

Using the parameterization of Eq.~(\ref{eq:gpdmodel}) for the GPD $H$ in terms of a double distribution and a D-term part, the evaluation of the amplitude in Eq.~(\ref{eq:DDVCS}) involves a principle-value integral which can be evaluated numerically, for the case $0 < \xi^\prime < \xi$, as:
\begin{eqnarray}
&&{\rm P.V.} \int_{0}^{1}dx \, \frac{H^{\rm{singlet}}(x, \xi, t)}{x - \xi^\prime} \nonumber \\
&&= \int_{0}^{1}dx \, 
\frac{H_{DD}^{\rm{singlet}}(x, \xi, t) - H_{DD}^{\rm{singlet}}(\xi^\prime, \xi, t)}{x - \xi^\prime} \nonumber \\
&&+ 2 \int_{0}^{\xi}dx \, 
\frac{D(x/\xi, t) - D(\xi^\prime/\xi, t)}{x - \xi^\prime} \nonumber \\
&&+ \ln\left( \frac{1 - \xi^\prime}{\xi^\prime}\right) \, H_{DD}^{\rm{singlet}}(\xi^\prime, \xi, t) \nonumber \\
&&+ \ln\left( \frac{\xi - \xi^\prime}{\xi^\prime} \right) 2 D(\xi^\prime/\xi, t),
\end{eqnarray}
with the singlet GPD defined as:
\begin{eqnarray}
H^{\rm{singlet}}(x, \xi, t ) \equiv 
H(x, \xi, t) - H(-x, \xi, t).
\end{eqnarray}

\section{Virtual soft-photon corrections}
\label{sec:virt_corr}

In this work, we evaluate all one-loop virtual photon radiative corrections to the $e^- p \to e^- p l^-l^+$ process in the soft-photon approximation. This limit is defined by the scaling of the loop momenta: we only account for the regions of integration where the loop momentum $l$ scales as:
\begin{equation}
    l\sim \lambda,
\end{equation}
where $\lambda$ is a small parameter compared to all external scales. We then calculate all contributions only up to order $\lambda$. The resulting corrections factorize in terms of the tree-level amplitude, which shows that this is a gauge-invariant subset of the full one-loop corrections.

From all soft-photon contributions, one can then further distinguish between three gauge invariant subsets:
\begin{itemize}
    \item class (a): soft photon attached to the beam electron line
    \item class (b): soft photon attached to the di-lepton pair
    \item class (c): soft photon connecting the beam electron line with the di-lepton line
\end{itemize}
We give analytical expressions for the corrections of all three types. In order to regularize the infrared divergences coming from the integration over the soft-photon loop momentum $l$ we use dimensional regularization~\cite{tHooft:1972tcz}. We therefore perform the loop integration in $D=4-2\epsilon$ dimensions. Infrared (IR) divergences manifest themselves as $1/\epsilon_{\mathrm{IR}}$ poles in the regularized expressions. We are using the on-shell renormalization scheme. 
In addition to the diagrams with virtual soft photons we also have to consider infrared divergent counter terms. Those counter terms are introduced to regularize ultraviolet (UV) divergences (which manifest themselves as $1/\epsilon_{\mathrm{UV}}$ poles), which due to the on-shell renormalization condition can also carry IR divergences. Those need to be included in the calculation in order to get a finite result in the end.

We will subsequently discuss the virtual radiative corrections to the spacelike Bethe-Heitler process, the timelike Bethe-Heitler process, and the double virtual Compton process. 

\subsection{Corrections to the spacelike Bethe-Heitler process\label{Sec:correction-sl}}

\subsubsection{Contributions of class (a)}
\begin{figure}
\includegraphics[scale=0.55]{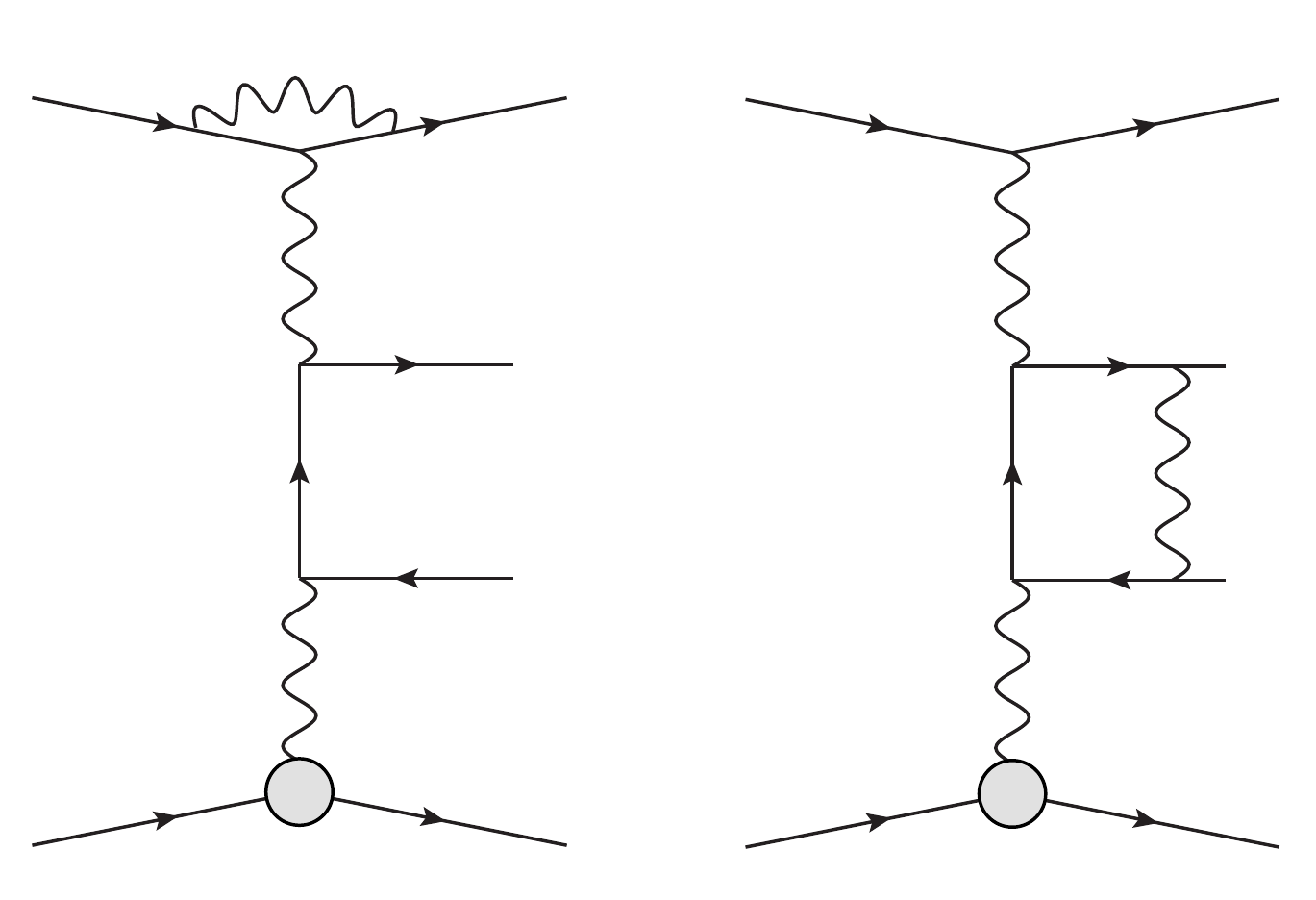}
  \caption{Virtual corrections of class (a) (left) and class (b) (right) to the spacelike BH process with two virtual photons. One also has to consider the corresponding counter term diagrams. The crossed diagrams with $l_-$ and $l_+$ interchanged yield the same correction.}
    \label{fig:virdiagramsslab}
\end{figure}

In this section we calculate the soft-photon corrections for which the soft-photon is attached to the electron line. This corresponds to the left diagram in Fig. \ref{fig:virdiagramsslab}. In the following we suppress helicity states in all spinors to make the formulas more compact and better readable. The first diagram in Fig. \ref{fig:virdiagramsslab} is given by, using Feynman gauge:
\begin{align}
\label{eq:matrixelement2}
&\mathcal{M}_{a}^\text{BH,SL}=-\frac{e^6}{Q^2 t}\Bar{N}\left( p' \right) \Gamma^\alpha\left( p',p \right) N\left(  p \right)  \mu^{4-D}  \nonumber\\
&\phantom{=}\times  \int_{}^{} \frac{d^Dl}{\left( 2\pi \right)^D}   \frac{ \Bar{u}\left( k' \right)\gamma^\beta (\slashed{k}^\prime+\slashed{l}+m) \gamma^\mu (\slashed{k}+\slashed{l}+m) \gamma_\beta u\left( k \right)  }{\left[(k'+l)^2-m^2\right]\left[(k+l)^2-m^2\right]\left[l^2\right] }  \nonumber\\
&\phantom{=}\times \frac{\Bar{u}\left( l_- \right)\gamma_\mu \left(  \slashed{l}_- -\slashed{q}+m_l\right)\gamma_\alpha v\left( l_+  \right)}{\left[ ( l_- - q )^2 -m_l^2 \right]}  ,
\end{align}
which reduces in the soft-photon approximation to
\begin{align}
    &\mathcal{M}_{a}^\text{BH,SL}=- i e^2 4 (k \cdot k^\prime) \mathcal{M}^\text{BH,SL}_0\mu^{4-D}\nonumber\\
    &\phantom{=}\times\int\frac{d^D l}{(2\pi)^D}\frac{1}{\left[l^2+2 k \cdot l\right]\left[l^2\right]\left[l^2+2k^\prime \cdot l\right]}\nonumber\\
    &= e^2 4 (k \cdot k^\prime) \frac{\pi^{D/2}}{\left( 2\pi \right)^D \Gamma\left( 1-\epsilon \right)} \nonumber\\
    &\phantom{=}\times C_0  \left( m^2, \left( k - k^\prime \right)^2, m^2;0,m^2,m^2 \right)\mathcal{M}^\text{BH,SL}_0.
\nonumber \\
    \label{eq:virspacecor}
\end{align}
Here and in the following $C_0$ denotes the scalar one-loop three-point function. We give an analytic expression of that function for the two different cases we need in this work in Appendix~\ref{app:threepoint}.
\begin{figure}
    \centering
    \includegraphics[scale=0.5]{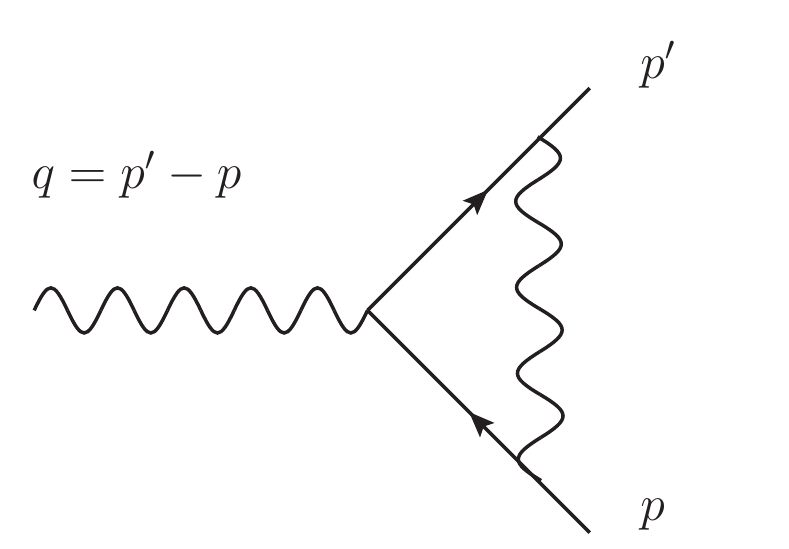}
    \includegraphics[scale=0.5]{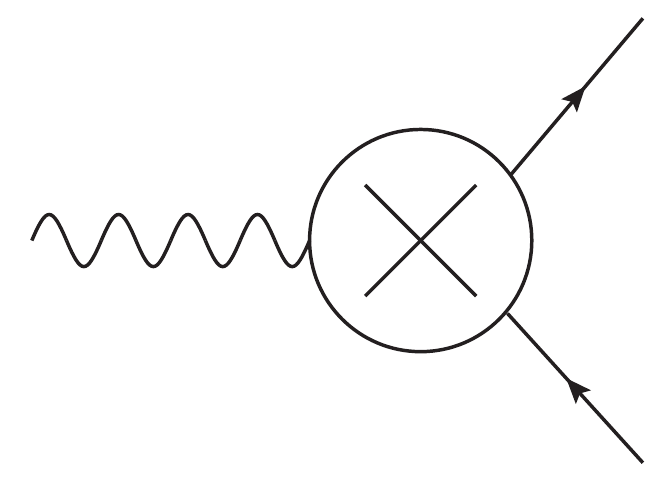}
    \caption{Left panel: one-loop vertex diagram. Right panel: vertex counter term.}
    \label{fig:one_loop_vertex}
\end{figure}

In addition to the contribution of Eq. \eqref{eq:virspacecor}, we also have to include the vertex counter term, which we show in Fig. \ref{fig:one_loop_vertex}. We are using the on-shell subtraction scheme, in which the counter term is defined to fix the electron charge $e$ at $q^2=0$. In the soft-photon approximation one has to extract only the IR divergent piece of the full expression, as has been done in Ref.~\cite{Heller:2018ypa}. To calculate the vertex counter term we consider the decomposition into the two form factors $F^e_D$ and $F^e_P$,
\begin{align}
    &\bar{u}(k')\Gamma^\mu u(k) \nonumber\\ &=\bar{u}(k')\left[(1+F^e_D(q^2))\gamma^\mu-iF^e_P(q^2)\sigma^{\mu\nu}\frac{q_\nu}{2m}\right]u(k),
\end{align}
where $q=k-k'$. In this decomposition only $F^e_D(q^2)$ is divergent. The renormalization constant $Z_1$ of the vertex is therefore given by
\begin{align}
Z_1&=1-F^e_D(0)=\nonumber\\
&=1-\frac{\alpha_{\rm{em}}}{4\pi}\left\{\left[\frac{1}{\epsilon_\text{UV}}-\gamma_E+\ln\left(\frac{4\pi\mu^2}{m^2}\right)\right]\right.\nonumber\\
&\phantom{=}\left.+2\left[\frac{1}{\epsilon_\text{IR}}-\gamma_E+\ln\left(\frac{4\pi\mu^2}{m^2}\right)\right]+4\right\},
\nonumber \\
\end{align}
yielding for the renormalized vertex $\tilde{\Gamma}$
\begin{equation}
    \tilde{\Gamma}^\mu=\Gamma^\mu +(Z_1-1)\gamma^\mu.
\end{equation}

Since we work in the soft-photon approximation, we only extract the infrared divergent part of the full one-loop renormalized vertex which can be found for example in Ref. \cite{Vanderhaeghen:2000ws} and find
\begin{equation}
    \tilde{\Gamma}^\mu_s = -\frac{\alpha_{\rm{em}}}{2\pi} \gamma^\mu \left[\frac{1}{\epsilon_\text{IR}}-\gamma_E+\ln\left(\frac{4\pi\mu^2}{m^2}\right)\right].\label{eq:vertex_CT_IR}
\end{equation}
After adding the vertex counter term to Eq.~\eqref{eq:virspacecor} and evaluating the three-point function, the infrared divergent part of the  virtual correction to the cross section of the spacelike process is given by
\begin{align}
    \nonumber
    \delta^\text{BH,SL}_{a,\text{IR}}=&-\frac{\alpha_{\rm{em}}}{\pi}\left[\left(\frac{1+\beta_Q^2}{2\beta_Q}\right)\ln \left(\frac{\beta_Q-1}{\beta_Q+1}\right)+1\right]\\ &\times\left[\frac{1}{\epsilon_\text{IR}}-\gamma_E+\ln\left(\frac{4\pi\mu^2}{m^2}\right)\right],\label{eq:class(a)IR}
\end{align}
and the finite part by
\begin{align}
    \nonumber
    \delta^\text{BH,SL}_a=&-\frac{\alpha_{\rm{em}}}{\pi}\left(\frac{1+\beta_Q^2}{2\beta_Q}\right)\bigg\{2\;\text{Li}_2\left(\frac{\beta_Q-1}{2\beta_Q}\right)\\&+\ln^2\left(\frac{\beta_Q-1}{2\beta_Q}\right)-\frac{1}{2}\ln^2\left(\frac{\beta_Q-1}{\beta_Q+1}\right)-\frac{\pi^2}{6}\bigg\},\label{eq:class(a)}
\end{align}
where $\beta_Q=\sqrt{1+\frac{4m^2}{Q^2}}$.

Note that here and in the following, we define $\delta$ to be the correction on the level of the cross section, not the amplitude. This corresponds to taking twice the real part of the correction on the level of the amplitude. 

For the crossed diagrams with $l_-$ and $l_+$ interchanged we find the same result as in Eqs. \eqref{eq:class(a)IR} and \eqref{eq:class(a)}. In the limit of a small electron mass, i.e. $m\ll Q^2$, the correction simplifies to
\begin{align}
    \delta^\text{BH,SL}_{a,\rm{IR}}=&\frac{\alpha_{\rm{em}}}{\pi}\left[\ln\frac{Q^2}{m^2}-1\right]\left[\frac{1}{\epsilon_\text{IR}}-\gamma_E+\ln\left(\frac{4\pi\mu^2}{m^2}\right)\right],\nonumber\\
    \delta^\text{BH,SL}_a=&\frac{-\alpha_{\rm{em}}}{\pi}\left[\frac{1}{2}\ln^2\frac{Q^2}{m^2}-\frac{\pi^2}{6}\right].
    \nonumber \\
\end{align}

\subsubsection{Contributions of class (b)}
 
 Here we calculate all contributions to the spacelike process, for which the soft-photon is attached to the di-lepton line. The Feynman diagram corresponding to this correction is shown in Fig.~\ref{fig:virdiagramsslab} on the right. The matrix element is given by
\begin{align}
\label{eq:matrixelement}
&\mathcal{M}_b^\text{BH,SL}=-\frac{e^6}{Q^2 t}\bar{N}\left( p' \right) \Gamma^\alpha\left( p',p \right) N\left(  p \right)  \bar{u}\left(k'\right)\gamma^\mu u\left( k\right) \mu^{4-D} \nonumber\\
 &\phantom{=}\times   \int_{}^{} \frac{d^Dl}{\left( 2\pi \right)^D} \frac{\Bar{u}\left( l_- \right)\gamma^\beta   \left(-\slashed{l} +\slashed{l}_-  +m_l \right)}{\left[ \left( l-l_- \right)^2-m_l^2 \right]}  \nonumber\\ 
 &\phantom{=}\times \frac{\gamma_\mu \left(   -\slashed{l} + \slashed{l}_- -\slashed{q}+m_l\right)\gamma_\alpha(-\slashed{l}-\slashed{l}_++m_l) \gamma_\beta v\left( l_+  \right)}{[ \left( l - l_-  + q \right)^2 -m_l^2 ][(l+l_+)^2-m_l^2]\left[l^2\right]} ,
\end{align}
which in the soft-photon approximation reduces to
\begin{align}
    &\mathcal{M}_b^\text{BH,SL}= i e^2 4 (l_- \cdot l_+) \mathcal{M}^\text{BH,SL}_0\mu^{4-D} \nonumber\\
    &\phantom{=}\times\int\frac{d^D l}{(2\pi)^D}\frac{1}{\left[l^2-2l_- \cdot l\right]\left[l^2\right]\left[l^2+2l_+ \cdot l\right]}\nonumber\\
    &=- e^2 4 (l_- \cdot l_+)  \frac{\pi^{D/2}}{\left( 2\pi \right)^D \Gamma\left( 1-\epsilon \right)} \nonumber\\
    &\phantom{=}\times  C_0  \left( m_l^2, \left(l_- + l_+ \right)^2, m_l^2;0,m_l^2,m_l^2 \right)\mathcal{M}^\text{BH,SL}_0. 
    \nonumber\\
    \label{eq:realspacecor}
\end{align}

\begin{figure}
\centering
\begin{minipage}[b]{0.45\linewidth}
\includegraphics[scale=0.5]{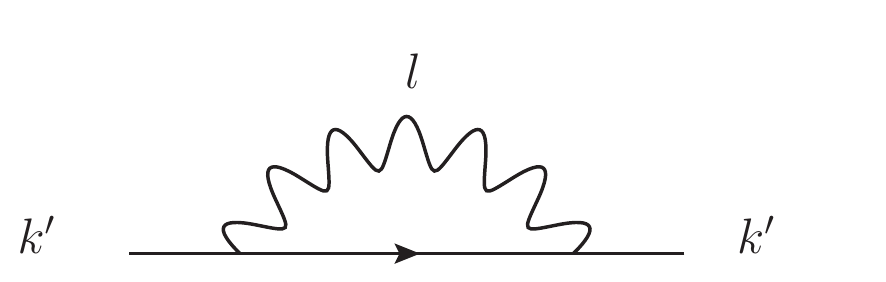}
\end{minipage}
\quad
\begin{minipage}[b]{0.45\linewidth}
\vspace{-3cm} \includegraphics[scale=0.5,clip]{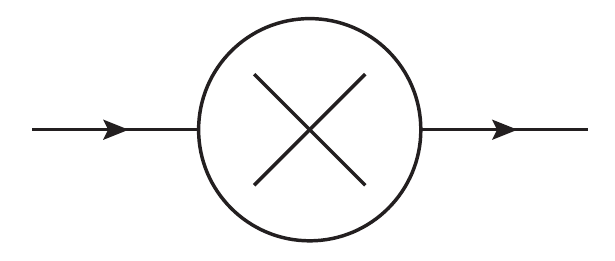}
\end{minipage}
\caption{Left panel: fermion self-energy at one-loop order. Right panel: counter term for fermion self energy}
\label{fig:se_ct}
\end{figure}

As for the class (a) contribution we need to include counter terms. In addition to the infrared divergent piece of the vertex counter term as given by Eq. \eqref{eq:vertex_CT_IR}, we also need the counter term of the fermion self energy, which is shown in Fig. \ref{fig:se_ct}. To first order in $\alpha_{\rm{em}}$, the self-energy of a fermion with mass $m_f$ and momentum $k'$ is calculated as
\begin{equation}
-i\Sigma(\sla{k'})=-e^2\mu^{4-D}\int\frac{d^D l}{(2\pi)^D}\frac{\gamma^{\alpha}(\sla{k}'+\sla{l}+m_f)\gamma_{\alpha}}{[(k'+l)^2-m_f^2]\;[l^2]}.\label{eq:fermion_SE}
\end{equation}
Eq. \eqref{eq:fermion_SE} has a UV divergence, which needs to be subtracted by an appropriate counter term. In the on-shell scheme this counter term is fixed by requiring that the fermion self-energy $\Sigma(k')$ has a pole at $k^{\prime 2}=m_f^2$ with residue equal to one. This fixes the wave-function renormalization constant $Z_2$ and the mass renormalization constant $Z_{m_f}$:
\begin{align}
Z_2&=1+\left.\frac{d\;\Sigma (\sla{k'})}{d\sla{k'}}\right|_{\sla{k'}\;=m_f},\\
(1-Z_{m_f})Z_2m_f&=\Sigma(m_f).
\end{align}
The evaluation of $\Sigma(k')$ and its derivative, results in the renormalization constants
\begin{align}
Z_2&=1-\frac{\alpha_{\rm{em}}}{4\pi}\left\{\left[\frac{1}{\epsilon_\text{UV}}-\gamma_E+\ln\left(\frac{4\pi\mu^2}{m_f^2}\right)\right]\right.\nonumber\\
&\left.+2\left[\frac{1}{\epsilon_\text{IR}}-\gamma_E+\ln\left(\frac{4\pi\mu^2}{m_f^2}\right)\right]+4\right\}, \nonumber \\
\\
Z_2Z_{m_f}&=1-\frac{\alpha_{\rm{em}}}{4\pi}\left\{4\left[\frac{1}{\epsilon_\text{UV}}-\gamma_E+\ln\left(\frac{4\pi\mu^2}{m_f^2}\right)\right]\right.\nonumber\\
&\left.+2\left[\frac{1}{\epsilon_\text{IR}}-\gamma_E+\ln\left(\frac{4\pi\mu^2}{m_f^2}\right)\right]+8\right\}.
\nonumber \\
\end{align}
The renormalized self-energy is then given by
\begin{equation}
\tilde{\Sigma}(k')=\Sigma(k')-(Z_2-1)\sla{k'}+(Z_2Z_m-1)m_f,
\end{equation}
and in the soft-photon limit, in which we only extract the IR divergence, we find
\begin{equation}
\tilde{\Sigma}_s(k')=\frac{\alpha_{\rm{em}}}{2\pi}\left(\sla{k'}-m_f\right)\left[\frac{1}{\epsilon_\text{IR}}-\gamma_E+\ln\left(\frac{4\pi\mu^2}{m_f^2}\right)\right].
\end{equation}

Adding the counter terms of the vertex and fermion self-energy to Eq.~\eqref{eq:realspacecor}, we find for the total contribution the infrared divergent part
\begin{align}
 \delta^{\text{BH,SL}}_{b,\text{IR}}=&\frac{ \alpha_{\rm{em}} }{ \pi } \left[ \frac{ 1 +\beta_{s_{ll}}^2 }{ 2\beta_{s_{ll}} }\ln\left( \frac{ 1+\beta_{s_{ll}} }{ 1-\beta_{s_{ll}} } \right) -1\right]\nonumber\\
    &\times  \left[ \frac{ 1 }{ \epsilon_{\rm{IR}} } -\gamma_E+\ln\left( \frac{ 4\pi\mu^2 }{m_l^2  }\label{eq:class(b)IR} \right)\right],
    \nonumber \\
\end{align}
and the finite part
\begin{align}
\delta^{\text{BH,SL}}_b=&-\frac{\alpha_{\rm{em}}}{\pi}
\left( \frac{1+\beta_{s_{ll}}^2}{2\beta_{s_{ll}}} \right)
\bigg[2\;\text{Li}_2\left(\frac{2\beta_{s_{ll}}}{\beta_{s_{ll}}+1}\right)\nonumber \\ &\hspace{2.cm}+\frac{1}{2}\ln^2\left(\frac{1-\beta_{s_{ll}}}{1+\beta_{s_{ll}}}\right)-\pi^2\bigg]. \nonumber \\
\label{eq:class(b)}
\end{align}
In the limit of small lepton masses, i.e. $m_l^2 \ll s_{ll}$, we find
\begin{align}
    \delta^{\text{BH,SL}}_{b,\text{IR}}&=\frac{\alpha_{\rm{em}}}{\pi}\left[\ln\frac{s_{ll}}{m_l^2}-1\right]\left[\frac{1}{\epsilon_\text{IR}}-\gamma_E+\ln\left(\frac{4\pi\mu^2}{m_l^2}\right)\right],\nonumber\\
    \delta^{\text{BH,SL}}_{b}&=-\frac{\alpha_{\rm{em}}}{\pi}\left[\frac{1}{2}\ln^2\frac{s_{ll}}{m_l^2}-\frac{2}{3}\pi^2\right]. 
    \nonumber \\
\end{align}

\subsubsection{Contributions of class (c)}

\begin{figure}[h]
\includegraphics[scale=0.5]{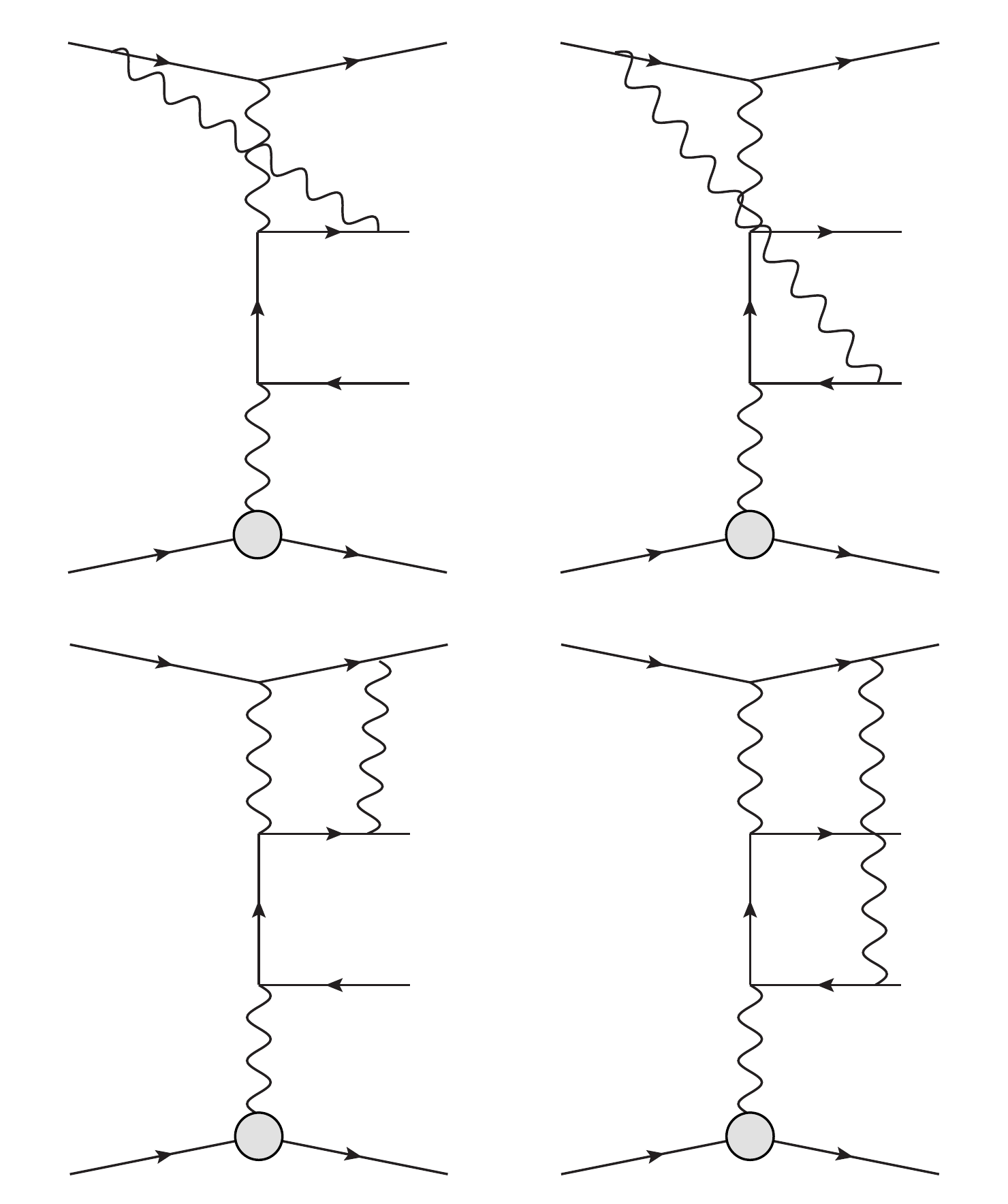}
  \caption{Virtual photon corrections of class (c) to the spacelike BH proces. The crossed diagrams with $l_-$ and $l_+$ interchanged yield the same correction.}
    \label{fig:virdiagramsslc}
\end{figure}

 In this section we calculate all diagrams, in which a soft-photon connects the electron line with the di-lepton line. We show the contributing diagrams of this class in Fig.~\ref{fig:virdiagramsslc}. For the contribution of class (c) no counter term diagrams have to be considered. 

 The first diagram in Fig.~\ref{fig:virdiagramsslc} is calculated as
\begin{align}
&\mathcal{M}^\text{BH,SL}_{c_1}=\frac{e^6}{t}\Bar{N}\left( p' \right) \Gamma^\alpha\left( p',p \right) N\left(  p \right)  \mu^{4-D} \nonumber\\
&\times \int_{}^{} \frac{d^Dl}{\left( 2\pi \right)^D} \frac{ \Bar{u}\left( k' \right) \gamma^\mu \left( \slashed{l} + \slashed{k} +m \right) \gamma^\beta u\left( k \right)  }{ \left[ ( l + k )^2 -m^2 \right] \left[ ( l + k-k' )^2 \right] \left[ l^2 \right]}\nonumber \\
&\times \frac{\Bar{u}\left(l_- \right) \gamma_\beta \left( \slashed{l}_- + \slashed{l}  + m_l \right) \gamma_\mu \left( \slashed{l}_-- \slashed{q} + m_l  \right) \gamma_\alpha v \left( l_+\right)}{ \left[ ( l+l_-)^2 -m_l^2 \right]\left[ ( l_- - q )^2 -m_l^2 \right] },
\end{align}
which in the soft-photon limit can be reduced to
\begin{eqnarray}
   \mathcal{M}^\text{BH,SL}_{c_1}&=&- ie^2 4 (k \cdot l_-) \mu^{4-D} \mathcal{M}^\text{BH,SL}_0 \nonumber \\ 
   &\times& \int_{}^{} \frac{d^Dl}{\left( 2\pi \right)^D} \frac{1}{ \left[  l^2\right] \left[ l^2 + 2 k \cdot l \right] \left[  l^2 +2l_- \cdot l\right]}\nonumber\\
   &=& e^2 4 (k \cdot l_-) \frac{\pi^{D/2}}{\left( 2\pi \right)^D \Gamma\left( 1-\epsilon \right)} \nonumber \\
   &\times& 
   C_0  \left( m^2, \left( k - l_- \right)^2, m_l^2;0,m^2,m_l^2 \right)\mathcal{M}^\text{BH,SL}_0. \nonumber \\
   \label{eq:slc1}
\end{eqnarray}
Evaluating the three-point function $C_0$, we find that the infrared-divergent part is given by:
\begin{align}
    \delta^\text{BH,SL}_{c_1,\text{IR}}=& \frac{\alpha_{\rm{em}}}{\pi}\frac{k \cdot l_-}{\lambda_{kl_-} (k-l_-)^2} \ln \left(\frac{\gamma_{kl_-} ^- (1-\gamma_{kl_-} ^+)}{(1-\gamma_{kl_-}
   ^-) \gamma_{kl_-} ^+}\right) \nonumber \\ \times & \left[ \frac{ 1 }{ \epsilon_{\rm{IR}} } -\gamma_E+\ln\left( \frac{ 4\pi\mu^2 }{m^2} \right)\right],
   \nonumber \\
   \label{eq:c1contributionIR}
\end{align}
and the finite part is given by:
\begin{widetext}
\begin{align}
    \delta^\text{BH,SL}_{c_1}=&
    \frac{\alpha_{\rm{em}}}{\pi}\frac{k \cdot  l_-}{\lambda_{kl_-}(k-l_-)^2 } \bigg\{ -\ln(-\lambda_{kl_-}) \left[ \ln \left(\frac{\gamma^+_{kl_-}-1}{\gamma^+_{kl_-}} \right)  + \ln \left(\frac{\gamma^-_{kl_-}}{\gamma^-_{kl_-}-1} \right)\right] + \frac{1}{2}    \ln^2 \left(  -\gamma_{kl_-}^+  \right)  -\frac{1}{2} \ln^2 \left(  \gamma_{kl_-}^-  \right) \nonumber \\ 
    &-  \frac{1}{2}  \ln^2 \left( 1 - \gamma_{kl_-}^+ \right)  + \frac{1}{2} \ln^2 \left(  \gamma_{kl_-}^- -1    \right)  - \text{Li}_2\left( \frac{1- \gamma_{kl_-}^-}{\lambda_{kl_-}}\right) - \text{Li}_2 \left(\frac{\gamma_{kl_-}^+}{\lambda_{kl_-}} \right) + \text{Li}_2 \left( \frac{\gamma_{kl_-}^+ -1} {\lambda_{kl_-}} \right) + \text{Li}_2 \left( \frac{\gamma_{kl_-}^-}{\lambda_{kl_-}} \right)\nonumber \\ 
    &  - \ln \left(\frac{\gamma_{kl_-} ^- (1-\gamma_{kl_-} ^+)}{(1-\gamma_{kl_-}
   ^-) \gamma_{kl_-} ^+}\right) \ln\left( \frac{-(k-l_-)^2}{m^2}\right)\Bigg\},
    \label{eq:c1contribution}
\end{align}
\end{widetext}
where
\begin{align}
    &\lambda_{kl_-}=\frac{2\sqrt{(k \cdot l_-)^2-m^2m_l^2}}{(k-l_-)^2},\nonumber \\
    &\gamma^{\pm}_{kl_-}=\left[\frac{m_l^2-k \cdot l_-}{(k-l_-)^2}\pm\frac{\lambda_{kl_-}}{2}\right].
\end{align}

We now consider two limits for this correction, in which the expressions simplify. The first limit corresponds to the case where the electron mass is small compared to all other scales. In this limit we find for the infrared divergent contribution
\begin{align}
    \delta^\text{BH,SL}_{c_1,\text{IR}}=&\frac{\alpha_{\rm{em}}}{\pi}\ln\left(\frac{2 k \cdot l_- }{m\; m_l}\right)\left[\frac{1}{\epsilon_\text{IR}}-\gamma_E+\ln\left(\frac{4\pi\mu^2}{m^2}\right)\right],
\end{align}
and for the finite contribution
\begin{align}
    &\delta^\text{BH,SL}_{c_1} = \frac{\alpha_{\rm{em}}}{2 \pi}\bigg\{\frac{1}{2}\ln^2\left(\frac{m^2}{2k \cdot  l_-}\right)-\frac{1}{2}\ln^2\left(\frac{2k \cdot l_-}{2k \cdot l_- -m_l^2}\right)\nonumber\\
    &+\frac{1}{2}\ln^2\left(\frac{m_l^2}{2k \cdot l_- -m_l^2}\right)-\ln\left(\frac{4(k \cdot l_-)^2}{m^2 m_l^2}\right)\ln\left(\frac{2 k \cdot l_-}{m^2}\right)\nonumber\\
    &+\text{Li}_2 \left( \frac{2k \cdot l_--m_l^2}{2k \cdot l_-} \right)-\text{Li}_2 \left( \frac{m_l^2}{2k \cdot l_-} \right)+\frac{\pi^2}{6}\bigg\}.
\end{align}
If in addition to $m^2\ll k \cdot l_-$, also $m_l=m$, i.e. considering electron pair production, we find
\begin{align}
    \delta^\text{BH,SL}_{c_1,\text{IR}} =& \frac{\alpha_{\rm{em}}}{\pi}\ln\left(\frac{2 k \cdot l_-}{m^2}\right)\left[\frac{1}{\epsilon_\text{IR}}-\gamma_E+\ln\left(\frac{4\pi\mu^2}{m^2}\right)\right],\nonumber\\
    \delta^\text{BH,SL}_{c_1} = -&\frac{\alpha_{\rm{em}}}{ \pi}\bigg\{\frac{1}{2}\ln^2\frac{2k \cdot l_-}{m^2}-\frac{\pi^2}{6}\bigg\}.\label{eq:c1}
\end{align}

The second diagram in the first row of Fig.~\ref{fig:virdiagramsslc} can be related to the previous one using Eq.~\eqref{eq:slc1} with the replacement $l_-\to l_+$ together with a sign change,
\begin{eqnarray}
\mathcal{M}^\text{BH,SL}_{c_2}&=&- e^2 4 (k \cdot l_+) \frac{\pi^{D/2}}{\left( 2\pi \right)^D \Gamma\left( 1-\epsilon \right)} \nonumber \\
&\times&  C_0  \left( m^2, \left( k - l_+ \right)^2, m_l^2;0,m^2,m_l^2 \right)\mathcal{M}^\text{BH,SL}_0. 
\nonumber \\
\end{eqnarray}
Therefore the correction on the level of the cross section is given by
\begin{align}
    \delta^\text{BH,SL}_{c_2,\text{IR}}=&-\delta^\text{BH,SL}_{c_1,\text{IR}}\Big\rvert_{l_- \to l_+},\quad \delta^\text{BH,SL}_{c_2}=&-\delta^\text{BH,SL}_{c_1}\Big\rvert_{l_- \to l_+}.\label{eq:c2}
\end{align}

The first diagram in the second row of Fig.~\ref{fig:virdiagramsslc} is given by
\begin{align}
&\mathcal{M}^\text{BH,SL}_{c_3}=\frac{e^6}{t}\bar{N}\left( p' \right) \Gamma^\alpha\left( p',p \right) N\left(  p \right)  \mu^{4-D} \nonumber
\\
&\phantom{=}\times \int_{}^{} \frac{d^Dl}{\left( 2\pi \right)^D} \frac{ \Bar{u}\left( k' \right) \gamma^\beta \left( \slashed{l} + \slashed{k'} +m \right) \gamma^\mu u\left( k \right)  }{ \left[ ( l + k' )^2 -m^2 \right] \left[ ( l -k+k' )^2 \right] \left[ l^2 \right]} \nonumber\\ 
&\phantom{=}\times\frac{\Bar{u}\left( l_- \right)\gamma_\beta  \left( \slashed{l}_- -\slashed{l}  +m_l \right)\gamma_\mu \left(  \slashed{l}_- -\slashed{q}+m_l\right) \gamma_\alpha v\left( l_+  \right)}{\left[ ( l-l_- )^2-m_l^2 \right]\left[ ( l_- - q )^2 -m_l^2 \right]},
\end{align}
which reduces to
\begin{eqnarray}
\mathcal{M}_{c_3}^\text{BH,SL} &=& - i e^2 4 (k' \cdot l_-) \mu^{4-D} \mathcal{M}^\text{BH,SL}_0 \nonumber \\ 
&\times& \int_{}^{} \frac{d^Dl}{\left( 2\pi \right)^D} \frac{  1}{\left[ l^2 + 2 k' l \right] \left[ l^2 - 2 l_- l \right]\left[ l^2 \right]} \nonumber \\ 
&=& e^2 4 (k' \cdot l_-) \frac{\pi^{D/2}}{\left( 2\pi \right)^D \Gamma\left( 1-\epsilon \right)}\nonumber \\ 
&\times& C_0  \left( m^2, \left( k' + l_- \right)^2, m_l^2;0,m^2,m_l^2 \right)\mathcal{M}^\text{BH,SL}_0.
\nonumber \\
\end{eqnarray}
In this case, the second argument of the three-point function is positive. Therefore, an analytic continuation of this function to the timelike region has to be performed. This yields
\begin{align}
    \delta^\text{BH,SL}_{c_3,\text{IR}}=&  \frac{\alpha_{\rm{em}}}{\pi}\frac{ (k' \cdot l_-)}{ \tilde{\lambda}_{k'l}(k^\prime+l_-)^2 } \ln \left(\frac{\tilde{\gamma}_{k'l_-} ^- (1-\tilde{\gamma}_{k'l_-} ^+)}{(1-\tilde{\gamma}_{k'l_-}
   ^-) \tilde{\gamma}_{k'l_-} ^+}\right) \nonumber \\ \times&  \left[ \frac{ 1 }{ \epsilon_{\rm{IR}} } -\gamma_E+\ln\left( \frac{ 4\pi\mu^2 }{m^2} \right)\right], 
\end{align}
and
\begin{widetext}
\begin{align}
    \delta^\text{BH,SL}_{c_3}= & \frac{\alpha_{\rm{em}}}{\pi}\frac{ (k' \cdot l_-)}{ \tilde{\lambda}_{k'l}(k^\prime+l_-)^2 }  \bigg\{\frac{1}{2} \ln ^2\left(\frac{\tilde{\lambda}_{k'l} }{ 1-\tilde{\gamma}_{k'l_-} ^+}\right)+\ln ^2\left(1-\tilde{\gamma}_{k'l_-} ^-\right)-\ln^2\left(\tilde{\gamma}_{k'l_-} ^-\right)-\ln ^2\left(1-\tilde{\gamma}_{k'l_-} ^+\right)+\ln ^2\left(\tilde{\gamma}_{k'l_-}^+\right)  \nonumber \\ 
    & +\text{Li}_2\left(\frac{\tilde{\lambda}_{k'l} }{1-\tilde{\gamma}_{k'l_-}^-}\right)+\text{Li}_2\left(-\frac{ \tilde{\gamma}_{k'l_-} ^-}{\tilde{\lambda}_{k'l}}\right)+\frac{1}{2} \ln ^2\left(\frac{\tilde{\lambda}_{k'l} }{\tilde{\gamma}_{k'l_-} ^-}\right)+\text{Li}_2\left(\frac{ \tilde{\gamma}_{k'l_-} ^+-1}{\tilde{\lambda}_{k'l}
   }\right) +\text{Li}_2\left(\frac{\tilde{\lambda}_{k'l} }{\tilde{\gamma}_{k'l_-} ^+}\right)-\frac{5 \pi ^2}{3}  \nonumber \\ 
    & -\ln \left(\frac{\tilde{\gamma}_{k'l_-} ^- (1-\tilde{\gamma}_{k'l_-} ^+)}{(1-\tilde{\gamma}_{k'l_-}
   ^-) \tilde{\gamma}_{k'l_-} ^+}\right)\ln\left(\frac{(k'+l)^2}{m^2}\right)\Bigg\},
\end{align}
\end{widetext}
where
\begin{align}
    &\tilde{\lambda}_{k^\prime l_-}=\frac{2\sqrt{(k^\prime \cdot l_-)^2-m^2m_l^2}}{(k^\prime+l_-)^2}, \nonumber \\
    &\tilde{\gamma}^{\pm}_{k^\prime l_-}=\left[\frac{m_l^2+k^\prime \cdot  l_-}{(k^\prime+l_-)^2}\pm\frac{\tilde{\lambda}_{k^\prime l_-}}{2}\right].
\end{align}
We consider the two limits like before. In the limit of a small electron mass, we find for the infrared divergent contribution:
\begin{align}
    \delta^\text{BH,SL}_{c_3,\text{IR}}&=\frac{-\alpha_{\rm{em}}}{\pi}\ln\left(\frac{2k' \cdot l_-}{m\; m_l}\right)\left[\frac{1}{\epsilon_\text{IR}}-\gamma_E+\ln\left(\frac{4\pi\mu^2}{m^2}\right)\right],
\end{align}
and for the finite contribution:
\begin{widetext}
\begin{align}
    \delta^\text{BH,SL}_{c_3}&=\frac{\alpha_{\rm{em}}}{2\pi}\bigg\{\frac{1}{2}\ln^2\left(\frac{2k'l_-}{m_l^2}\right)-\ln^2\left(\frac{2k'l_-}{m^2}\right)+\ln^2\left(\frac{2k'l_-}{2k'l_-+m_l^2}\right)+\frac{1}{2}\ln^2\left(\frac{4(k'l_-)^2}{m^2(2k'l_-+m_l^2)}\right)\nonumber\\
    &\phantom{=}-\ln^2\left(\frac{m_l^2}{2k'l_-+m_l^2}\right)+\ln\left(\frac{m^2 m_l^2}{4(k'l_-)^2}\right)\ln\left(\frac{m^2}{2k'l_-+m_l^2}\right)+\text{Li}_2\left(-\frac{m_l^2}{2k'l_-}\right)+\text{Li}_2\left(\frac{2k'l_-}{2k'l_-+m_l^2}\right)-\frac{3}{2}\pi^2\bigg\}.
\end{align}
\end{widetext}
Considering electron production, $m_l=m$, we find
\begin{align}
    &\delta^\text{BH,SL}_{c_3,\text{IR}} = -\frac{\alpha_{\rm{em}}}{\pi}\ln\left(\frac{2k' \cdot l_-}{m^2}\right)\left[\frac{1}{\epsilon_\text{IR}}-\gamma_E+\ln\left(\frac{4\pi\mu^2}{m^2}\right)\right],\nonumber\\
    &\delta^\text{BH,SL}_{c_3} = \frac{\alpha_{\rm{em}}}{ \pi}\bigg\{\frac{1}{2}\ln^2\frac{2k' \cdot l_-}{m^2}-\frac{2}{3}\pi^2\bigg\}.
    \label{eq:c3}
\end{align}

For the second diagram in the second row of Fig.~\ref{fig:virdiagramsslc} we can derive the correction in the soft-photon approximation from the previous result, leading to:
\begin{align}
\delta^\text{BH,SL}_{c_4}&=-\delta^\text{BH,SL}_{c_3}\Big\rvert_{l_- \to l_+}.\label{eq:c4}
\end{align}

Note that from Eqs. \eqref{eq:c1}, \eqref{eq:c2}, \eqref{eq:c3} and \eqref{eq:c4} we can see that the sum of all class (c) corrections is \textit{anti-symmetric} with respect to interchanging $l^+ \leftrightarrow l^-$. This is in contrast to the contributions of class (a) and (b), which are \textit{symmetric} with respect to the interchange of $l^+$ and $l^-$.

\subsection{Corrections to the timelike Bethe-Heitler process}

In this section we calculate the soft-photon corrections for the timelike process. We show that they lead to exactly the same corrections as in the spacelike process.
\begin{figure}[h]
\includegraphics[scale=0.48]{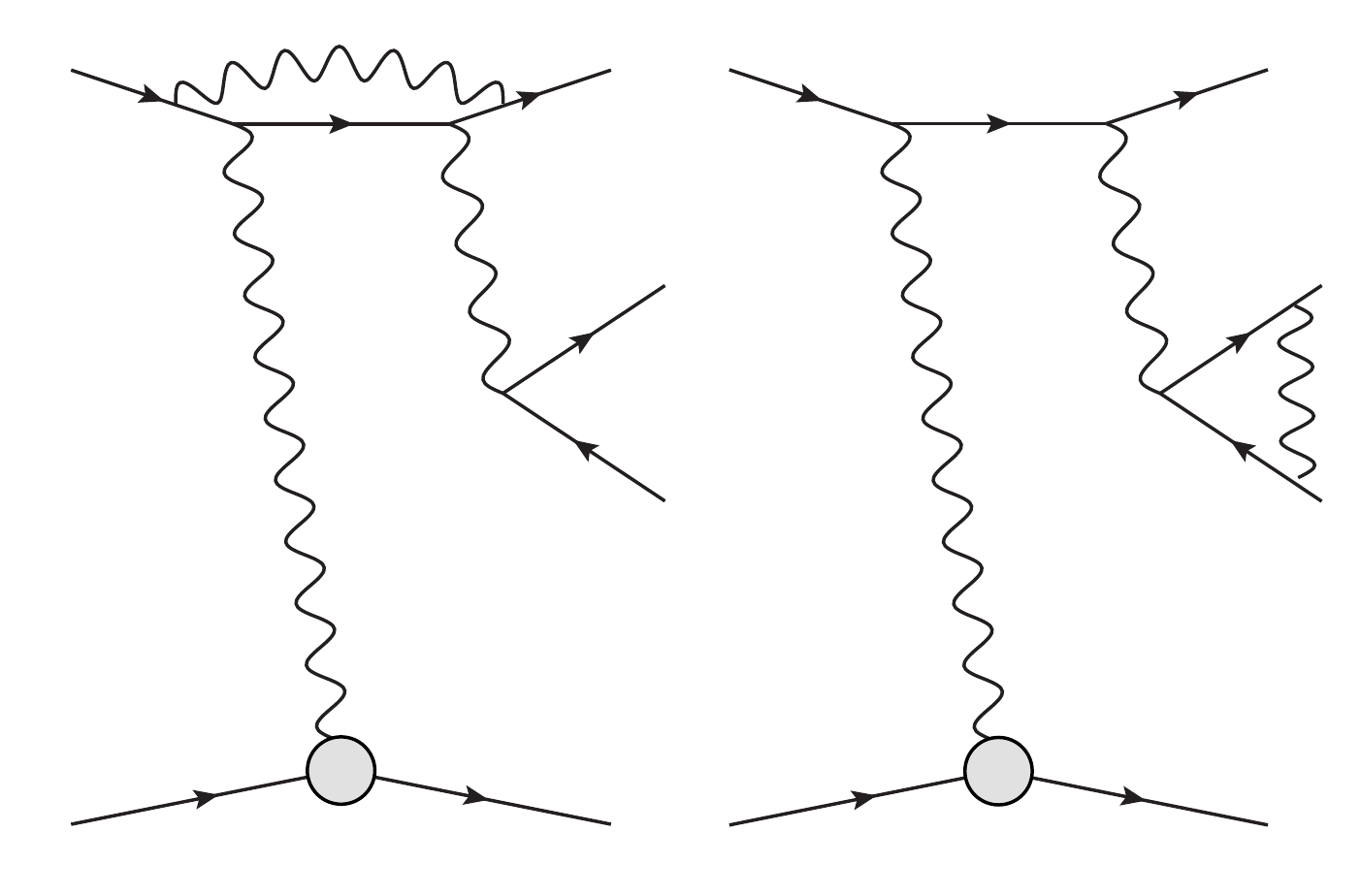}
  \caption{Virtual corrections of class (a) (left) and class (b) (right) to the timelike BH process. One also has to consider the corresponding counter-term diagrams. For the correction, the crossed diagrams with $\Delta$ and $q'$ interchanged yield the same result.}
    \label{fig:virdiagramstlab}
\end{figure}

The first diagram in Fig.~\ref{fig:virdiagramstlab} is given by:
\begin{align}
&\mathcal{M}^{\text{BH,TL}}_{a}=\frac{e^6}{ s_{ll} t}\Bar{N}\left( p' \right) \Gamma^\alpha\left( p',p \right) N\left(  p \right)  \Bar{u}\left( l_- \right)\gamma^\mu v\left( l_+ \right) \mu^{4-D} \nonumber\\ 
&\phantom{=}\times  \int_{}^{} \frac{d^Dl}{\left( 2\pi \right)^D}  \frac{\Bar{u}\left(k' \right) \gamma_\beta \left( \slashed{k'} 
+\slashed{l} + m \right) }{\left[ ( k' + l )^2 -m^2 \right]  } \nonumber \\ 
&\phantom{=}\times\frac{\gamma_\mu\left( \slashed{k}' + \slashed{q}'+ \slashed{l} + m  \right) \gamma_\alpha \left( \slashed{k} + \slashed{l} + m  \right) \gamma^\beta u \left( k \right)}{\left[ ( k' + q' + l )^2 -m^2 \right]\left[ ( k +l )^2 -m^2 \right] \left[l^2 \right]},
\nonumber \\
\end{align}
which in the soft-photon limit reduces to:
\begin{eqnarray}
\mathcal{M}_a^\text{BH,TL}&=&-ie^2 4 (k \cdot k') \mu^{4-D} \mathcal{M}^\text{BH,TL}_0 \nonumber\\
    &\times& \int_{}^{} \frac{d^Dl}{\left( 2\pi \right)^D} \frac{1}{ \left[  l^2\right] \left[ l^2 + 2 k \cdot l \right] \left[  l^2 +2k' \cdot l\right]}\nonumber\\
    &=& e^2 4 (k \cdot k') \frac{\pi^{D/2}}{\left( 2\pi \right)^D \Gamma\left( 1-\epsilon \right)} \nonumber\\  
    &\times& C_0  \left( m^2, \left(k - k' \right)^2, m^2;0,m^2,m^2 \right)\mathcal{M}^\text{BH,TL}_0.
    \nonumber \\
\end{eqnarray}
After adding the counter terms, on the level of the cross section the same correction as for the spacelike process is found:
\begin{align}
      \delta^\text{BH,TL}_{a}= \delta^\text{BH,SL}_{a}.
\end{align}

By the same argument, the second diagram in Fig.~\ref{fig:virdiagramstlab}, including counter terms, yields the same correction as for the spacelike process from Eqs.~\eqref{eq:class(b)IR} and \eqref{eq:class(b)}:
\begin{align}
      \delta^\text{BH,TL}_{b}=\delta^\text{BH,SL}_{b}.
\end{align}

\begin{figure}
\centering
\includegraphics[scale=0.45]{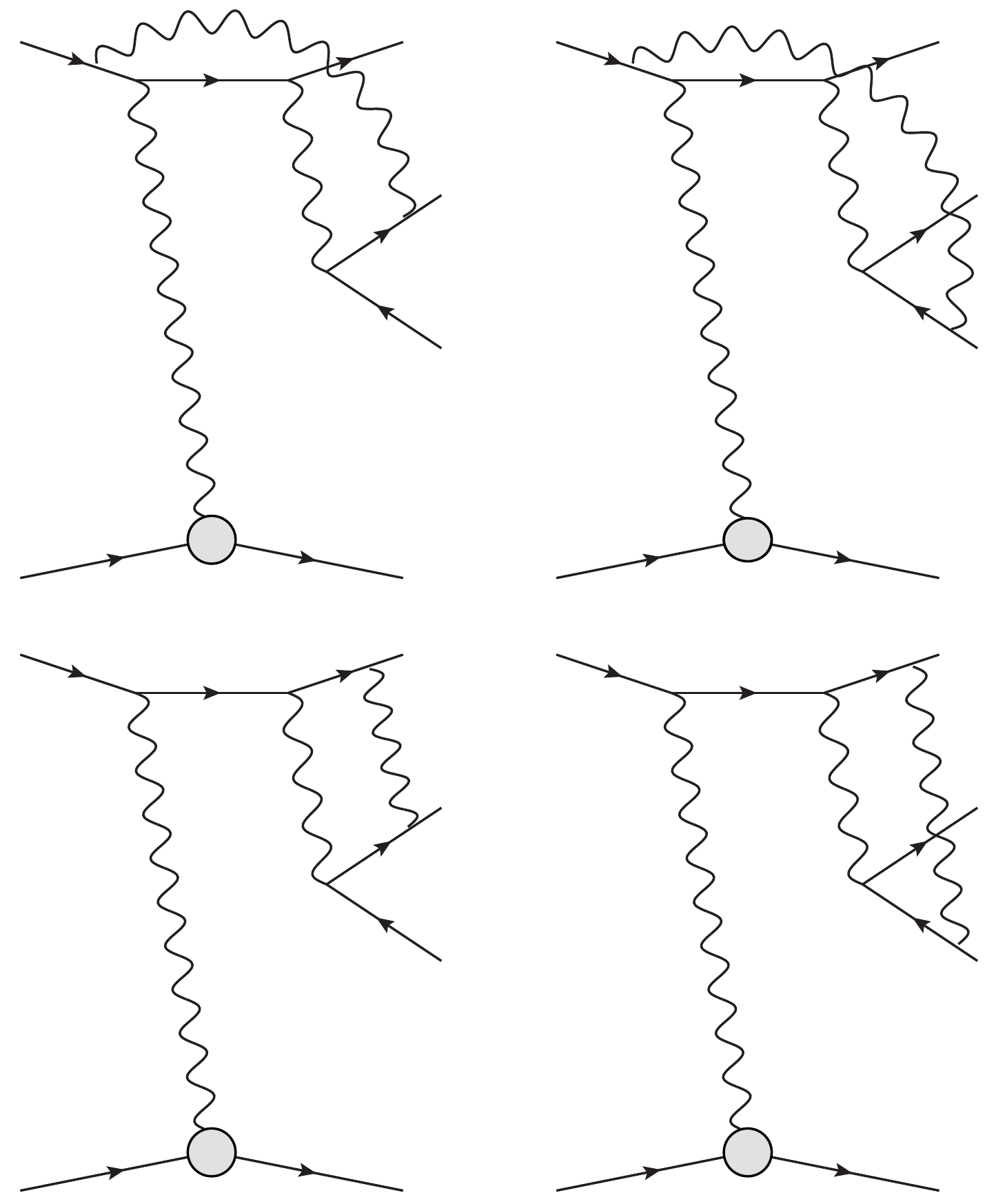}
\caption{Contributing diagrams from class (c) for the timelike Bethe-Heitler process. The crossed diagrams with $\Delta$ and $q'$ interchanged yield the same result.\label{fig:virdiagramstlc}} 
\end{figure}

The same argument also applies to the four diagrams of class (c), shown in  Fig.~\ref{fig:virdiagramstlc}, which yield the same correction as for the spacelike process: 
\begin{eqnarray}
      \delta^\text{BH,TL}_{c_1}&=& \delta^\text{BH,SL}_{c_1}, 
      \quad
      \delta^\text{BH,TL}_{c_2}= \delta^\text{BH,SL}_{c_2}, 
      \nonumber \\
      \delta^\text{BH,TL}_{c_3}&=& \delta^\text{BH,SL}_{c_3}, 
      \quad
      \delta^\text{BH,TL}_{c_4}= \delta^\text{BH,SL}_{c_4}.
\end{eqnarray}

\subsection{Corrections to the Compton process}

In this section we calculate the corrections for the Compton scattering, which in the soft-photon limit lead again to the same results as before for space-like and time-like Bethe-Heitler process. Therefore, on the level of the cross section, the correction in the soft-photon approximation 
can be factorized for the total process, and is given by:
\begin{eqnarray}
\delta^\text{dVCS}_{a} &=& \delta^\text{BH,SL}_{a}, \quad 
\delta^\text{dVCS}_{b} = \delta^\text{BH,SL}_{b}, \nonumber \\
\delta^\text{dVCS}_{c1} &=& \delta^\text{BH,SL}_{c1}, 
\quad
\delta^\text{dVCS}_{c2} = \delta^\text{BH,SL}_{c2}, \nonumber \\
\delta^\text{dVCS}_{c3} &=& \delta^\text{BH,SL}_{c3}, 
\quad \delta^\text{dVCS}_{c4}= \delta^\text{BH,SL}_{c4}.
\end{eqnarray}

\begin{figure}[h]
\includegraphics[scale=0.48]{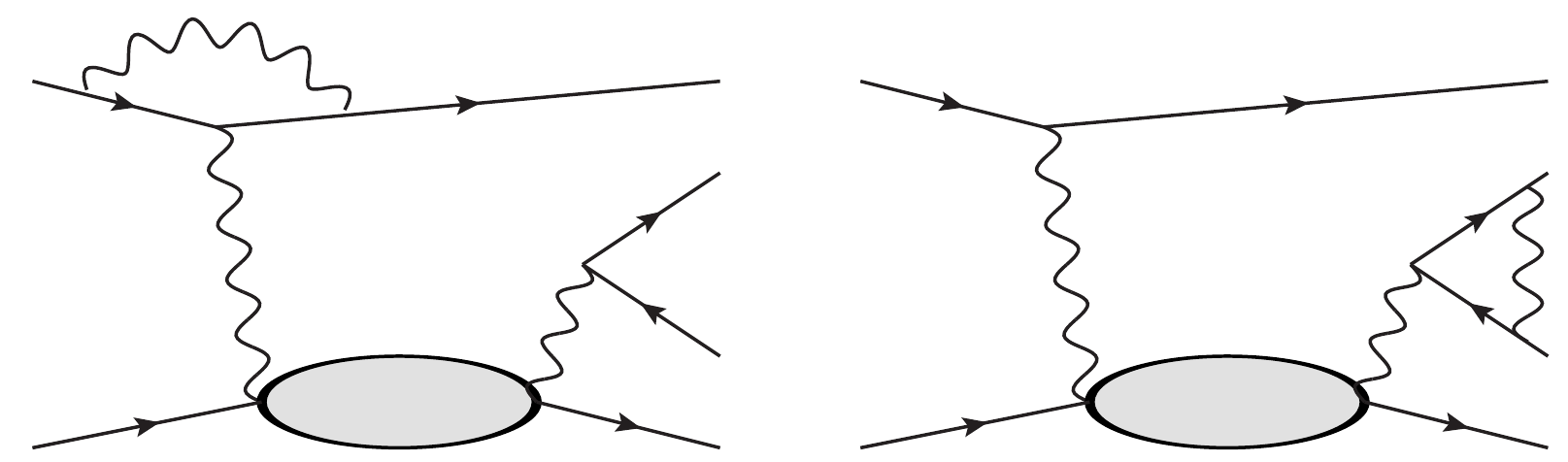}
\includegraphics[scale=0.48]{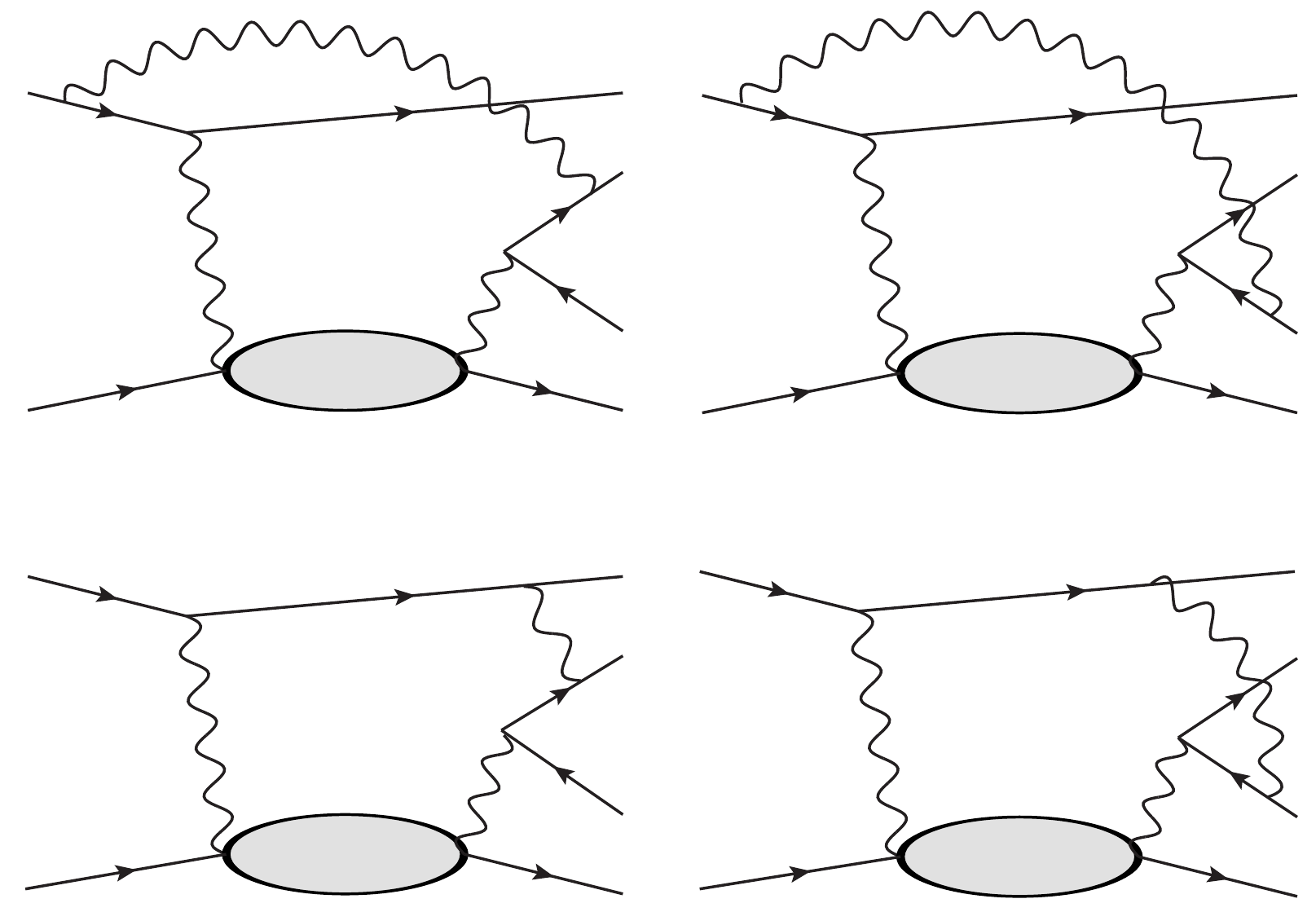}
  \caption{Virtual corrections of class (a) (top left), class (b) (top right), and class (c) (lower two rows) to the dVCS process. One also has to consider the corresponding counter-term diagrams. The crossed diagrams with $q$ and $q'$ interchanged yield the same result.}
    \label{fig:virdiagramscs}
\end{figure}

\subsection{Sum of all virtual soft-photon corrections}

Adding all contributions from classes (a), (b) and (c), we define the virtual soft-photon corrections on the cross section as:
\begin{equation}
    d\sigma_{s;v}=d\sigma_0 (1+\delta_{s;v}).
\end{equation}
The correction can be separated in the IR divergent contribution:
\begin{eqnarray}
    \delta_{s;v}^{\text{IR}}&=&\frac{\alpha_{\rm{em}}}{\pi}\bigg\{\bigg[ \ln \left( \frac{Q^2}{m^2} \right) +\ln\left(\frac{2 k \cdot l_-}{m\;m_l}\right)-\ln\left(\frac{2 k \cdot l_+}{m\;m_l}\right) \nonumber \\
    &&\hspace{1cm}-\ln\left(\frac{2 k' \cdot l_-}{m\;m_l}\right)+\ln\left(\frac{2 k' \cdot l_+}{m\;m_l}\right)-1\bigg] \nonumber\\
    &&\hspace{1cm}\times\bigg[\frac{1}{\epsilon_\text{IR}}-\gamma_E+\ln\left(\frac{4\pi\mu^2}{m^2}\right)\bigg]
    \nonumber \\
    &&\hspace{1cm}+\left[\left(\frac{1+\beta_{s_{ll}}}{1-\beta_{s_{ll}}}\right)\ln\left(\frac{1+\beta_{s_{ll}}}{1-\beta_{s_{ll}}}\right)-1\right] \nonumber \\
    &&\hspace{1cm}\times \left[\frac{1}{\epsilon_\text{IR}}-\gamma_E+\ln\left(\frac{4\pi\mu^2}{m_l^2}\right)\right]\bigg\}, 
    \nonumber \\
    \label{eq:IR_virt}
\end{eqnarray}
and a finite contribution:
\begin{equation}
    \delta_{s;v}=\delta_a+\delta_b+\delta_{c_1}+\delta_{c_2}+\delta_{c_3}+\delta_{c_4}.
    \label{eq:virt_finitesum}
\end{equation}
For convenience of the reader, we summarize all formulas derived in the previous sections:
\begin{widetext}
\begin{align}
     \delta_a &=-\frac{\alpha_{\rm{em}}}{\pi}\left\{\frac{1}{2}\ln^2 \left(\frac{Q^2}{m^2} \right)-\frac{\pi^2}{6}\right\},
     \label{eq:virta}\\
     \delta_b&=-\frac{\alpha_{\rm{em}}}{\pi} 
     \left(\frac{1+\beta_{s_{ll}}^2}{2\beta_{s_{ll}}} \right)
\bigg\{2\;\text{Li}_2\left(\frac{2\beta_{s_{ll}}}{\beta_{s_{ll}}+1}\right)+\frac{1}{2}\ln^2\left(\frac{1-\beta_{s_{ll}}}{1+\beta_{s_{ll}}}\right)-\pi^2\bigg\},
    \label{eq:virtb} \\
    \delta_{c_1} &= \frac{\alpha_{\rm{em}}}{2 \pi}\bigg\{\frac{1}{2}\ln^2\left(\frac{m^2}{2k \cdot  l_-}\right)-\frac{1}{2}\ln^2\left(\frac{2k \cdot l_-}{2k \cdot l_- -m_l^2}\right)+\frac{1}{2}\ln^2\left(\frac{m_l^2}{2k \cdot l_- -m_l^2}\right)-\ln\left(\frac{4(k \cdot l_-)^2}{m^2 m_l^2}\right) \ln\left(\frac{2 k \cdot l_-}{2 k \cdot l_- - m_l^2}\right) \nonumber \\
    &+\text{Li}_2 \left( \frac{2k \cdot l_--m_l^2}{2k \cdot l_-} \right)-\text{Li}_2 \left( \frac{m_l^2}{2k \cdot l_-} \right)+\frac{\pi^2}{6}\bigg\},
    \label{eq:virtc1} \\
    \delta_{c_2}&=-\delta_{c_1}\Big\rvert_{l_- \to l_+}, 
    \label{eq:virtc2}\\
    \delta_{c_3}&=\frac{\alpha_{\rm{em}}}{2\pi}\bigg\{\frac{1}{2}\ln^2\left(\frac{2k' \cdot l_-}{m_l^2}\right)-\ln^2\left(\frac{2k' \cdot l_-}{m^2}\right)+\ln^2\left(\frac{2k' \cdot l_-}{2k' \cdot l_-+m_l^2}\right)+\frac{1}{2}\ln^2\left(\frac{4(k' \cdot l_-)^2}{m^2(2k' \cdot l_-+m_l^2)}\right)\nonumber\\
    &-\ln^2\left(\frac{m_l^2}{2k'\cdot l_-+m_l^2}\right)+\ln\left(\frac{m^2 m_l^2}{4(k' \cdot l_-)^2}\right)\ln\left(\frac{m^2}{2k' \cdot l_-+m_l^2}\right)+\text{Li}_2\left(\frac{-m_l^2}{2k' \cdot l_-}\right)+\text{Li}_2\left(\frac{2k' \cdot l_-}{2k' \cdot l_-+m_l^2}\right)-\frac{3}{2}\pi^2\bigg\}, \quad \quad 
    \label{eq:virtc3} \\
    \delta_{c_4}&=-\delta_{c_3}\Big\rvert_{l_- \to l_+}. 
    \label{eq:virtc4}
\end{align}

For the electroproduction of an $e^-e^+$ pair, the formula simplify as:
\begin{eqnarray}
\delta_{s;v}&=&-\frac{\alpha_{\rm{em}}}{2\pi}\bigg\{\ln^2\left(\frac{Q^2}{m^2}\right)+\ln^2 \left(\frac{s_{ll}}{m^2} \right) -\ln^2\left(\frac{2k'\cdot l_-}{m^2}\right)
+\ln^2\left(\frac{2k' \cdot l_+}{m^2}\right) 
+\ln^2\left(\frac{2k\cdot l_-}{m^2}\right)-\ln^2\left(\frac{2k \cdot l_+}{m^2}\right)-\frac{5}{3}\pi^2\bigg\}. 
\nonumber \\
\end{eqnarray}
\end{widetext}

\section{Soft-photon bremsstrahlung}
\label{sec:bremsstrahlung}

To cancel the IR divergences of the virtual soft-photon corrections, we need to include soft real radiation (soft bremstrahlung). On the level of cross section the IR divergences cancel, resulting in a finite physical result. The contribution due to soft bremstrahlung stems from Feynman diagrams in which an additional soft photon is emitted from an external fermion line. Denoting the momentum of the fermion line with $l$ and the momentum of the soft photon by $k_\gamma$, this corresponds to the amplitude:
\begin{equation}
    \mathcal{M}_s=\pm e Q_f\frac{\varepsilon^\ast \cdot l}{k_\gamma \cdot l}\mathcal{M}_0,\label{eq:real_soft_amp}
\end{equation}
with the $+$ sign, if the fermion is outgoing and the $-$ sign if it is incoming, where $Q_f$ denotes the charge of the lepton, and where $\mathcal{M}_0$ denotes the amplitude without soft-photon emission. 

The evaluation of the bremstrahlung contribution requires integrating over the momentum of the unobserved soft photon up to an energy cutoff $\Delta E_s$. The integration has to be performed in a reference frame in which the dependence of the integral on the photon momentum is isotropic. The choice of this frame depends on the experimental condition. In the present work, we consider the process $e^- p\rightarrow e^- l^- l^+ p$ where the dilepton momenta $l^-$ and $l^+$ are measured and the scattered proton with momentum $p'$ remains unobserved. Thus, the bremstrahlung integral has to be performed in a system in the rest frame of the unobserved proton and soft-photon. Defining the missing momentum $p_m \equiv p' + k_\gamma$, this frame is defined by the condition $\vec{p}_m=0$. The bremstrahlung contribution to the cross section in this frame is given by:
\begin{widetext}
\begin{align}
 d\sigma_{s;r}=d\sigma_0 &\frac{-e^2}{(2\pi)^3}\int_{|\vec{k}_\gamma|<\Delta E_s}\frac{d^3\vec{k}_\gamma}{2k_\gamma^0}\left\{\frac{m_l^2}{(k_\gamma \cdot l_+)^2}+\frac{m_l^2}{(k_\gamma \cdot l_-)^2}+\frac{m^2}{(k_\gamma \cdot k)^2}+\frac{m^2}{(k_\gamma \cdot k')^2}-\frac{2l_+ \cdot l_-}{(k_\gamma \cdot l_+)(k_\gamma \cdot l_-)}\right.\nonumber\\
 &\left.-\frac{2k'\cdot l_+}{(k_\gamma\cdot k')(k_\gamma\cdot l_+)}+\frac{2k'\cdot l_-}{(k_\gamma\cdot k')(k_\gamma\cdot l_-)}-\frac{2k'\cdot k}{(k_\gamma\cdot k)(k_\gamma\cdot k')}-\frac{2k\cdot l_-}{(k_\gamma\cdot k)(k_\gamma\cdot l_-)}+\frac{2k\cdot l_+}{(k_\gamma\cdot k)(k_\gamma\cdot l_+)}
 \right\},\label{eq:softphotonint}
\end{align}
\end{widetext}
where the maximal soft-photon energy in that frame is denoted by $\Delta E_s$. The expression after performing the integration in Eq.~\eqref{eq:softphotonint} is lengthy and complicated in the general case. Here, we give explicit results only in the limit of a small electron mass, i.e. we only keep the logarithmic dependence on $m$. 
For the calculation, one considers the basic integral $I_{ij}$ corresponding to the interference of two terms  like Eq.~\eqref{eq:real_soft_amp} from two fermion lines. This basic integral has been worked out in \cite{tHooft:1978jhc} for a generic lepton mass. In the limit $m\rightarrow 0$, we find:
\begin{eqnarray}
    I_{ij} &\equiv& \int_{|\vec{k}_\gamma|<\Delta E_s}\frac{d^3\vec{k}_\gamma}{k_\gamma^0}\frac{p_i\cdot p_j}{(k_\gamma\cdot p_i)(k_\gamma\cdot p_j) }\nonumber\\
    &=&-2\pi\bigg\{\frac{1}{4}\left[\ln^2\left(\frac{m^2}{4E_i^2}\right)+\ln^2\left(\frac{m^2}{4E_j^2}\right)\right]\nonumber\\
    &+&\text{Li}_2\left(1-\frac{2E_i E_j}{p_i\cdot p_j}\right)+\frac{\pi^2}{3}\nonumber\\
    &+&\ln\left(\frac{2 p_i\cdot p_j}{m^2}\right)\left[\frac{1}{\epsilon_\text{IR}}-\gamma_E+\ln \left(\frac{4\pi\mu^2}{4\Delta E^2}\right)\right]\bigg\}. \quad
\end{eqnarray}
Using this expression and the general one for finite lepton masses from \cite{tHooft:1978jhc}, we can now easily perform the integration in Eq.~\eqref{eq:softphotonint}. Let us stress again that we only keep the dependence of the electron mass $m$ in the logarithms while for the lepton mass $m_l$ we keep the full dependence of the soft-photon integral. For the infrared divergent contribution we then find:
\begin{widetext}
\begin{align}
\delta^\text{IR}_\text{s;r}=&-\frac{\alpha_{\rm{em}}}{\pi}\bigg\{\bigg[\ln \left(\frac{Q^2}{m^2}\right)+\ln\left(\frac{2k\cdot l_-}{m\;m_l}\right)-\ln\left(\frac{2k'\cdot l_-}{m\;m_l}\right)+\ln\left(\frac{2k'\cdot l_+}{m\;m_l}\right)-\ln\left(\frac{2k\cdot l_+}{m\;m_l}\right)-1\bigg]
\bigg[\frac{1}{\epsilon_\text{IR}}-\gamma_E+\ln\left(\frac{4\pi\mu^2}{m^2}\right)\bigg]
\nonumber \\
&+\left[\left(\frac{1+\beta_{s_{ll}}^2}{2\beta_{s_{ll}}}\right)\ln\left(\frac{1+\beta_{s_{ll}}}{1-\beta_{s_{ll}}}\right)-1\right]\left[\frac{1}{\epsilon_\text{IR}}-\gamma_E+\ln\left(\frac{4\pi\mu^2}{m_l^2}\right)\right]\bigg\}, \label{eq:IR_real}
\end{align}
while for the finite part we find: 
\begin{align}
  \delta_\text{s;r}\equiv& \delta^\text{s;r}_a+\delta^\text{s;r}_b+\delta^\text{s;r}_c,  
  \label{eq:finite_real} \\
\delta^\text{s;r}_a=&-\frac{\alpha_{\rm{em}}}{\pi}\Bigg\{\ln\left(\frac{4 (\Delta E_s)^2}{m^2}\right)\left[1-\ln \left(\frac{Q^2}{m^2} \right)\right]
+\frac{1}{2}\ln \left(\frac{m^2}{4\tilde{E'}^2}\right)
+\frac{1}{2}\ln \left(\frac{m^2}{4\tilde{E}^2}\right)
+\frac{1}{4}\ln^2\left(\frac{4\tilde{E'}^2}{m^2}\right)+\frac{1}{4}\ln^2\left(\frac{4\tilde{E}^2}{m^2}\right)
\nonumber\\
 &+\text{Li}_2\left(1-\frac{4\tilde{E}\tilde{E'}}{Q^2}\right)+\frac{\pi^2}{3}\Bigg\},
 \label{eq:reala}
\\
\delta^\text{s;r}_b=&-\frac{\alpha_{\rm{em}}}{\pi}\left\{\ln\left(\frac{4 (\Delta E_s)^2}{m_l^2}\right)\left[1-\left(\frac{1+\beta_{s_{ll}}^2}{2\beta_{s_{ll}}}\right)\ln\left(\frac{1+\beta_{s_{ll}}}{1-\beta_{s_{ll}}}\right)\right]\right.+ \frac{1}{2 \tilde \beta_-}\ln\left(\frac{1- \tilde \beta_-}{1+ \tilde \beta_-}\right)
 + \frac{1}{2 \tilde \beta_+}\ln\left(\frac{1- \tilde \beta_+}{1+ \tilde \beta_+}\right)
 \nonumber \\
 &- \left(\frac{1+\beta_{s_{ll}}^2}{2\beta_{s_{ll}}}\right)
 \left[ \frac{1}{4} \ln^2 \left(\frac{1-\tilde \beta_-}{1+ \tilde \beta_-}\right)
 - \frac{1}{4} \ln^2 \left(\frac{1-\tilde \beta_+}{1+ \tilde \beta_+}\right)
 \right. + \text{Li}_2\left(1 - \left(\frac{1+\beta_{s_{ll}}}{1-\beta_{s_{ll}}}\right) \frac{\tilde E_-}{v} (1 + \tilde \beta_-) \right) \nonumber \\
 &+ \text{Li}_2\left(1 - \left(\frac{1+\beta_{s_{ll}}}{1-\beta_{s_{ll}}}\right) \frac{\tilde E_-}{v} (1 - \tilde \beta_-) \right) - \text{Li}_2\left(1 - \frac{\tilde E_+}{v} (1 + \tilde \beta_+) \right)  \left. \left. - \text{Li}_2\left(1 - \frac{\tilde E_+}{v} (1 - \tilde \beta_+) \right) 
 \right]\right\},
 \label{eq:realb}
\\
\delta^\text{s;r}_c=&-\frac{\alpha_{\rm{em}}}{\pi}\Bigg\{\ln\left(\frac{4 (\Delta E_s)^2}{m^2}\right)\ln\left(\frac{k' \cdot l_-}{k \cdot l_-}\right) \nonumber \\
&-\Bigg[\text{Li}_2\left(1-\frac{\tilde{E'}}{k'\cdot l_-}\left(\tilde E_--\sqrt{\tilde E_-^2-m_l^2}\right)\right)-\text{Li}_2\left(1-\frac{\tilde{E}}{k \cdot l_-}\left(\tilde E_--\sqrt{\tilde E_-^2-m_l^2}\right)\right)\nonumber\\
&+\text{Li}_2\left(1-\frac{\tilde{E}'}{k' \cdot l_-}\left(\tilde E_-+\sqrt{\tilde E_-^2-m_l^2}\right)\right)-\text{Li}_2\left(1-\frac{\tilde{E}}{k \cdot l_-}\left(\tilde E_-+\sqrt{\tilde E_-^2-m_l^2}\right)\right)\Bigg]\Bigg\}-\bigg(l_-\rightarrow l_+\bigg),
\label{eq:realc}
\end{align}
\end{widetext}
where
\begin{eqnarray}
    v &\equiv& \frac{\beta_{s_{ll}} s_{ll}}{2 (\tilde E_- - \frac{1 - \beta_{s_{ll}}}{1 + \beta_{s_{ll}}} \tilde E_+)}, \\
    \tilde \beta_\mp &=& \left(1 - m_l^2 / \tilde E_\mp^2 \right)^{1/2}, 
\end{eqnarray}
$\tilde{E}_{\pm}$ denotes the energy of the lepton with momentum $l^\pm$ in the rest frame of the soft-photon and recoil proton, and where $\tilde E$ ($\tilde{E}'$) denotes the energy of the electron with momentum $k$ ($k'$) in the same system.

Adding Eqs. \eqref{eq:IR_virt} and \eqref{eq:IR_real}, we verify that the IR divergences from real and virtual soft-photon corrections cancel on the level of cross section.

As mentioned before, the integration of the soft-photon bremsstrahlung is performed up to a small energy cut-off $\Delta E_s$. This cut-off can be related to the experimental resolution of the detector. In the frame $\vec{p}_m=0$ we find
\begin{equation}
    \Delta E_s= \Delta \left(\frac{p_m^2-M^2}{2\sqrt{p_m^2}}\right)\approx \frac{\Delta p_m^2}{2M},\label{eq:delta_Es}
\end{equation}
where to first order we have used $p_m^2 \approx M^2$ in the denominator, and where $\Delta p_m^2$ denotes the resolution in the missing mass squared. In order to express $\Delta E_s$ in terms of Lab quantities, one needs to calculate the missing mass in that frame. Neglecting the lepton masses, we find
\begin{eqnarray}
    p_m^2&=&(q-q'+p)^2 \nonumber \\
        &=&M^2+s_{ll}-Q^2-2q \cdot q'+2p \cdot (q-q')\nonumber\\
        &=& \Big[ M^2+ 2 M q^0 
        + 4 E_- E_+ \sin^2 \theta_{ll}/2 \nonumber \\
&&-4 E E' \sin^2 \theta_{kk'}/2 
+2  \lvert\pvec{q}\rvert \lvert\pvec{q}'\rvert \cos\theta_{\gamma\gamma}  \nonumber \\ 
&&-2 (q^0 + M) (E_+ + E_-) \Big]_{\text{Lab}},
\label{eq:p_m}
\end{eqnarray}
where all quantities on the rhs have to be given in the Lab frame, where $\theta_{kk'}$ denotes the scattering angle between the incoming electron with momentum $k$ and the outgoing with momentum $k'$, 
and where $\theta_{ll}$ denotes the Lab angle between the lepton pair momenta. Eqs.~\eqref{eq:delta_Es} and \eqref{eq:p_m} allow one to express the maximal soft-photon energy $\Delta E_s$ (defined in the system $\vec{p}_m=0$) in terms of Lab quantities and detector resolutions.

In the following it will be convenient to express the energies $\tilde{E}_\pm$, $\tilde E$ and $\tilde E'$ in terms of kinematic invariants. For the case of a large lepton mass, for which the formulas are lengthy and complicated, we use the formulas given in Appendix \ref{appendix:kin} and then boost to the rest frame of the recoiled proton and soft-photon to calculate the energies numerically. 
In the case of electron-pair production in which we can neglect the mass $m$ compared to other quantities, the formulas become more compact. In that case, we also find more compact expressions for the bremstrahlung corrections. We find
\begin{widetext}
\begin{align}
    \delta_{s;r}=&-\frac{\alpha_{\rm{em}}}{\pi}\Bigg\{\ln\left(\frac{4(\Delta E_s)^2}{m^2}\right)\bigg[2-\ln \left(\frac{Q^2}{m^2} \right)
    \nonumber \\
    &-\ln\frac{s_{ll}}{m^2}+\ln\frac{k'\cdot l_-}{k\cdot l_-}-\ln\frac{k'\cdot l_+}{k\cdot l_+}\bigg]-\frac{1}{2}\bigg[\ln\frac{4\tilde E_+^2}{m^2}+\ln\frac{4\tilde E_-^2}{m^2}+\ln\frac{4\tilde{E}^2}{m^2}+\ln\frac{4\tilde{E}'^2}{m^2}\bigg]\nonumber\\
    &+\frac{1}{4}\bigg[\ln^2\frac{4\tilde E_-^2}{m^2}+\ln^2\frac{4\tilde E_+^2}{m^2}+\ln^2\frac{4\tilde E^2}{m^2}+\ln^2\frac{4\tilde E'^2}{m^2}\bigg]+\text{Li}_2\left(1-\frac{4\tilde E_- \tilde E_+}{s_{ll}}\right)+\text{Li}_2\left(1-\frac{4\tilde E \tilde E'}{Q^2}\right)\nonumber\\
    &+\text{Li}_2\left(1-\frac{2\tilde E' \tilde E_+}{k'\cdot l_+}\right)-\text{Li}_2\left(1-\frac{2\tilde E' \tilde E_-}{k'\cdot l_-}\right)+\text{Li}_2\left(1-\frac{2\tilde E \tilde E_-}{k\cdot l_-}\right)-\text{Li}_2\left(1-\frac{2\tilde E \tilde E_+}{k\cdot l_+}\right)+\frac{2}{3}\pi^2\Bigg\}.
    \label{eq:softreal}
\end{align}
The energies $\tilde{E}_\pm$, $\tilde E$ and $\tilde E'$ in the rest frame of the recoil proton + soft photon are given by:
\begin{eqnarray}
\tilde E_\mp &=&
 \frac{p_m \cdot l_\mp}{\sqrt{p_m^2}} \approx 
\frac{1}{M} (q + p - q') \cdot l_\mp
\nonumber \\
& =& \frac{1}{4 M} \left\{ 
 (W^2 - M^2 - s_{ll})\pm  
[(W^2 - M^2 - s_{ll})^2 - 4 M^2 s_{ll}]^{1/2}
\beta_{s_{ll}} \cos \theta_l^\ast \right\}, 
\label{Eq:leptonenergies}\\
\tilde{E}&=& 
\frac{p_m \cdot k}{\sqrt{p_m^2}} \approx 
\frac{1}{M} (q + p - q') \cdot k
\nonumber \\
&=&\frac{1}{2 M \left( \left(W^2 - M^2 + Q^2\right)^2 + 4 M^2 Q^2\right)}\bigg\{
Q^2 \bigg[t (W^2 + M^2 + Q^2) - 2 M^2 (Q^2 + s_{ll}) \bigg] \nonumber \\
&+& (s - M^2) \bigg[(W^2-M^2)(W^2 - M^2 + Q^2 - s_{ll}+t) + Q^2 (-s_{ll} - t + 4 M^2) \bigg]
\nonumber \\
&+& 2 Q \cos (\Phi ) \bigg[s \left(s-M^2-Q^2\right) - W^2 \left(s-M^2\right) 
\bigg]^{1/2}\nonumber\\
&\times& \bigg[
-t (W^2 - M^2) (W^2 - M^2 + Q^2 - s_{ll} + t) 
-M^2 \bigg( (Q^2 - s_{ll}+t)^2 + 4 s_{ll} Q^2 \bigg) \bigg]^{1/2}
\bigg\}, \\
\tilde{E}'&=&\tilde{E}-\frac{W^2 - M^2-s_{ll}+t}{2M}.
\end{eqnarray}
\end{widetext}

\newpage

\section{Results\label{sec:results}}
\subsection{Observables\label{sec:result_observable}}

We use our setup to study the $e^- p \to e^- p l^-l^+ $ process, including the first-order radiative corrections 
in both the low- and high-energy regimes. For both cases we study the effect of these corrections in the soft-photon approximation on the cross section, on the forward-backward asymmetry $A_{FB}$, as well as on the beam-spin asymmetry $A_{\odot}$. These asymmetries are respectively defined as
\begin{align}
    A_{FB}&=\frac{d\sigma_{\theta_l^\ast,\phi_l^\ast}-d\sigma_{\pi - \theta_l^\ast,\phi_l^\ast + \pi}}{d\sigma_{\theta_l^\ast,\phi_l^\ast}+d\sigma_{\pi - \theta_l^\ast,\phi_l^\ast + \pi}},
    \label{eq:asymmfb}\\
    A_{\odot}&=\frac{d\sigma^+-d\sigma^-}{d\sigma^+ +d\sigma^-},\label{eq:asymmhel}
\end{align}
where $d\sigma_{\theta_l^\ast,\phi_l^\ast}$ in $A_{FB}$ stands for the unpolarized cross section measured at lepton angles $\theta_l^\ast$ and $\phi_l^\ast$ (defined in the $l^-l^+$ rest frame), 
and where $d\sigma^{\pm}$ in $A_{\odot}$ stand for the polarized cross sections for a polarized electron beam with helicity $\pm 1/2$ respectively. In the following we show plots ranging from $\theta_l^\ast=-180^\circ$ to $\theta_l^\ast=+180^\circ$. This allows us to show forward and backward cross sections economically in one plot, since $d\sigma(\theta_l^\ast,\phi_l^\ast+\pi)=d\sigma(-\theta_l^\ast,\phi_l^\ast)$. The forward-backward asymmetry can therefore also be written as
\begin{align}
    A_{FB}&=\frac{d\sigma(\theta_l^\ast,\phi_l^\ast)-d\sigma(\pi-\theta_l^\ast,\pi+\phi_l^\ast)}{d\sigma(\theta_l^\ast,\phi_l^\ast)+d\sigma(\pi-\theta_l^\ast,\pi+\phi_l^\ast)}\nonumber\\
    &=\frac{d\sigma(\theta_l^\ast,\phi_l^\ast)-d\sigma(\theta_l^\ast-\pi,\phi_l^\ast)}{d\sigma(\theta_l^\ast,\phi_l^\ast)+d\sigma(\theta_l^\ast-\pi,\phi_l^\ast)},
    \nonumber \\
\end{align}
and, including radiative correction explicitly, it is given by:
\begin{align}
    A_{FB}&=\frac{d\sigma_0(\theta_l^\ast)(1+\delta(\theta_l^\ast))-d\sigma_0(\theta_l^\ast-\pi)(1+\delta(\theta_l^\ast-\pi))}{d\sigma_0(\theta_l^\ast)(1+\delta(\theta_l^\ast))+d\sigma_0(\theta_l^\ast-\pi)(1+\delta(\theta_l^\ast-\pi))}.\label{eq:AFB_rad_corr}
\end{align}

From Eq. \eqref{eq:AFB_rad_corr} one can see that corrections that are symmetric under the interchange $l^- \leftrightarrow l^+$, corresponding with $\theta_l^\ast \leftrightarrow \theta_l^\ast - \pi$, drop out in the ratio. Therefore, to first order, only corrections of class (c) give a contribution to the asymmetry. 
\begin{eqnarray}
    A_{FB}
    &=& \frac{A^0_{FB} + \delta_c /(1 + \delta_a + \delta_b)}
    {1 + A_{FB}^0 \, \delta_c /(1 + \delta_a + \delta_b)} 
    \nonumber \\
    &\approx& A^0_{FB} + \delta_c \left( 1 - (A^0_{FB})^2 \right),
    \nonumber \\
\end{eqnarray}
where $A^0_{FB}$ denotes the uncorrected asymmetry. 

On the other hand the radiatively corrected beam-spin asymmetry (BSA), given by
\begin{equation}
    A_{\odot}=\frac{d\sigma^+ (1+\delta^+) - d\sigma^- (1+\delta^-)}{d\sigma^+ (1+\delta^+) + d\sigma^- (1+\delta^-)},
\end{equation}
does not get modified in the soft-photon approximation, since the corrections are the same for both helicity cross sections, i.e. $\delta^+ =\delta^-$, and therefore drop out in the ratio.

In the following, we show our numerical results for the $e^- p \to e^- p l^- l^+$ observables including the first order soft-photon radiative corrections.

\subsection{Results for dVCS observables in the $\Delta(1232)$ region}

In this section we show our results  in the low-energy regime in which we choose the dVCS center-of-mass energy $W=1.25$ GeV. We model the dVCS amplitude in terms of the Born amplitude and the first proton excitation, the $\Delta(1232)$ resonance. As was found in Ref. \cite{Pauk:2020gjv}, this model can reproduce the full calculation based on empirical structure functions from Ref.~\cite{Gryniuk:2015eza} with an accuracy in the few per-cent range for the process $\gamma p \rightarrow e^- e^+ p$ (i.e. for a real photon). Therefore, we can safely assume that the Born + $\Delta$-pole model describes the dVCS amplitude sufficiently well also in the virtual-photon process for sufficiently small photon virtualities. 

\begin{figure}[h]
    \centering
    \includegraphics[width=0.48\textwidth]{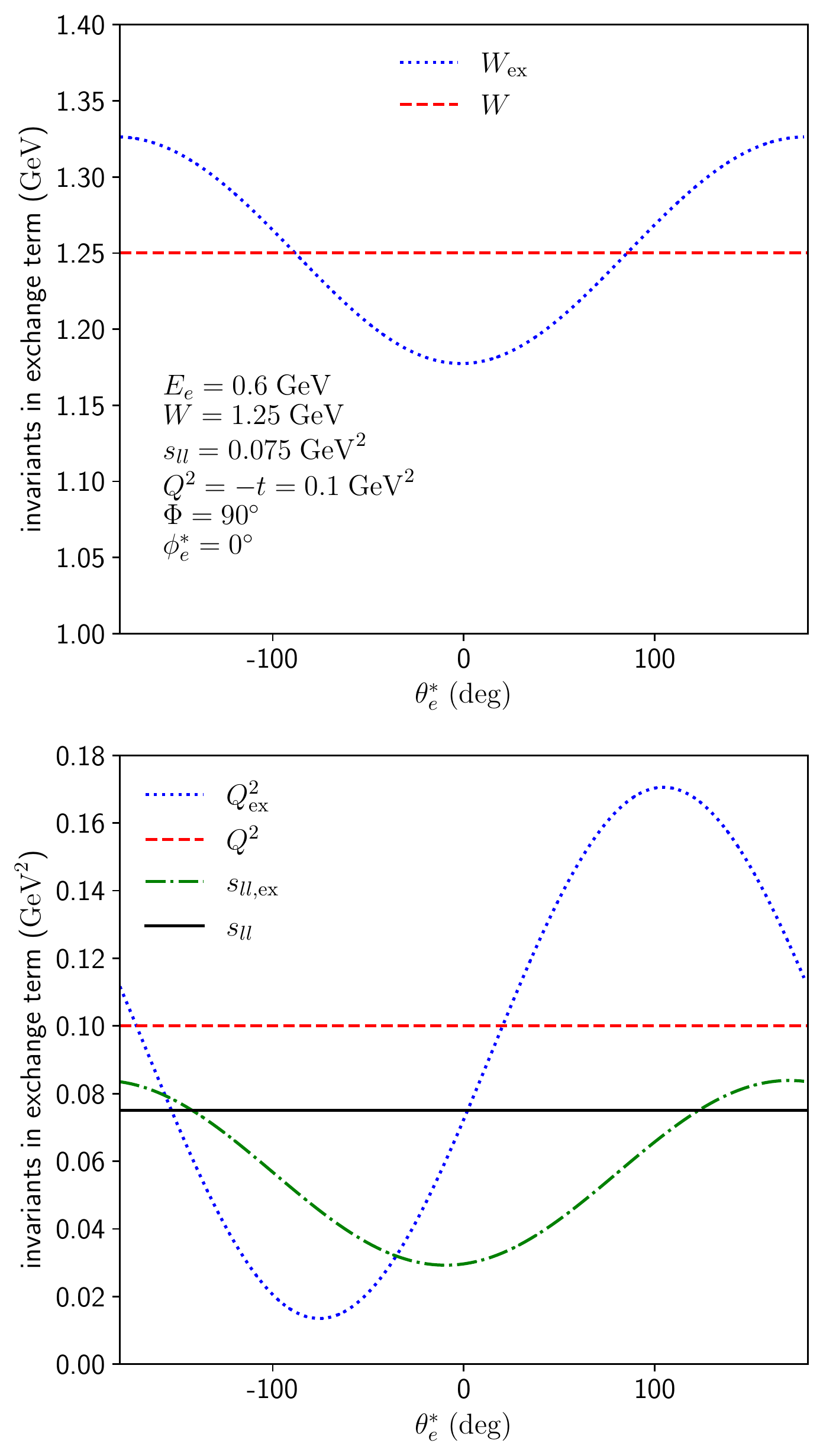}
    \caption{Kinematic quantities entering the exchange dVCS amplitude for the $e^- p \to e^- p e^-e^+$ process in the $\Delta(1232)$ region. In the upper panel we show the c.m. energy $W_{ex}$, as function of $\theta_e^\ast$,  compared with the value $W=1.25$ GeV of the direct process. In the lower panel we compare the $\theta_e^\ast$ dependence of both photon virtualities in the exchange dVCS amplitude with their constant values for the direct dVCS amplitude. }
    \label{fig:low_energy_ex_kin}
\end{figure}

In order for the dVCS model to be also accurate for the $e^- p \to e^- p e^-e^+$ process, in which we need to anti-symmetrize the full amplitude under exchange of both final electrons as given by Eq.~(\ref{eq:antisymm}), we choose the kinematics in such a way that also for the exchange dVCS amplitude the c.m. energy $W_{\rm{ex}}$ remains in the $\Delta(1232)$ resonance region, and the photon virtualities entering the exchange process remain sufficiently small. As can be seen from Fig.~\ref{fig:low_energy_ex_kin} (upper panel), for the choice of an electron beam of $0.6$ GeV, we find that $W_{\rm{ex}}$ (blue dotted curve) is roughly of the same magnitude as $W$ (dashed red curve), varying between $1.18$ and $1.33$ GeV as function of $\theta_e^\ast$. 
Note that a larger electron beam energy leads to a larger  value of $W_{\rm{ex}}$. 
From the lower panel of Fig.~\ref{fig:low_energy_ex_kin}, we furthermore see that both photon virtualities in the exchange dVCS amplitude, denoted by $Q_{\rm{ex}}$ (blue dotted curve) and $s_{ll,\rm{ex}}$ (green dash-dashed curve), are both below $0.18$ GeV$^2$ for the full range of the lepton angle $\theta_e^\ast$. We are thus in a kinematic regime where we can  study the sensitivity of the full amplitude to the low-energy constant $b_{3,0}$ described in Sec.~\ref{sec:dvcs_low_energy}. 

Having studied the appropriate kinematics to describe both the direct and the exchange dVCS amplitude within the same model, we next explore the sensitivity of the $e^- p \to e^- p l^-l^+$ observables on 
the low-energy constant $b_{3,0}$ introduced in  the low-energy expansion of Eq.~(\ref{lexdvcs}). This low-energy constant is the main unknown in the determination of the ${\cal O}(Q^4)$ term of the subtraction function $\bar T_1(0,Q^2)$ entering the theoretical calculation of the $\mu H$ Lamb shift.  

\begin{figure*}[h]
    \centering
    \includegraphics[width=0.45\textwidth]{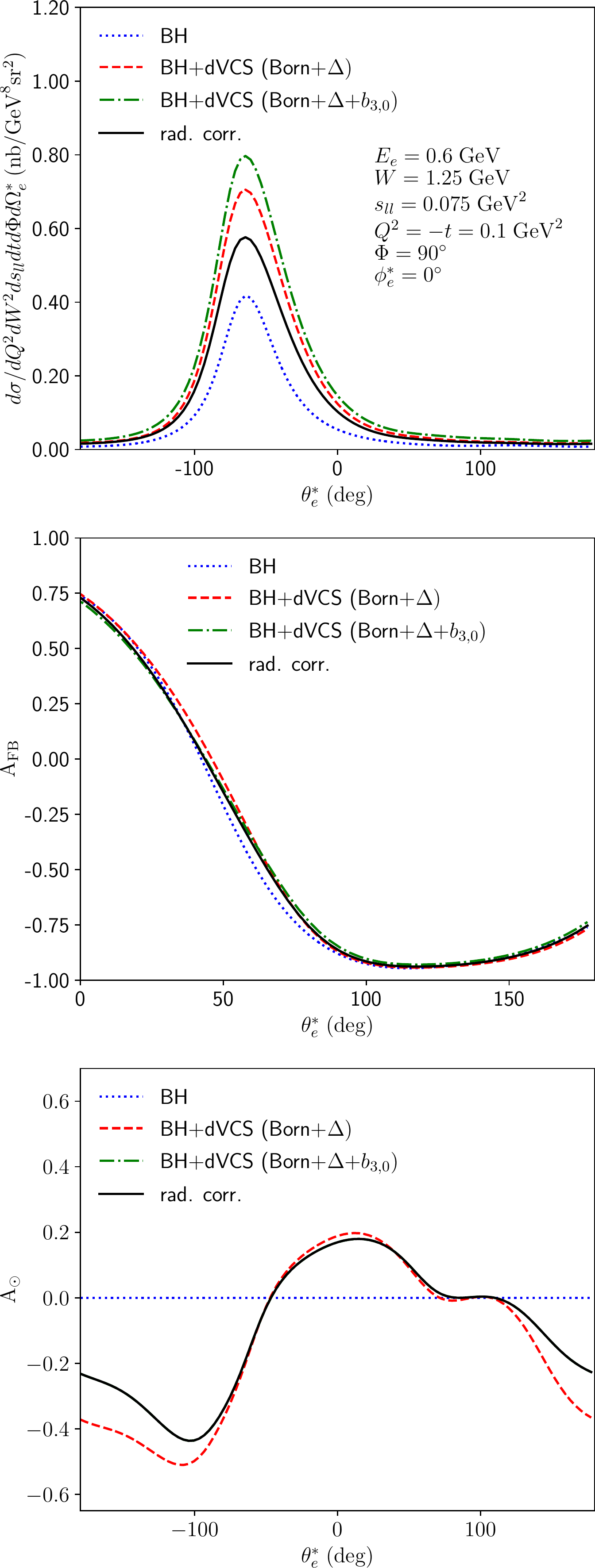}\hspace{0.5cm}
    \includegraphics[width=0.45\textwidth]{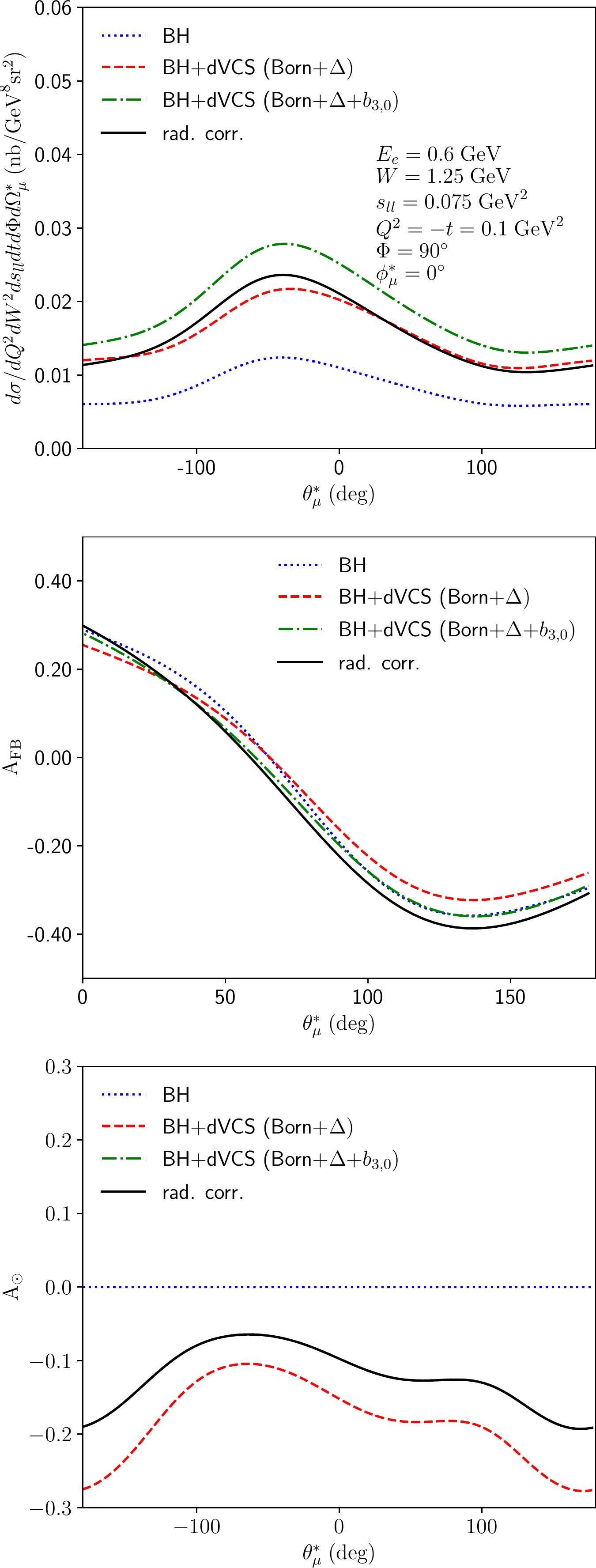}
    \caption{$\theta_l^\ast$ dependence of the $e^- p \to e^- p l^-l^+$ cross section (upper panels), forward-backward asymmetry (middle panels) and beam-spin asymmetry (lower panels) for $e^-e^+$ production (left panels) and $\mu^- \mu^+$ production (right panels) in the $\Delta(1232)$ region, for $\Phi = 90^\circ$. The curves show the predictions for BH and BH + dVCS for two models showing the sensitivity to the low-energy constant $b_{3,0}$. The black solid curves show the effect of the radiative corrections, for the hadronic model of the green dashed-dotted curves (these curves exactly coincide in the lower panels).}
    \label{fig:low_energy_asyms}
\end{figure*}

In Fig.~\ref{fig:low_energy_asyms}, we show the dependence on the lepton angle $\theta_l^\ast$ of the $e^- p \to e^- p l^-l^+$ differential cross section (upper panels), the forward-backward asymmetry (middle panels) as well as the beam-spin asymmetry (lower panels) for both $e^-e^+$ and $\mu^- \mu^+$ production (left and right panels respectively). We choose the kinematics as in Fig.~\ref{fig:low_energy_ex_kin}. As can be seen from the upper panel, the interference between the dVCS process with the BH process amplifies the cross section for both $e^-e^+$ and $\mu^- \mu^+$ production by roughly a factor of two as compared with the BH process itself. 
Furthermore, the spread between the different theoretical estimates for the low-energy constant $b_{3,0}$, as shown in Table~\ref{tab:subtraction}, increases the cross sections additionally by approximately $15\%$ in both cases.

We also study the effect of the soft-photon radiative corrections on the cross section, as given by  \cref{eq:virt_finitesum,eq:virta,eq:virtb,eq:virtc1,eq:virtc2,eq:virtc3,eq:virtc4,eq:finite_real,eq:reala,eq:realb,eq:realc,eq:softreal}. For the real soft-photon emission correction,  we choose the soft-photon energy cut-off of $\Delta E_s=0.01$ GeV, which corresponds to approximately $1.5 \%$ of the lepton beam energy.  
As can be seen from Fig.~\ref{fig:low_energy_asyms},  
the effect of the first-order radiative corrections is found to be quite sizeable on the level of cross sections. 
In the case of $e^-e^+$ production the effect leads to a decrease of the cross section by around $30\%$, whereas for $\mu^- \mu^+$ production it leads to a decrease of the order of $15 \%$. 
Therefore, although the cross section by itself has a relatively high sensitivity on the low-energy constant $b_{3,0}$, for an experimental extraction of $b_{3,0}$ the inclusion of the radiative corrections is imperative. A comparable importance of the radiative corrections was also found in the extraction of the proton generalized polarizabilities from the cross sections of the VCS process $e^- p \to e^- p \gamma$~\cite{Vanderhaeghen:2000ws,Fonvieille:2019eyf}.

\begin{figure*}
\centering
    \includegraphics[width=0.47\textwidth]{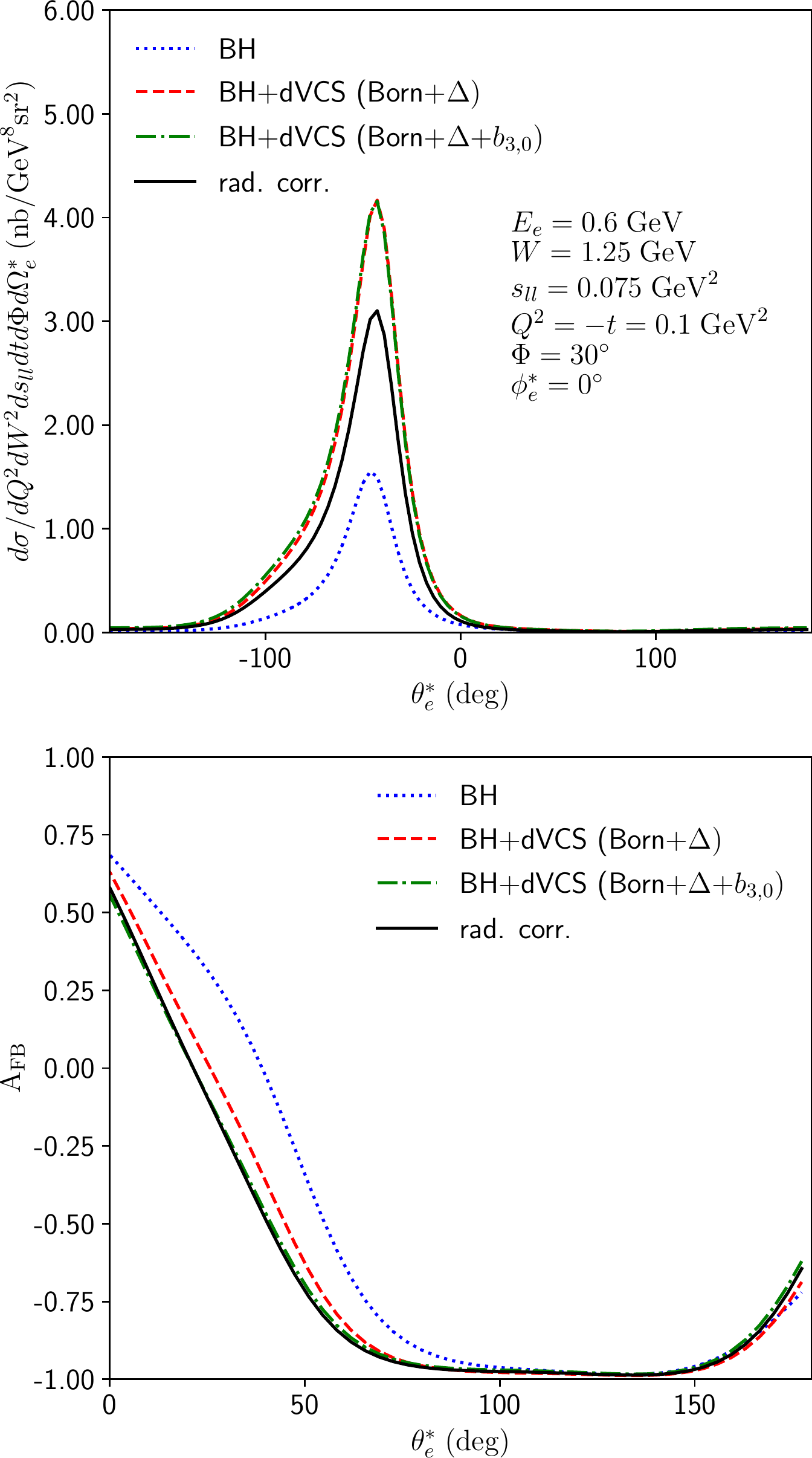}\hspace{0.5cm}
    \includegraphics[width=0.47\textwidth]{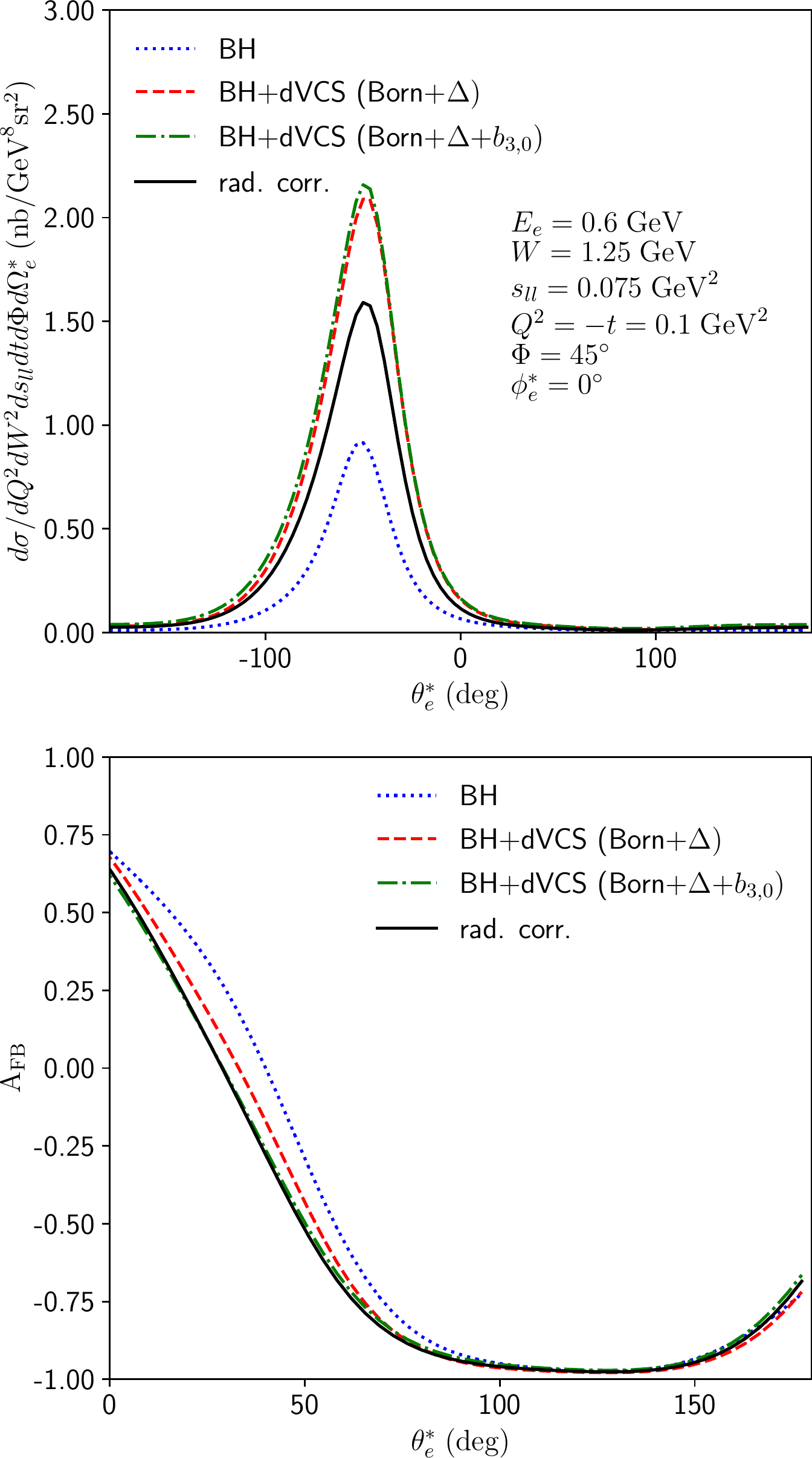}
    \caption{$\theta_e^\ast$ dependence of the $e^- p \to e^- p e^-e^+$ cross section (upper panels) and forward-backward asymmetry (lower panels) 
    in the $\Delta(1232)$ region for $\Phi=30^\circ$ (left panels) and $\Phi=45^\circ$ (right panels). 
    Curve conventions as in Fig.~\ref{fig:low_energy_asyms}.}
    \label{fig:low_energy_afb}
\end{figure*}

The situation is different for the asymmetries. For the forward-backward asymmetry $A_{FB}$, 
we find for the kinematics of Fig.~\ref{fig:low_energy_asyms} only a small sensitivity to the dVCS amplitude and its underlying hadronic model. However, this is mainly due to the choice of $\Phi = 90^\circ$, for which the forward-backward asymmetry is completely dominated by the BH process. The sensitivity can be increased by varying $\Phi$. In Fig.~\ref{fig:low_energy_afb} we show the cross sections and forward-backward asymmetries for the same kinematics, but for smaller angles between the $(\vec k, \vec k')$ and $(\vec q, \vec q\, ')$ scattering planes in  Fig.~\ref{fig:kin_plane}: $\Phi=30^\circ$ and $\Phi=45^\circ$. For these cases we find a $20\%$ shift of the forward-backward asymmetry for the case including the $\Delta$-resonance compared to the BH process by itself. Including the range of theoretical values for the dVCS low-energy constant $b_{3,0}$, we find a further shift of the asymmetry of up to $5\%$ on $A_{FB}$, while the inclusion of radiative corrections is found to have a very small effect, around or below the $1\%$ range on $A_{FB}$.

For the beam-spin asymmetry $A_\odot$, we find a significantly higher sensitivity on $b_{3,0}$ than for the forward-backward asymmetry, as shown in the lower panels of Fig.~\ref{fig:low_energy_asyms}. 
Note that the result for Born + $\Delta$-pole + $b_{3,0}$ (green dashed-dotted curves) and the result which in addition also includes the radiative corrections (black solid curves), coincide, since in the soft-photon approximation the radiative corrections drop out in the ratio of cross sections calculated for the BSA, as discussed above. 
Including the range of theoretical values for the dVCS low-energy constant $b_{3,0}$, leads to a shift in the BSA up to around $15\%$ for $e^-e^+$ production and up to around $10\%$ for $\mu^- \mu^+$ production.

As the BSA and $A_{FB}$ are basically not affected by the radiative corrections, a combined analysis of the cross section, the $A_{FB}$, and the BSA holds promise to extract the 
dVCS low-energy constant $b_{3,0}$.

\begin{figure}
\centering
\includegraphics[width=0.45\textwidth]{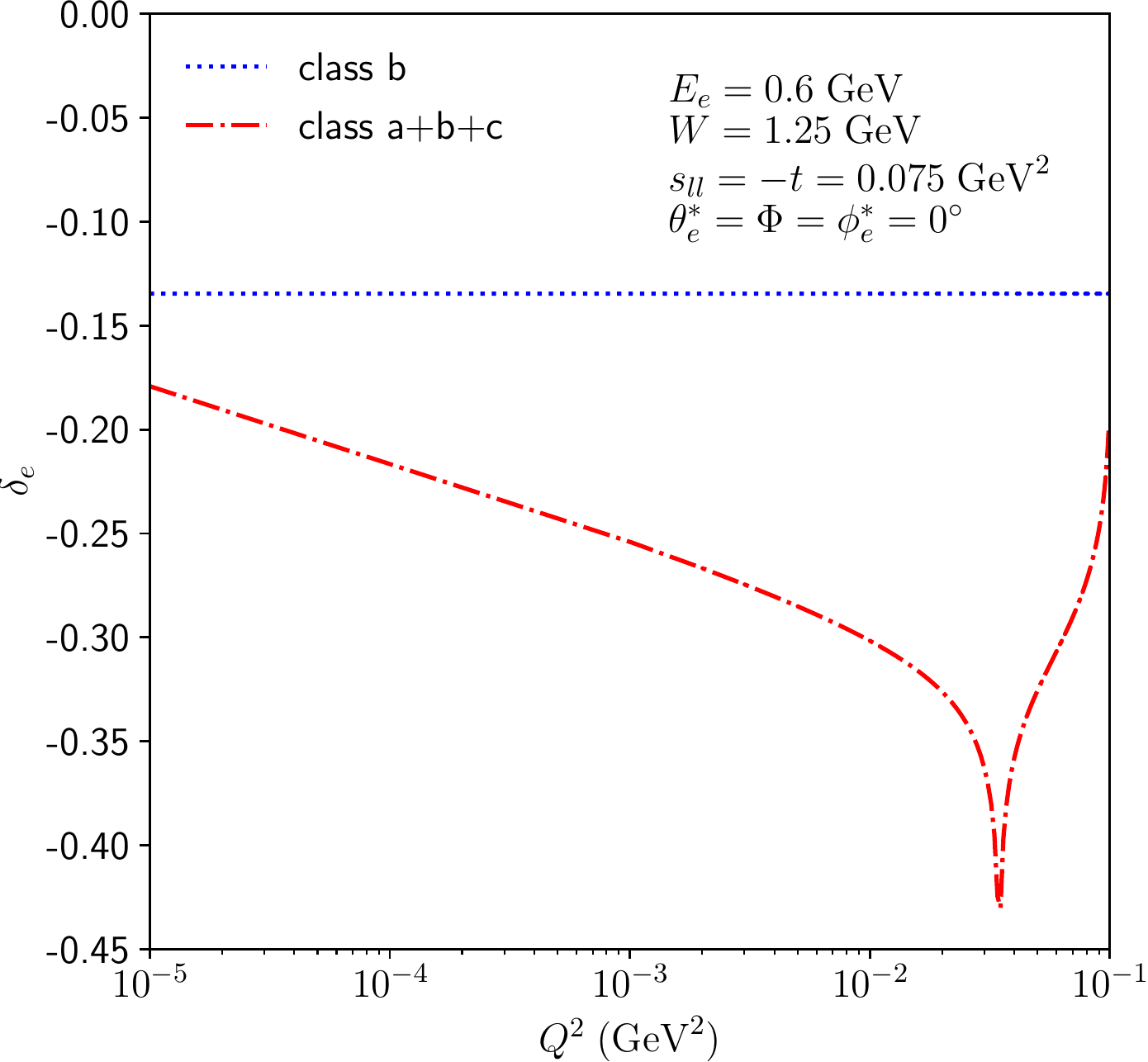}
\caption{Radiative corrections for the $e^- p \to e^- p e^-e^+ $ process in the $\Delta(1232)$ region in the limit $Q^2 \to 0$ for the comparable kinematic setup as was studied before for the $\gamma p \to e^-e^+ p$ process in Ref. \cite{Heller:2020lnm}. The corrections are for soft-photon cut-off energy of $\Delta E_s = 0.01$~GeV.}
\label{fig:low_energy_delta_limit}
\end{figure}

In Fig.~\ref{fig:low_energy_delta_limit} we show in more detail how the radiative corrections to the $e^- p \to e^- p e^-e^+$ process vary when the intial photon approaches the real photon  limit, i.e. $Q^2\rightarrow 0$. In this limit only the class (b) corrections contribute. We choose the kinematics comparable to Ref.~\cite{Heller:2020lnm}, in which we studied the effect of radiative corrections for the process $\gamma p \rightarrow e^- e^+ p$. In that study we found corrections of roughly $8\%$ for the full set of one-loop QED corrections for the kinematics shown in Fig.~\ref{fig:low_energy_delta_limit}. In the soft-photon approximation the corrections are somewhat over-estimated, as can be seen from the blue dotted curve corresponding with a correction of roughly $13\%$. Comparing the blue dotted curve with the red dashed-dotted one, we see that for quasi-real photons with a virtuality of $10^{-4}$ to $10^{-3}$ GeV$^2$ the inclusion of all corrections is important, and the description as a real photon underestimates the corrections by $10 - 15 \%$. Note that in the region around $Q^2 \approx 0.03$ GeV$^2$ the two outgoing electrons with momenta $k'$ and $l_-$ are becoming collinear. This explains the spiked behavior in the red dashed-dotted curve, since the logarithm with the argument proportional to the scalar product $k'\cdot l_-$ is becoming large.

\begin{figure*}
\centering
\includegraphics[width=0.45\textwidth]{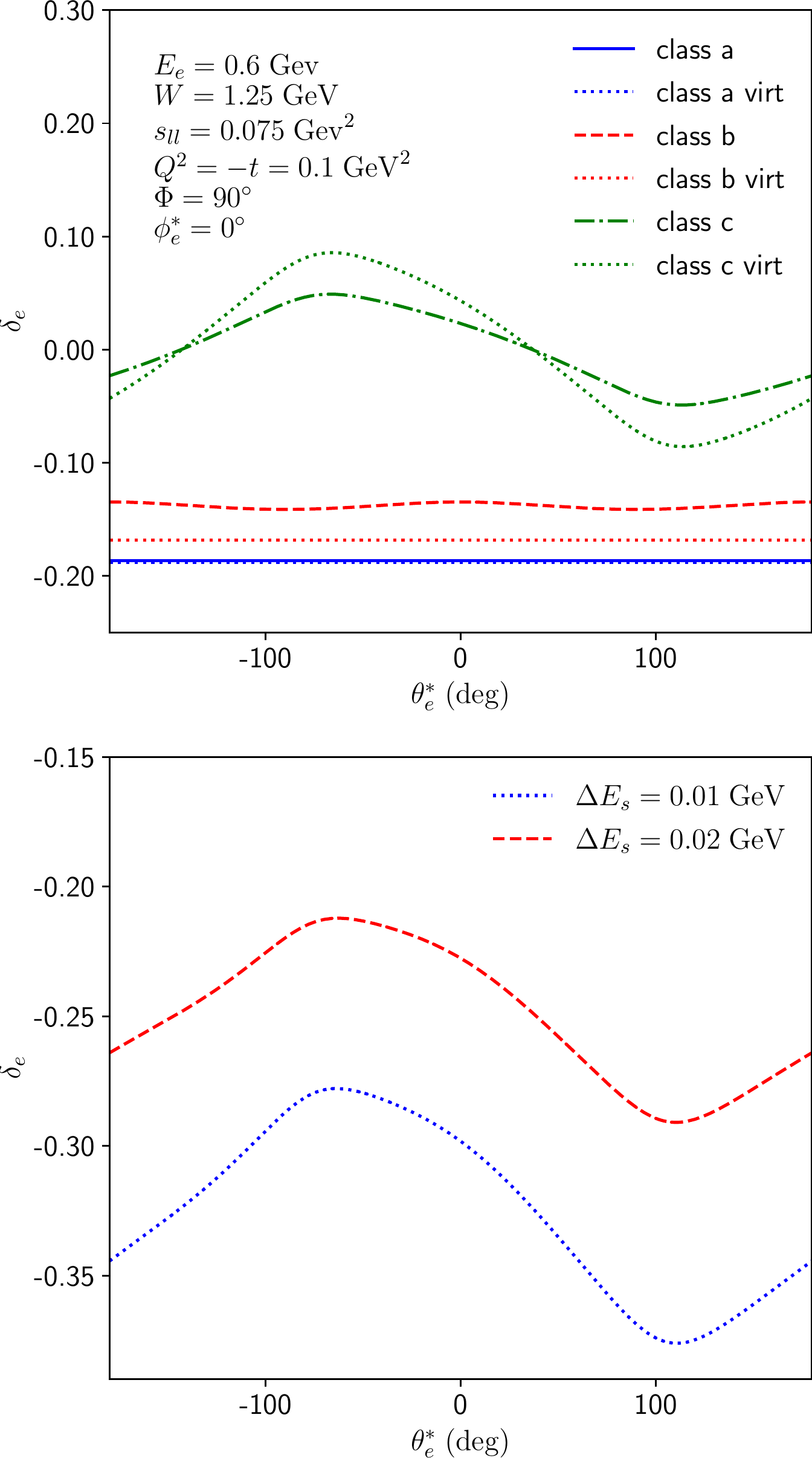}\hspace{.5cm}
\includegraphics[width=0.45\textwidth]{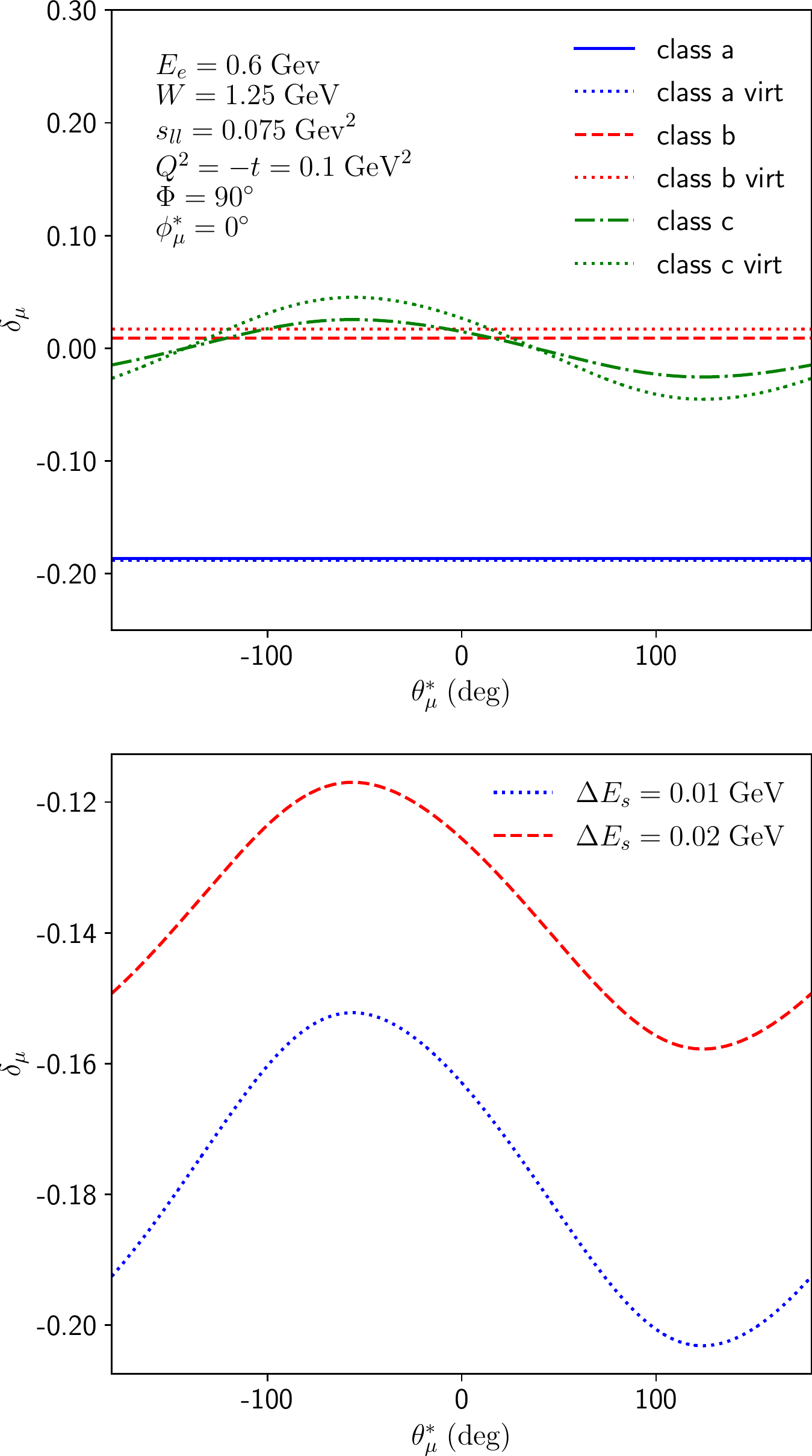}
\caption{Upper panels: $\theta_l^\ast$ dependence of the radiative corrections to the $e^- p \to e^- p l^-l^+$ cross section  in the $\Delta(1232)$ region for $e^-e^+$ production (left) and $\mu^-\mu^+$ production (right), for the different classes of radiative corrections for $\Delta E_s = 0.01$~GeV. Lower panels: $\theta_l^\ast$ dependence of the total radiative correction for different values of $\Delta E_s$ for both $e^-e^+$ production (left) and $\mu^-\mu^+$ production (right).}
\label{fig:low_energy_delta}
\end{figure*}

In Fig.~\ref{fig:low_energy_delta} (upper panels) we study in more detail the relative size of the radiative corrections due to the three different diagram classes (a), (b) and (c). While for $e^-e^+$ production the corrections due to class (a) and (b) are dominant and negative, for $\mu^-\mu^+$ production the main correction arises from class (a) as it involves the vertex correction on the beam electron, whereas the corrections between the produced $\mu^-\mu^+$ pair are small and positive. Comparing left and right panel one can clearly see that the biggest difference between $e^-e^+$ and $\mu^-\mu^+$ production is due to the corrections of class (b), 
which correspond with the corrections from the produced di-lepton pair. Furthermore Fig.~\ref{fig:low_energy_delta} illustrates, as mentioned above, that the corrections of class (a) and (b) are symmetric under the interchange of $l^+$ and $l^-$, corresponding to the angular shift $\theta_l^\ast \rightarrow \theta_l^\ast - \pi$, while class (c) is anti-symmetric. As only class (c) contributes to the forward-backward asymmetry to lowest order, the smallness of the corrections of class (c) also explains why the $A_{FB}$ is largely unaffected by the radiative corrections.

Furthermore in Fig.~\ref{fig:low_energy_delta} (lower panels), we show the sum of all three types of corrections also for a twice larger value of the soft-photon cut-off energy $\Delta E_s$. One notices the positive contribution to the cross section correction $\delta$ upon increasing the value of $\Delta E_s$.  

\subsection{Results for high-energy DDVCS observables}

In this section we show our results for the 
$e^- p \to e^- p l^-l^+$ observables in the 
high-energy regime, in which we use GPDs to model the dVCS amplitude in the deeply virtual regime, the so-called DDVCS process, as described in Section~\ref{sec:dvcs_high_energy}. 
We will explore the sensitivity of this process to the modeling of the GPDs, in particular the D-term contribution, and quantify the effect of the radiative corrections in the soft-photon approximation.

As is conventional in the high-energy regime, we give the differential cross section with respect to the Bjorken scaling variable instead of the c.m. energy $W$ describing the Compton process. Therefore, in this section we show differential cross sections with respect to the quantity $\xi$, which is related to the kinematical invariants through Eq.~\eqref{eq:xi}. The cross section differential  w.r.t. $\xi$ is related to the cross section differential w.r.t. $W^2$ as:
\begin{eqnarray}
    \left(\frac{\text{d}\sigma}{\text{d}Q^2\text{d}\xi\text{d}\Phi\text{d}t\text{d}s_{ll}\text{d}\Omega_l^\ast} \right) 
    &=& \left( \frac{Q^2 + s_{ll}}{2 \xi^2}\right) 
    \nonumber \\
    &\times& \left(\frac{\text{d}\sigma}{\text{d}Q^2\text{d}W^2\text{d}\Phi\text{d}t\text{d}s_{ll}\text{d}\Omega_l^\ast} \right). \nonumber \\
\end{eqnarray}

\begin{figure*}
\centering
   \includegraphics[width=0.45\textwidth]{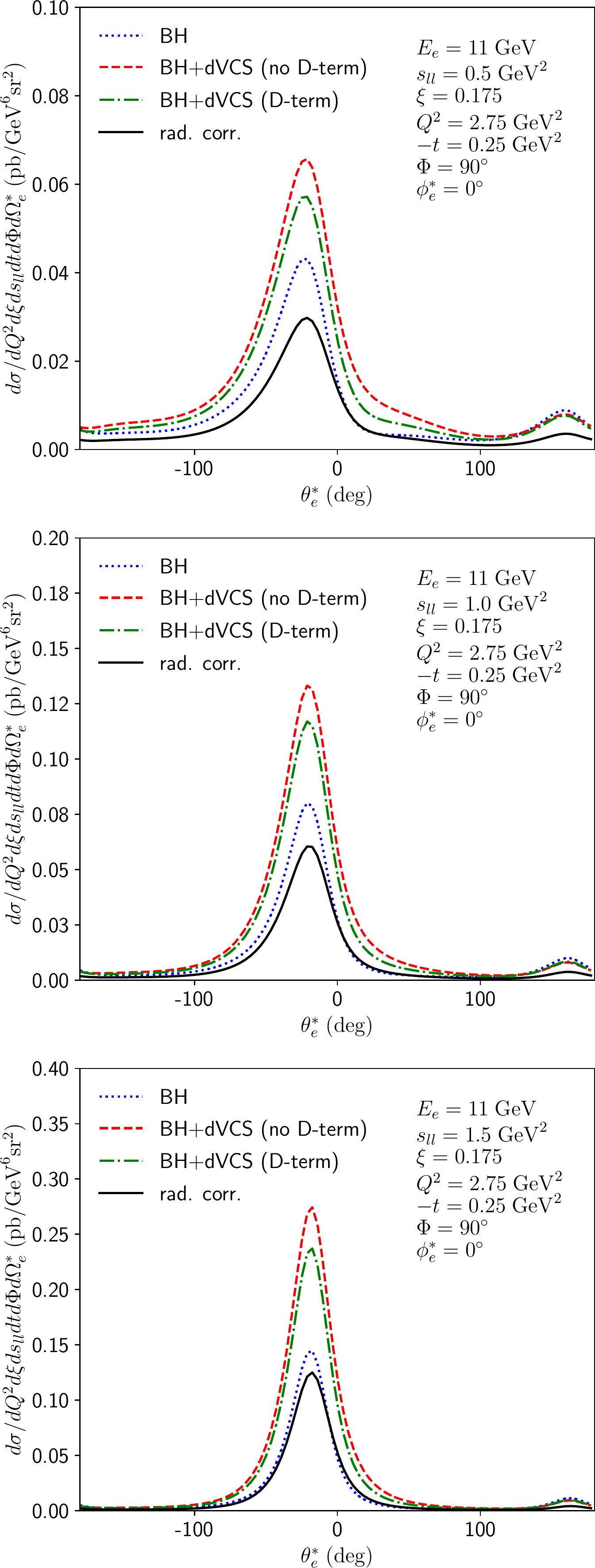} \hspace{.5cm}
    \includegraphics[width=0.46\textwidth]{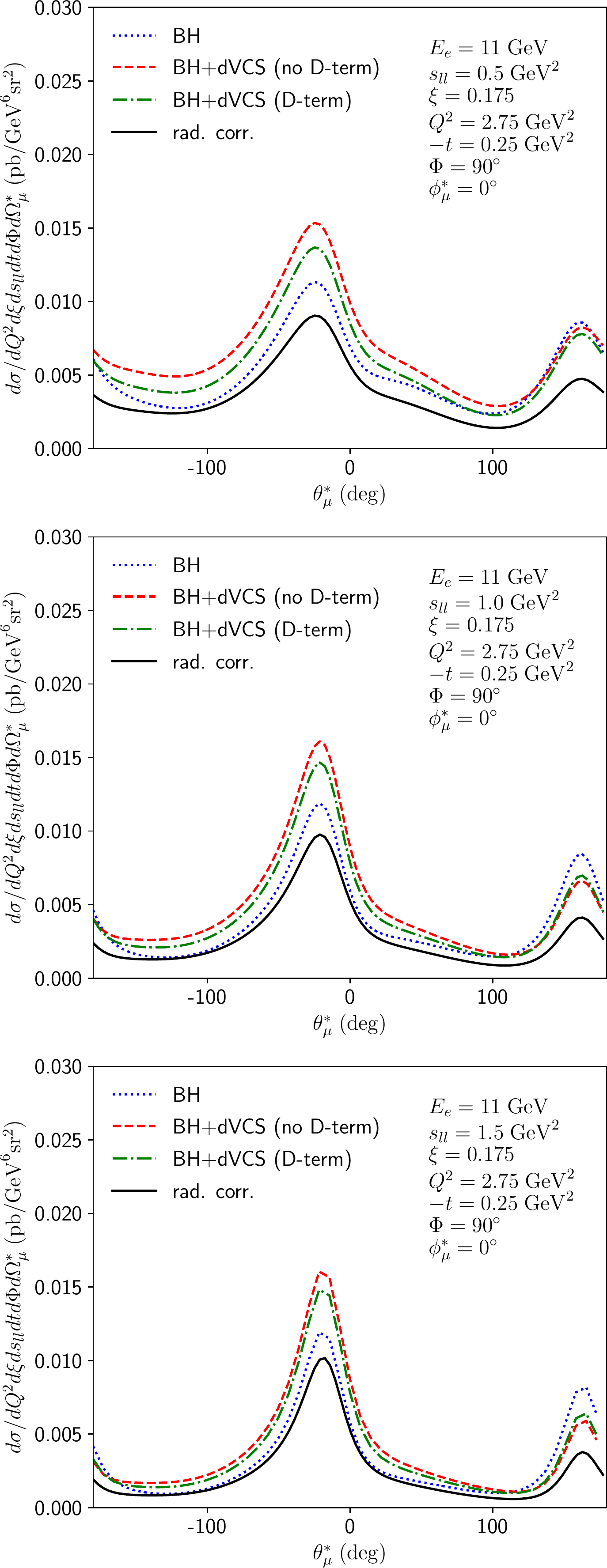}
\caption{$\theta_l^\ast$ dependence of the $e^- p \to e^- p l^-l^+$ cross section in the 
DDVCS regime 
for $e^-e^+$ production (left panels) and $\mu^- \mu^+$ production (right panels), for different values of the di-lepton invariant mass $s_{ll}$.
The curves show the predictions for BH and BH + dVCS for two models showing the sensitivity to the D-term in the GPD parameterization. The black solid curves show the effect of the radiative corrections, for the hadronic model of the green dashed-dotted curves.
    }
    \label{fig:cross_sections}
\end{figure*}

In Fig.~\ref{fig:cross_sections} we show a comparison of $e^- p \to e^- p l^-l^+$ cross sections for $e^-e^+$ production (left panels) vs $\mu^-\mu^+$ production  (right panels). The cross sections are shown for an incoming electron beam energy of $11$ GeV, which corresponds to the experimental setup of the CLAS12@JLab experiment and of the SoLID@JLab project. We show the cross section for $\xi=0.175$, $Q^2=2.75$ GeV$^2$, $-t=0.25$ GeV$^2$, $\Phi = 90^\circ$ and for three values of the di-lepton invariant mass $s_{ll}$. While for $e^-e^+$ production one observes a pronounced peak around $\theta_e^\ast \approx -20^\circ$ for all three cross sections, the cross sections for $\mu^-\mu^+$ production appear flatter. Furthermore, the cross sections for $e^-e^+$ production are $4$ to $15$ times larger than the cross sections for $\mu^-\mu^+$ production at the same value of $s_{ll}$. This significant difference is due to the contribution of the exchange diagrams, which only contribute to $e^-e^+$ production to satisfy the Pauli principle. Naively, one would expect the cross sections to be of roughly the same magnitude, since even for $s_{ll}=0.5$ GeV$^2$ the di-lepton invariant mass is $10$ times larger  than the $\mu^-\mu^+$ production threshold, such that effects from the lepton mass don't play a crucial role. However, the anti-symmetrization of the final-state electrons for $e^-e^+$ production yields a large contribution of the exchange diagrams, which increases with  increasing values of $s_{ll}$. 

Furthermore, one can see from Fig.~\ref{fig:cross_sections} 
that the BH process again serves as an amplifier of the dVCS process, and the BH+dVCS cross section is roughly 50\% larger than the BH cross section. 
The cross section therefore has a strong sensitivity to the underlying GPD model. In particular through such interference, the D-term contribution to the GPD, using the dispersive estimate of Ref.~\cite{Pasquini:2014vua}, 
decreases the cross section by up to approximately $20\%$ (green dashed-dotted curves vs red dashed curves). 

In order to use the $e^- p \to e^- p l^-l^+$ as a tool to access the DDVCS amplitude, it is important to quantify the radiative corrections, which is an aim of this work. In Fig.~\ref{fig:cross_sections} we show the impact of the radiative corrections on the cross section. 
For the real soft-photon emission correction,  we choose the soft-photon energy cut-off of $\Delta E_s=0.05$ GeV,  
which is roughly $1\%$ of the $e - p$ center-of-mass energy $\sqrt{s}=4.64$ GeV (corresponding to an electron beam of $11$ GeV).
We see from Fig.~\ref{fig:cross_sections} that the soft-photon radiative corrections are very sizeable in the DDVCS regime, decreasing the cross sections by up to $50\%$ for $e^-e^+$ production and by up to $35\%$ for $\mu^-\mu^+$ production (black solid curves vs green dashed-dotted curves).  

\begin{figure}
    \centering
    \includegraphics[width=0.47\textwidth]{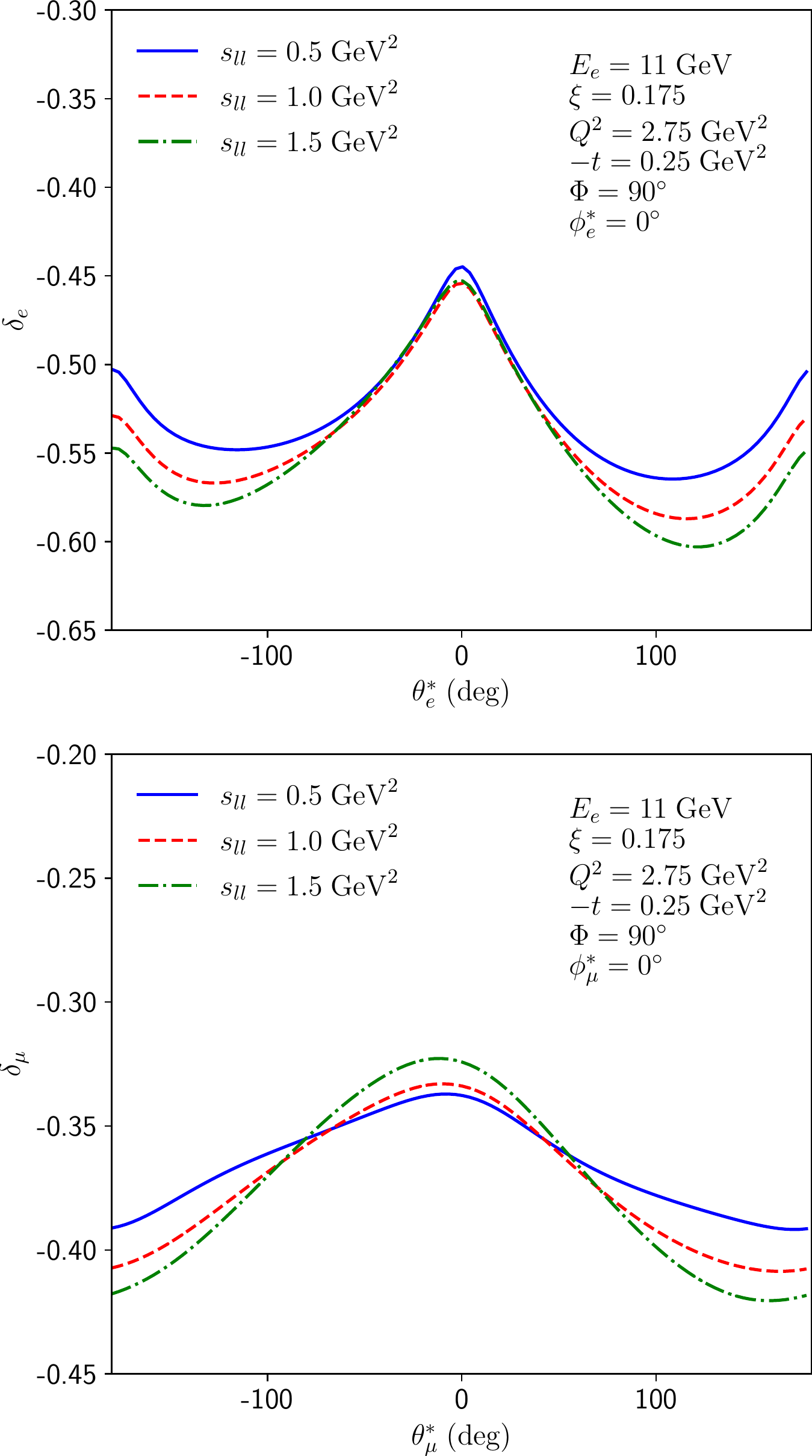}
    \caption{Upper panel: $\theta_l^\ast$ dependence on the soft-photon radiative corrections to the $e^- p\rightarrow e^- p l^- l^+$ cross section in the DDVCS regime, for different values of $s_{ll}$. Upper (lower) panel is for $e^-e^+$  ($\mu^- \mu^+$) production respectively. The corrections are for a soft-photon cut-off energy of $\Delta E_s = 0.05$~GeV.}
    \label{fig:panel_delta_xi_0175}
\end{figure}

In Fig. \ref{fig:panel_delta_xi_0175} we show the $\theta_l^\ast$ dependence of the soft-photon radiative correction factor on the $e^- p\rightarrow e^- p l^- l^+$ cross section in more detail, for the three values of $s_{ll}$, corresponding to the cross sections shown in Fig~\ref{fig:cross_sections}. As mentioned above, using $\Delta E_s = 0.05$~GeV, the corrections in the DDVCS regime vary between $-60\%$ and $-45\%$ for $e^-e^+$ production, while for $\mu^-\mu^+$ production they are slightly smaller and vary between $-45\%$ and $-30\%$. 

\begin{figure*}
\centering
    \includegraphics[width=0.47\textwidth]{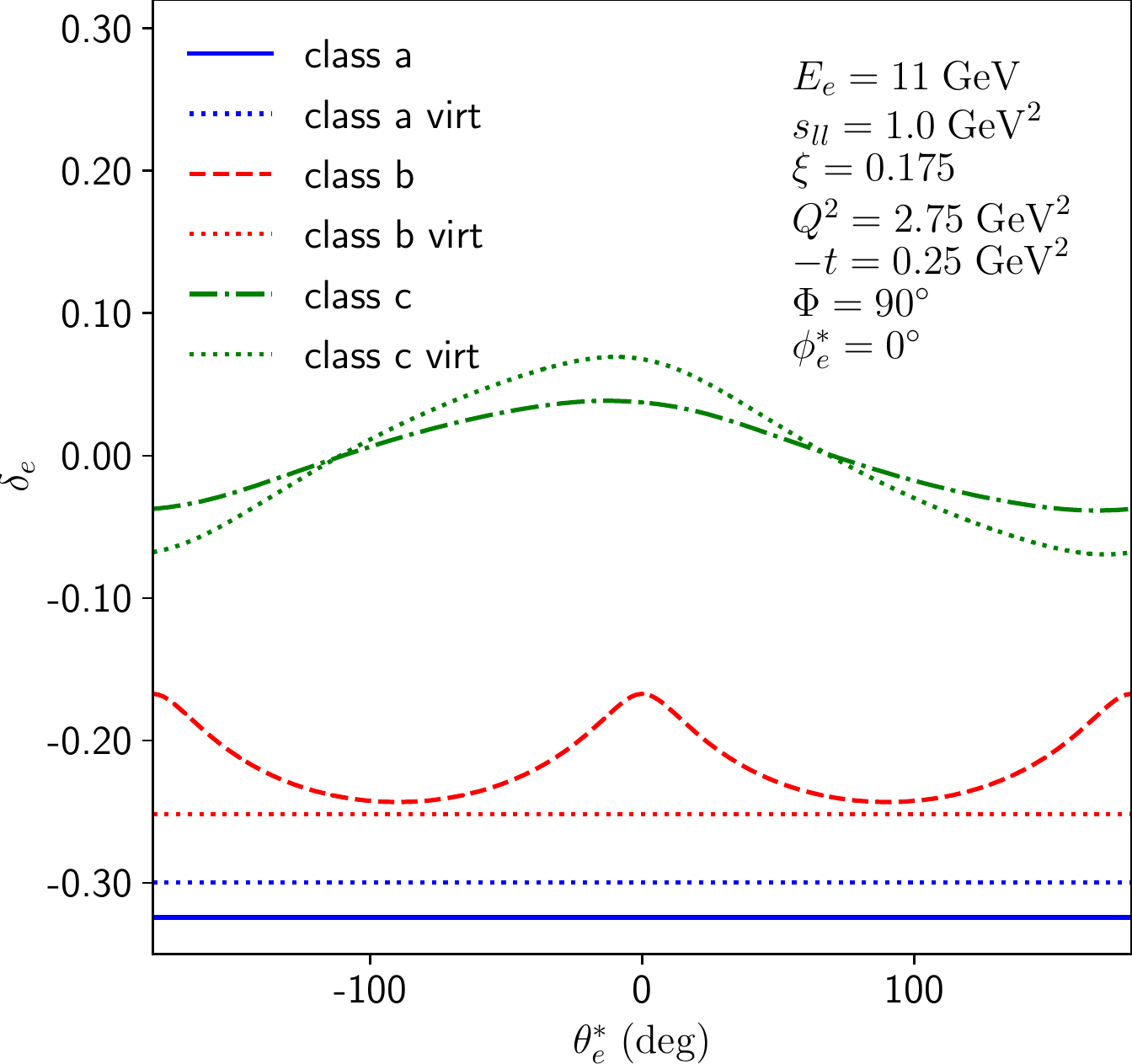}\qquad
    \includegraphics[width=0.47\textwidth]{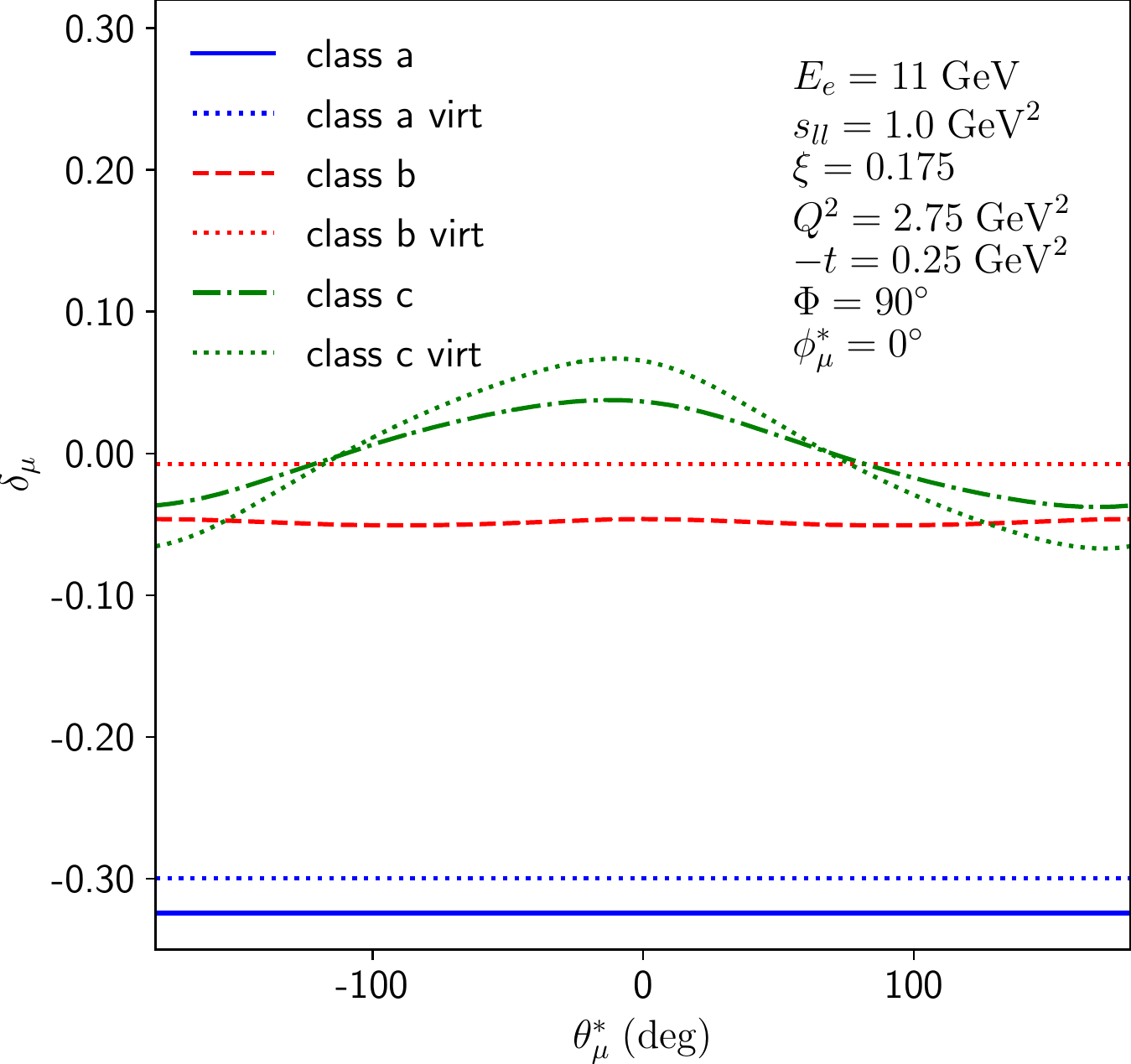}
    \includegraphics[width=0.47\textwidth]{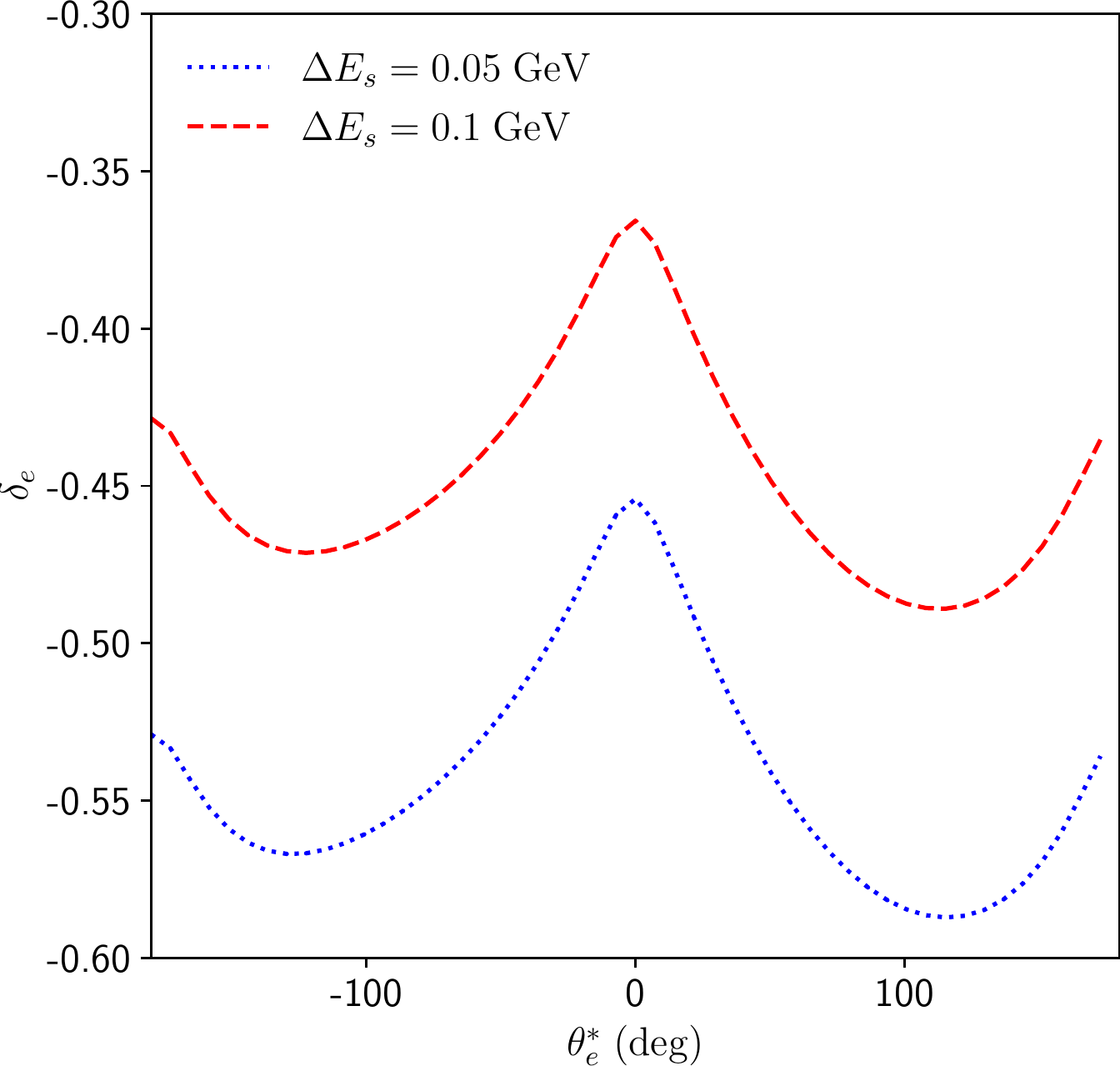}
    \qquad
    \includegraphics[width=0.47\textwidth]{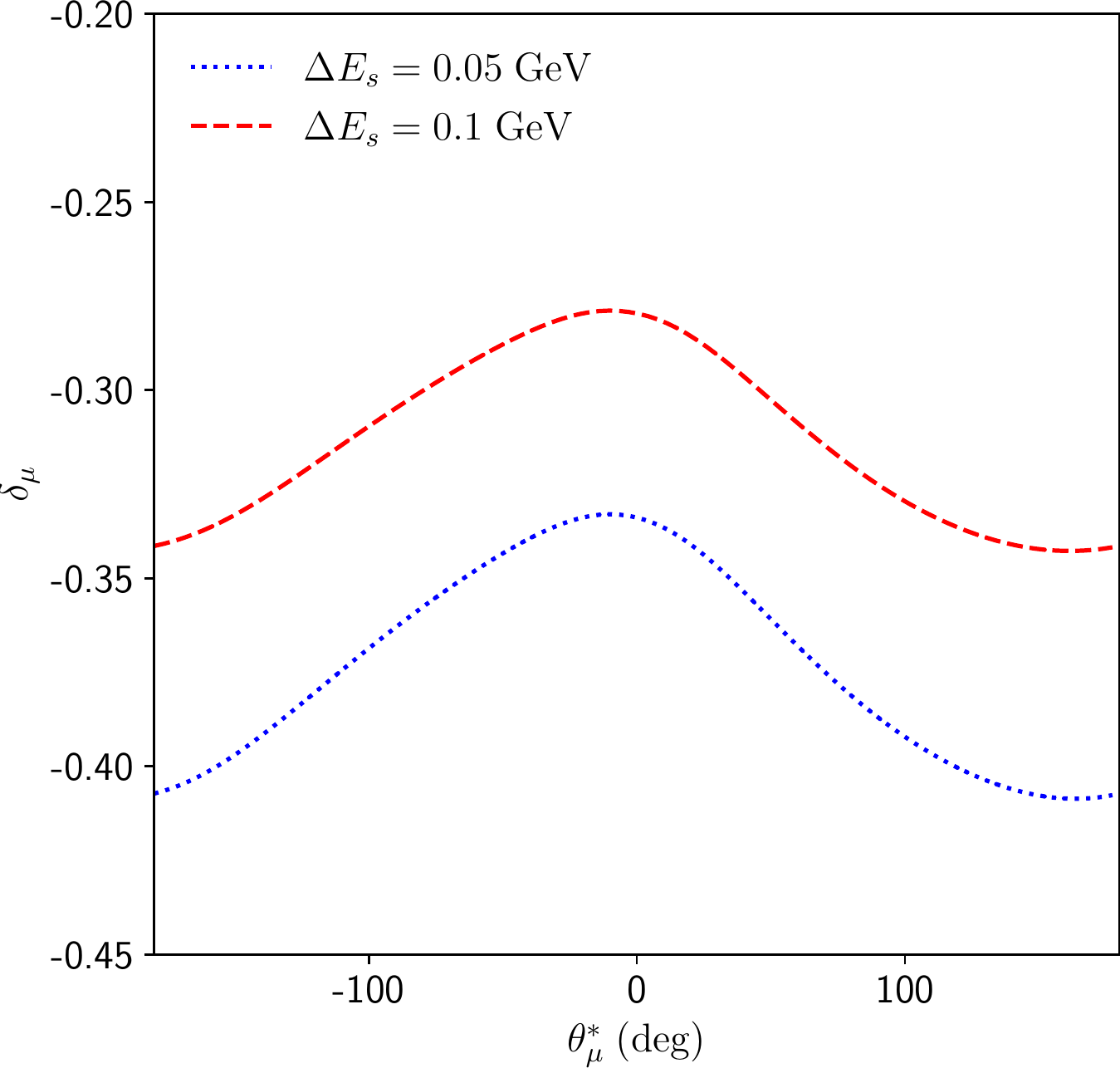}
    \caption{Upper panels: $\theta_l^\ast$ dependence of the radiative corrections to the $e^- p \to e^- p l^-l^+$ cross section  in the DDVCS regime for $e^-e^+$ production (left) and $\mu^-\mu^+$ production (right), for the different classes of radiative corrections for $\Delta E_s = 0.05$~GeV. Lower panels: $\theta_l^\ast$ dependence of the total radiative correction for different values of $\Delta E_s$ for both $e^-e^+$ production (left) and $\mu^-\mu^+$ production (right).}
    \label{fig:panel_delta_classes}
\end{figure*}

In Fig.~\ref{fig:panel_delta_classes} (upper panels) we show the relative size of the corrections to the unpolarized cross sections stemming from the three different classes of diagrams for the central value of the squared di-lepton mass, $s_{ll}=1.0$ GeV$^2$. 
We show both the virtual corrections (labeled "virt"), corresponding with 
\cref{eq:virt_finitesum,eq:virta,eq:virtb,eq:virtc1,eq:virtc2,eq:virtc3,eq:virtc4}, 
as well as the virtual + real soft-photon corrections, with the real corrections given by 
\cref{eq:finite_real,eq:reala,eq:realb,eq:realc,eq:softreal}.

While the corrections of class (a) are just constant and negative, the corrections of class (b) and (c) have a non-trivial dependence on the di-lepton scattering angle $\theta_l^\ast$. For class (b) this dependence is coming entirely from the real photon emission correction, since the di-lepton energies in the soft-photon frame depend on the lepton scattering angle $\theta_l^\ast$, 
see Eq.~\eqref{Eq:leptonenergies}. Comparing the soft-photon radiative corrections to $e^-e^+$ and $\mu^-\mu^+$ production we see that the largest difference is again coming from class (b), which is expected  since it is most sensitive to the lepton mass $m_l$. Let us note that,  as discussed before for the low-energy case, classes (a) and (b) are symmetric with respect to the interchange $l^+$ and $l^-$, while class (c) is anti-symmetric. Therefore to first order only class (c) contributes to the $A_{FB}$. 

In the lower panels of Fig.~\ref{fig:panel_delta_classes} we show the sum of all corrections for two different values of the soft-photon energy cutoff of $\Delta E_s = 0.05\;\rm{GeV}$ (blue dotted curve) and $\Delta E_s = 0.1\;\rm{GeV}$ (red dashed curve). The difference between both curves is a constant proportional to $\ln\Delta E_s^2/m^2$. For the twice higher value of the soft-photon energy we find in absolute values smaller corrections shifted by approximately $10\%$. Furthermore we see that the sum of all corrections is in absolute value smaller by more than $10\%$ for $\mu^-\mu^+$ production compared to $e^-e^+$ production. As mentioned above the difference is coming mainly from corrections of class (b) which are (in absolute value) smaller for $\mu^-\mu^+$ production.

\begin{figure}
\centering
    \includegraphics[width=0.47\textwidth]{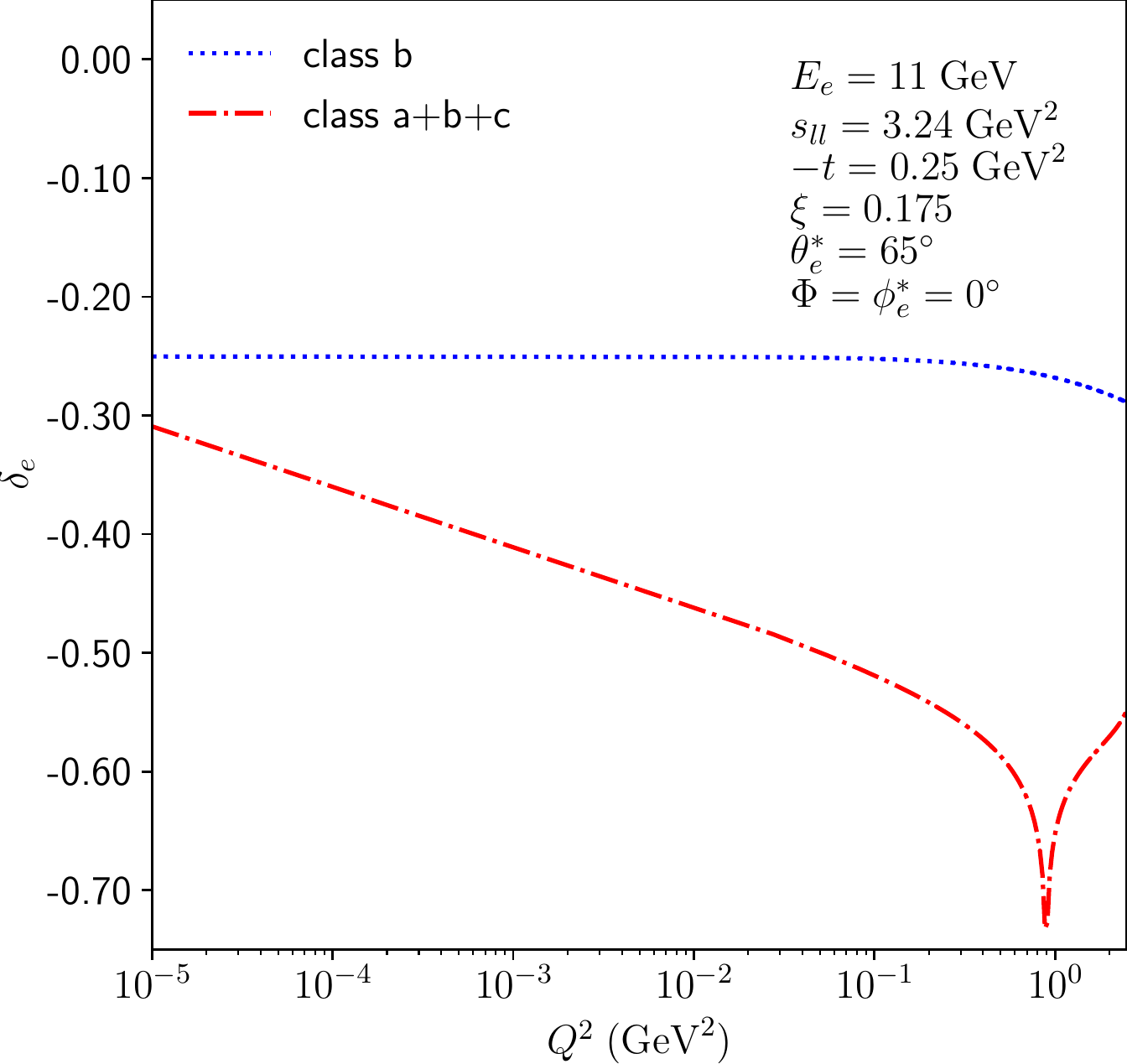}\qquad
    \caption{Radiative correction for the $e^- p \to e^- p e^-e^+$ process in the TCS limit $Q^2\rightarrow 0$ in a kinematic setup comparable to Ref.~\cite{Heller:2020lnm}. }
    \label{fig:panel_delta_real_limit}
\end{figure}

\begin{figure*}
\centering
    \includegraphics[width=0.47\textwidth]{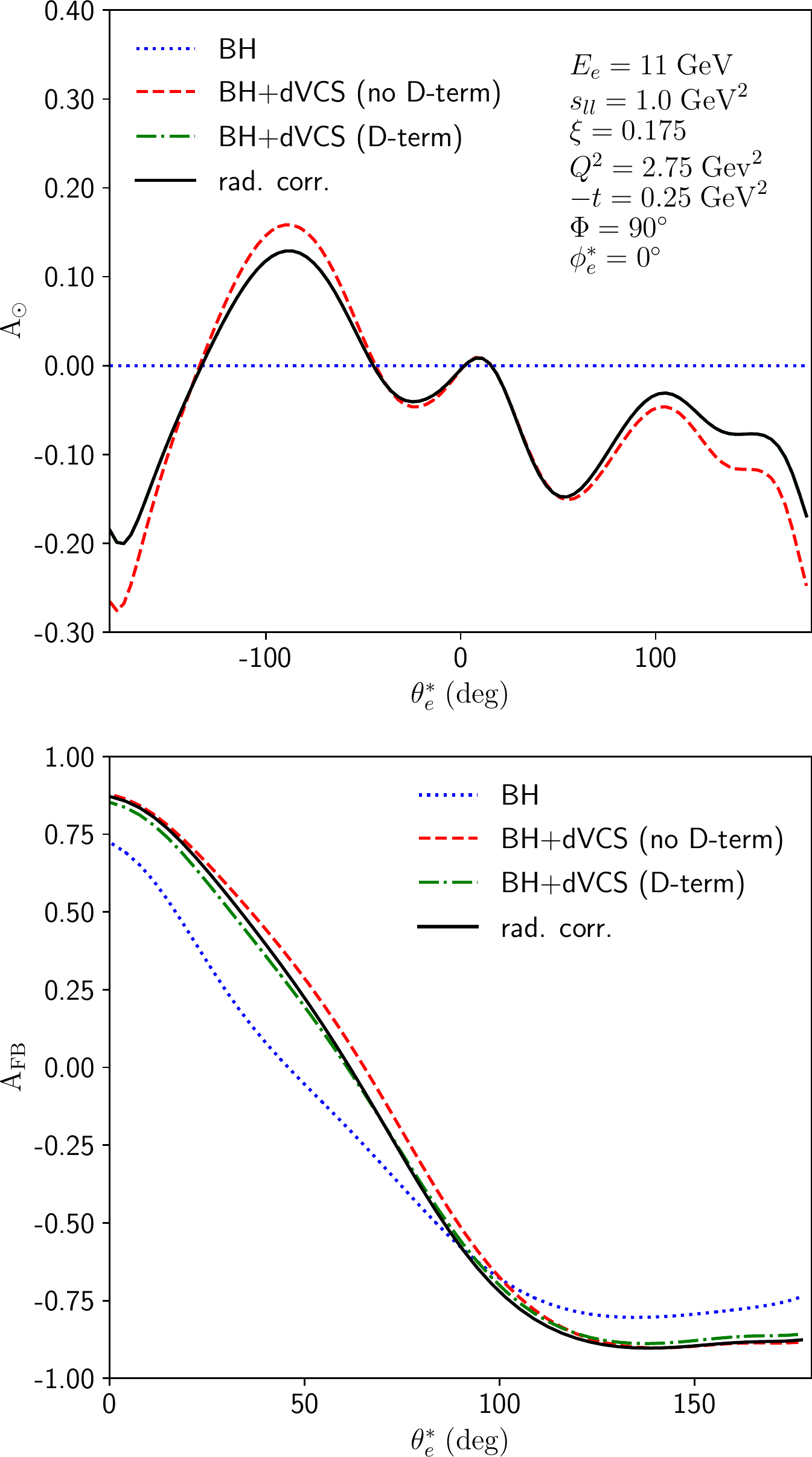}\qquad
    \includegraphics[width=0.47\textwidth]{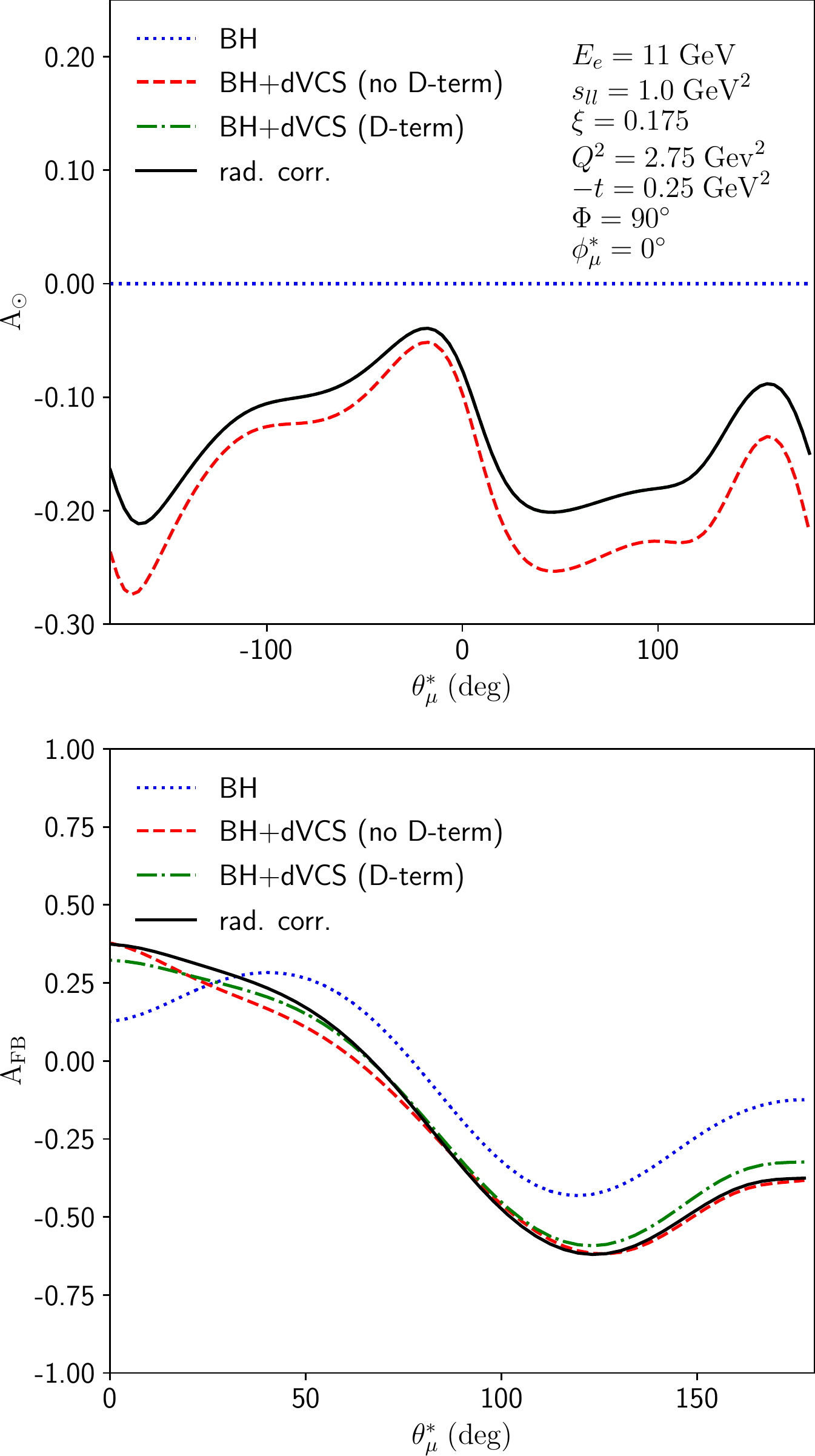}
    \caption{$\theta_l^\ast$ dependence of the $e^- p \to e^- p l^-l^+$ beam-spin asymmetry $A_\odot$ (upper panels) and forward-backward asymmetry $A_{FB}$ (lower panels)  in the DDVCS regime, for $e^-e^+$ production (left) and $\mu^- \mu^+$ production (right). 
    Curve conventions as in Fig.~\ref{fig:cross_sections}.}
    \label{fig:asym_muon_vs_electron}
\end{figure*}

In Fig.~\ref{fig:panel_delta_real_limit} we show the soft-photon radiative corrections for the 
$e^- p \to e^- p e^-e^+$ process,  
for roughly the same kinematics as in 
Ref.~\cite{Heller:2020lnm}, in which we studied the effect of radiative corrections for the timelike Compton scattering (TCS) process, $\gamma p \to e^-e^+ p$, with an on-shell incoming photon. In the soft-photon approximation the corrections to that process are equivalent to that of class (b) studied in the present work. Therefore, we find for corrections of class (b) the same order of magnitude of approximately -25~\% as in \cite{Heller:2020lnm} (blue dotted curve). Including also corrections of class (a) and (c) we find that the corrections increase with increasing $Q^2$ values, varying from -35~\% to -50~\% when varying $Q^2$ from $10^{-5}$ GeV$^2$ to $10^{-1}$ GeV$^2$ (red dashed curve).  Furthermore, one observes around $Q^2=1$ GeV$^2$ the same spiked behavior as in the low-energy case in Fig.~\ref{fig:low_energy_delta_limit}. The reason is again that in this kinematics the two outgoing electrons with momenta $k'$ and $l_-$ become collinear which leads to a large logarithm in the corrections of type (c). For an experiment such kinematic region should be avoided.

In Fig.~\ref{fig:asym_muon_vs_electron} we show a comparison of both, beam-spin and forward-backward asymmetries in the DDVCS regime, both for $e^-e^+$ and $\mu^-\mu^+$ production, and for a di-lepton invariant mass squared of $s_{ll}=1.0$ GeV$^2$. 

Studying the DDVCS process in this energy regime is of particular interest as it allows to extend the DVCS beam spin asymmetry measurements of GPDs into the so-called ERBL domain~\cite{Guidal:2002kt,Belitsky:2002tf}.
The BSA is proportional to the imaginary part of the DDVCS amplitude of Eq.~(\ref{eq:DDVCS}), and allows to access the GPDs directly unlike the real part of the amplitude which depends on a convolution integral over the GPDs. The numerator of the BSA directly yields for both the cases $\xi^\prime > 0$ ($Q^2 > q'^2$) and $\xi^\prime  < 0$ ($Q^2 < q'^2$):
\begin{eqnarray}
\sigma^+ - \sigma^- &=&  c \;  H^{\rm{singlet}}(\xi^\prime, \xi, t) + ...,
\label{eq:bsanum}
\end{eqnarray}
with $-\xi < \xi^\prime < \xi$. 
In Eq.~(\ref{eq:bsanum}) $c$ is a known factor, originating from the BH amplitude dependent on the nucleon elastic form factors, and the ellipses stand for the subdominant contribution of GPDs beyond $H^{\rm{singlet}}$.  

As $H^{\rm{singlet}}$ is an odd function in its first argument, we thus see that the BSA for the DDVCS process changes sign when 
crossing the point $\xi^\prime = 0$. The BSA for the DVCS and TCS limits have the same magnitude but opposite signs, expressing the fact that 
the GPD information content in both limits is the same. 

Given that the real BH process does not yield a BSA by itself, we see from Fig.~\ref{fig:asym_muon_vs_electron} that the BSA  has a significant sensitivity to the GPDs, yielding asymmetries between -25\% and +10\% for $e^-e^+$ production and between -25\% and -5\% for $\mu^- \mu^+$ production. The difference between both cases is mainly due to the effect of anti-symmetrization in both outgoing electrons for $e^-e^+$ production. Furthermore, unlike the DVCS and TCS cases, the BSA for DDVCS is also sensitive to the D-term contribution to the GPD, as it also yields a contribution to the imaginary part of the DDVCS amplitude. By comparing the red dashed and black solid curves in Fig.~\ref{fig:asym_muon_vs_electron}, we notice that the sensitivity to the D-term induces a change of the BSA by 5\% or more over a large angular range.  
As noticed above, the radiative correction drop out of the BSA in the soft-photon approximation. 

We also show the forward-backward asymmetry in Fig.~\ref{fig:asym_muon_vs_electron}, and notice that the anti-symmetrization induces already a large effect for the BH process itself (blue dotted curves in Fig.~\ref{fig:asym_muon_vs_electron}). 
Adding the dVCS contribution changes the forward-backward asymmetry by up to 25\% over a large angular range, while the effect of radiative corrections (black solid curves) is in the few percent range only. The sensitivity on the D-term for the forward-backward asymmetry is much smaller, comparing the curve including the D-term (green dot-dashed line) and the curve excluding the D-term (red dashed line) we find a difference of up to approximately $5\%$.

\begin{figure}
    \centering
    \includegraphics[width=0.47\textwidth]{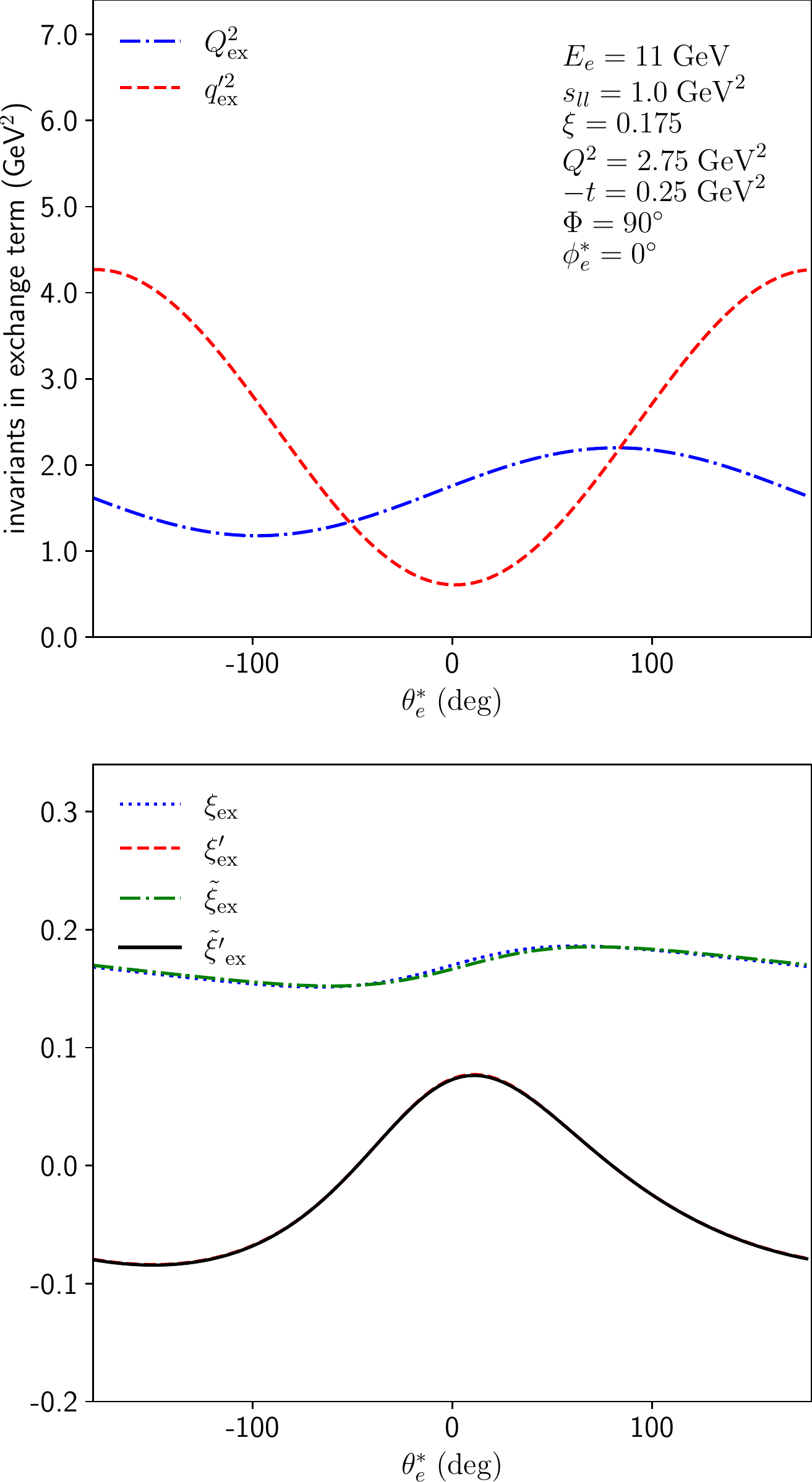}
    \caption{Upper panel: photon virtualities entering the dVCS amplitude in the exchange diagrams of the $e^- p \to e^- p e^-e^+$ process for the kinematics of Fig.~\ref{fig:asym_muon_vs_electron}. Lower panel: scaling variables entering the GPDs for the exchange diagrams.}
    \label{fig:panel_kin_xi}
\end{figure}

For the calculation of the $e^- p \to e^- p e^-e^+$ cross sections shown in Fig.~\ref{fig:cross_sections} (left panels) and the corresponding asymmetries shown in Fig.~\ref{fig:asym_muon_vs_electron} (left panels)  we have to ensure that the model used for the dVCS amplitude is applicable for both the direct and the exchange terms. In Fig.~\ref{fig:panel_kin_xi} we show the two photon virtualities entering the dVCS tensor for the exchange diagrams. One can see that in the kinematics considered one of the two virtualities is around or above 2 GeV$^2$ for nearly all lepton angles, which corresponds with the lower limit for which the QCD-factorization in terms of GPDs is expected to hold. This justifies the use of the handbag description of Sec.~\ref{sec:dvcs_high_energy} in terms of GPDs  also for the exchange term. 

Furthermore in Fig.~\ref{fig:panel_kin_xi} (lower panel) we show the scaling variables entering the DDVCS tensor for the exchange diagrams. While $\xi_{\rm{ex}}$ and $\tilde \xi_{\rm{ex}}$ both are roughly constant as function of the di-lepton angle, close to the value of $\xi = 0.175$ for the direct diagram, one notices a large angular variation for $\xi'_{\rm{ex}}$ and $\tilde\xi'_{\rm{ex}}$. Compared to the constant value $\xi' = 0.0758$ entering the direct diagram, $\xi'_{\rm{ex}}$ varies 
between $-0.08$ and $0.08$. 
Thus the $e^- p \to e^- p e^- e^+$ process has the unique feature, due to the anti-symmetrization in both outgoing electrons, that by varying the di-lepton angle $\theta_e^\ast$ one performs a systematic scan in the scaling variable $\xi'_{\rm{ex}}$ in the ERBL domain of the GPDs.

\begin{figure}
    \centering
    \includegraphics[width=0.47\textwidth]{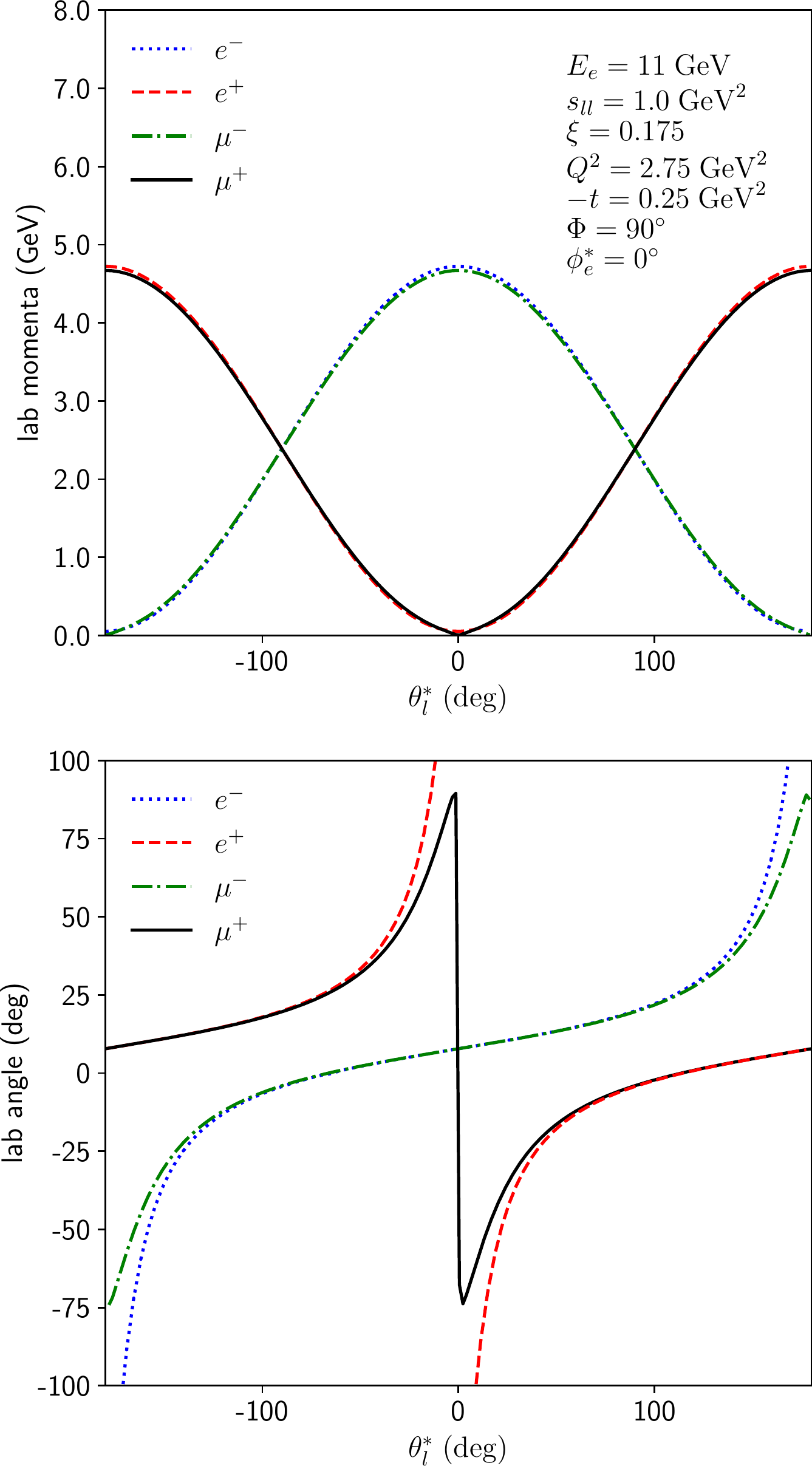}
    \caption{Lab momenta (upper panel) and  Lab scattering angles (lower panel) of the di-lepton pair as function of di-lepton rest frame angle $\theta_l^\ast$ for the kinematics of Fig.~\ref{fig:asym_muon_vs_electron}.}
    \label{fig:panel_LAB_kin}
\end{figure}

In Fig.~\ref{fig:panel_LAB_kin} we also show the di-lepton momenta and scattering angles measured in the Lab frame as function of the di-lepton rest frame angle $\theta_l^\ast$. From the upper panel of that figure it becomes clear, that most of the region is experimental accessible. Except for angles around $0^\circ$ and $\pm 180^\circ$, over most of the di-lepton angular range the lepton momenta are larger than $0.1$ GeV, which makes it quite feasible to detect the particles. The scattering angles measured in the Lab system are shown in the lower panel in Fig.~\ref{fig:panel_LAB_kin}. In the region where the Lab momenta of the di-lepton pair are larger than $0.1$ GeV, they range from $-50^\circ$ to $50^\circ$. This initial study seems promising for  a measurement of the $e^- p \to e^- p e^-e^+$ process over a large range of scattering angles.

\section{Conclusions\label{sec:conclusion}}

In this paper we studied the soft-photon radiative  corrections to the process $e^- p \rightarrow e^- p l^- l^+$, where $l=e$ or $l=\mu$. The process contains two distinct contributions: firstly the space- and time-like Bethe-Heitler processes which only depend on the nucleon elastic form factors, and secondly the double virtual Compton scattering process.  The latter is sensitive to the underlying hadronic model describing the virtual photon-nucleon interaction,  and a measurement of $e^- p \rightarrow e^- p l^- l^+ $ observables can therefore be used to test and study nucleon structure models for different energy regimes. In the present work we studied the $e^- p \rightarrow e^- p l^- l^+$ process in two different energy regimes. 

In the low-energy regime, in which the center-of mass energy is close to the $\Delta(1232)$-resonance, and in which both photon virtualities are typically below or around 0.1 GeV$^2$, we described the interaction using a $\Delta$-pole model together with a low-energy expansion of the dVCS amplitude. This regime is motivated to better constrain the hadronic corrections to precision atomic spectroscopy. In particular for the muonic Hydrogen Lamb shift, the main hadronic unknown to date results from a low-energy nucleon structure constant, denoted by $b_{3, 0}$, which enters the empirical determination of the ${\cal O}(Q^4)$ term in the subtraction function $T_1(0,Q^4)$ of the forward double virtual Compton amplitude.  
We found that the spread between the different theoretical estimates for the low-energy constant 
$b_{3,0}$ increases the $e^- p \rightarrow e^- p l^- l^+ $ cross section by approximately 15\% both for $e^-e^+$ and $\mu^-\mu^+$ production. 
Furthermore, we also found that the 
beam-spin asymmetry and the forward-backward asymmetry, resulting from an interchange in the kinematics of the produced di-lepton pair, are sensitive to the low-energy constant $b_{3,0}$. 
For the beam-spin asymmetry the range of theoretical values for this low-energy constant leads to a shift in the asymmetry up to 15\% for $e^-e^+$ production, and up to around 10\% for $\mu^-\mu^+$ production. 
A measurement of the $e^- p \rightarrow e^- p l^- l^+ $ observables in this low-energy regime is thus promising to 
extract the nucleon structure constant, which could help to reduce the main uncertainty in the theoretical $\mu H$ Lamb shift estimate. 

For the high-energy deeply-virtual regime, we modeled the dVCS amplitude in terms of GPDs. We studied the 
sensitivity of the $e^- p \rightarrow e^- p l^- l^+ $ process to the modeling of the GPDs, in particular the so-called D-term contribution. In kinematics of future experiments at JLab, we found that dispersive estimates for the D-term contribution to the GPDs induce around 20\% change in the 
$e^- p \rightarrow e^- p l^- l^+ $ cross section. 
Furthermore, we also found a large sensitivity to the GPD model for the beam-spin as well as the  forward-backward asymmetry. The beam-spin asymmetry is of particular interest as it does not involve any convolution over GPDs, but is directly proportional to the GPDs, mostly in a  linear way, through interference with the Bethe-Heitler process. For the $e^- p \rightarrow e^- p e^- e^+ $  process, the beam-spin asymmetry has the unique feature, due to the anti-symmetrization in both outgoing electrons, that by varying the di-lepton angle one performs a systematic scan in the average quark momentum fraction in the ERBL domain of the GPDs, due to the exchange term. 

In order to use the $e^- p \rightarrow e^- p l^- l^+ $ process in either the low-energy or high-energy regimes as a probe of nucleon structure, we also studied the QED radiative corrections on the observables, in the soft-photon approximation. We found that the radiative corrections have a large impact on the cross sections. In the low-energy regime we find that these corrections lead to a decrease of the cross section of up to 30\% for $e^-e^+$ production, and around 15\% for $\mu^-\mu^+$ production. 
In the high-energy deeply-virtual regime, the corrections even range up to 50\% for $e^-e^+$ production, and around 35\% for $\mu^-\mu^+$ production in JLab kinematics. For the forward-backward and beam-spin asymmetries the situation is different. For the $A_{FB}$ the radiative corrections were found to affect the asymmetry only around or below the 1\% level, whereas the beam-spin asymmetry is not affected at all in the soft-photon approximation. A combined analysis of the cross section and of both asymmetries thus holds promise to access the hadronic structure information in both regimes.

A next step to interpret future measurements of 
$e^- p \rightarrow e^- p l^- l^+ $ observables, 
would consist in performing a full one-loop radiative correction calculation, beyond the soft-photon approximation.  Such calculation can build upon the work of  Refs.~\cite{Heller:2019dyv,Heller:2020lnm}, in which such study was performed for the related $\gamma p \rightarrow l^- l^+ p$ process. The latter study has shown that the soft-photon approximation can be expected to somewhat over-estimate the full one-loop corrections on the cross sections, while the beam-spin and forward-backward asymmetries remain nearly unaffected by the radiative corrections.

\section*{Acknowledgements}
This work was supported by the Deutsche Forschungsgemeinschaft (DFG, German Research Foundation), in part through the Collaborative Research Center [The Low-Energy Frontier of the Standard Model, Projektnummer 204404729 - SFB 1044], and in part through the Cluster of Excellence [Precision Physics, Fundamental Interactions, and Structure of Matter] (PRISMA$^+$ EXC 2118/1) within the German Excellence Strategy (Project ID 39083149).

\appendix

\section{Kinematics in $\gamma^* p$ rest frame}
\label{appendix:kin}

In this appendix, we derive expressions for the four-momenta in the rest frame of proton and the momentum transfer of the scattered electron, i.e.
\begin{equation}
    \vec{q}=\vec{k}-\vec{k}^\prime=-\vec{p}.
\end{equation}
We align this system along the $z$-axis, such that the energy of the virtual photon with momentum $q$ is given by
\begin{equation}
    q^0_{cm}=\frac{W^2-M^2-Q^2}{2 W},
\end{equation}
and the z-component of the three-momentum by
\begin{align}
  q_{cm}  &\equiv q_z\nonumber\\
  &=\frac{1}{2W}\left[\left(\left(W+M\right)^2+Q^2\right)\left(\left(W-M\right)^2+Q^2\right)\right]^{1/2}.
\end{align}
The energy of the incoming electron with momentum $k$ is given by
\begin{equation}
    k^0=\frac{s-M^2-m^2-Q^2}{2 W}.
\end{equation}
In order to write down the three-momentum $\vec{k}$, we define $r_k$, which is the magnitude of the three-momentum in $x$- and $y$-direction:
\begin{eqnarray}
    r_k &=&  \frac{Q}{2W q_{cm}}\left[s (s - M^2 - Q^2) - W^2 (s-M^2) \right. \nonumber \\
    &-&\left.  m^2\left((W^2 - M^2)^2/Q^2 + 2s + W^2 + M^2\right)+m^4 \right]^{1/2}, \nonumber \\
    \end{eqnarray}
such that $\vec{k}$ is given by

\begin{align}
    k_x&=r_k \cos \Phi\nonumber\\
    k_y&=r_k \sin \Phi\nonumber\\
    k_z&=\frac{W Q^2+q_{cm}^0(s-m^2-M^2-Q^2)}{2q_{cm} W}.
\end{align}

Using these quantities, we can write down the momenta of the incoming and outgoing electron $k$ and $k'$ as
\begin{equation}
    k=\left(\begin{array}{c} k^0  \\ r_k \cos \Phi \\ r_k\sin \Phi \\ k_z \end{array}\right) \qquad k'=\left(\begin{array}{c} k^0-  q^0_{cm} \\ r_k \cos \Phi \\ r_k\sin \Phi \\ k_z - q_{cm} \end{array}\right).\vspace{0.5cm}
\end{equation}
The energy of the photon with momentum  $q'$ is given by
\begin{equation}
    q'^0_{cm}=\frac{W^2-M^2+s_{ll}}{2 W},
\end{equation}
and its three-momentum by
\begin{align}
    q'_{cm}\equiv& \lvert\vec{q'}_{cm}\rvert\nonumber\\
    =&\frac{1}{2W}\left[\left(\left(W+M\right)^2-s_{ll}\right)\left(\left(W-M\right)^2-s_{ll}\right)\right]^{1/2}.
\end{align}
The angle $\theta_{\gamma\gamma}$ is defined as the angle between the two virtual photons with momenta $q$ and $q^\prime$. It can be calculated in terms of invariants via:
\begin{eqnarray}
    2 q_{cm} q'_{cm} \cos(\theta_{\gamma\gamma})&=&\frac{(W^2-M^2-Q^2)(W^2-M^2+s_{ll})}{2W^2} \nonumber\\
    &+&t-s_{ll}+Q^2.
\end{eqnarray}
We can now write down the four-momentum of $l_-$:
\begin{widetext}
\begin{align}
    l_-=\left(\begin{array}{c}  \frac{ q'^0_{cm} }{2} \left( 1+ \frac{ q'_{cm} }{  q'^0_{cm}  } \beta_{s_{ll}} 
    \cos \theta_l^\ast \right)\\    \frac{ q'^0_{cm} }{2} \left( \beta_{s_{ll}} \cos \theta_l^\ast + \frac{ q'_{cm} }{  q'^0_{cm}  }  \right) \sin(\theta_{\gamma\gamma})+\frac{\sqrt{s_{ll}}}{2} \beta_{s_{ll}} \sin \theta_l^\ast \cos \phi_l^\ast \cos(\theta_{\gamma\gamma})\\ \frac{\sqrt{s_{ll}}}{2} \beta_{s_{ll}} \sin \theta_l^\ast \sin \phi_l^\ast \\ \frac{ q'^0_{cm} }{2} \left( \beta_{s_{ll}} \cos \theta_l^\ast + \frac{ q'_{cm} }{  q'^0_{cm}  }  \right) \cos(\theta_{\gamma\gamma})-\frac{\sqrt{s_{ll}}}{2} \beta_{s_{ll}} \sin \theta_l^\ast \cos \phi_l^\ast \sin(\theta_{\gamma\gamma}) \end{array}\right).
\end{align}
\end{widetext}
The momentum $l_+$ of the other lepton can be obtained via the transformation
\begin{align}
    &\cos \theta_l^\ast \rightarrow -\cos \theta_l^\ast, \quad \cos \phi_l^\ast \rightarrow -\cos \phi_l^\ast,\nonumber\\
    &\sin \phi_l^\ast \rightarrow -\sin \phi_l^\ast.
\end{align}
The momentum $p$ of the incoming proton is aligned to the $z$-axis. The energy and the $z$-component are given by
\begin{equation}
    p^0=\sqrt{M^2+q_{cm}^2},\quad p_z=-q_{cm}.
\end{equation}
The momentum $p'$ of the outgoing proton can be calculated using energy-momentum conservation. The energy and three-momentum are given by
\begin{equation}
   p'^0=\sqrt{M^2 + \lvert \vec{q'} \rvert ^2}, \qquad \vec{p'}=-\vec{q'} .
\end{equation}

Having derived all four-momenta in the $\gamma^* p$ center-of-mass frame, one can easily perform a Lorentz transformation to get the four-momenta in any other system. In particular, one can perform the boost to the recoil proton + soft-photon rest frame, which is needed to calculate the soft-photon integrals from Sec. \ref{sec:bremsstrahlung}.

\section{Three-point functions}
\label{app:threepoint}

In this appendix we give analytic expressions for the three-point function which we need for the virtual soft-photon corrections. The results are taken from Ref.~\cite{Ellis:2007qk}. The three-point function with equal masses is given by
\begin{align}
&C_0\left(m^2, s, m^2,0, m^2, m^2\right)=\frac{1}{s\beta}\left\{\frac{1}{\epsilon_\text{IR}}\ln \left(\frac{\beta-1}{\beta+1}\right)\right.\nonumber\\
&\hspace{1cm}\left.+2\;\text{Li}_2\left(\frac{\beta-1}{2\beta}\right)+\ln^2\left(\frac{\beta-1}{2\beta}\right)\right.\nonumber\\
&\hspace{1cm}\left.-\frac{1}{2}\ln^2\left(\frac{\beta-1}{\beta+1}\right)-\frac{\pi^2}{6}\right\},\label{C0-eucl}
\end{align}
where we defined
\begin{equation}
    \beta=\sqrt{1-\frac{4m^2}{s}}.
\end{equation}
The above expression is valid in the space-like region in which $s<0$.

The three-point function with two different masses $m$ and $m_l$ is given by
\begin{widetext}
\begin{align}
    C_0\left( m^2, s, m_l^2;0,m^2,m_l^2 \right)=&\frac{1}{2 \lambda}\left(-\frac{\mu ^2}{ s}\right)^{\epsilon_\text{IR} }\Bigg\{    \frac{1}{\epsilon_\text{IR}} \ln\left(x_+ x_- \right)    -\ln(-\frac{\lambda}{s}) \left( \ln(x_+) + \ln(x_-)\right) + \frac{1}{2}    \ln^2 \left(  -\gamma^+  \right) \nonumber \\ 
    &+ \frac{1}{2} \ln^2 \left(  \gamma^- -1    \right)  -\frac{1}{2} \ln^2 \left(  -\gamma^-  \right)   + \frac{1}{2}  \ln^2 \left( 1 - \gamma^+ \right) - \text{Li}_2\left( \frac{s- s\gamma^-}{\lambda}\right) \nonumber \\
    &- \text{Li}_2 \left(\frac{s\gamma^+}{\lambda} \right) + \text{Li}_2 \left( \frac{s\gamma^+ -s} {\lambda} \right) + \text{Li}_2 \left( \frac{s\gamma^-}{\lambda} \right) \Bigg\},\label{Eq:C02massEucl}
\end{align}
with
\begin{align}
    \lambda=\sqrt{(-s  + m^2 +m_l^2)^2-4m^2m_l^2},\quad
    \gamma_{\pm}=\frac{1}{2}\left[ 1+\frac{m_l^2-m^2}{s} \pm \frac{\lambda}{s} \right],\quad
    x_-=\frac{-\gamma_-}{1-\gamma_-},\qquad x_+=\frac{\gamma_+-1}{\gamma_+}.
\end{align}
\end{widetext}
As above, the expression is valid in the space-like region with $s<0$.

\cleardoublepage

\bibliography{biblio_dvcs}

\begin{thebibliography}{55}%
\makeatletter
\providecommand \@ifxundefined [1]{%
 \@ifx{#1\undefined}
}%
\providecommand \@ifnum [1]{%
 \ifnum #1\expandafter \@firstoftwo
 \else \expandafter \@secondoftwo
 \fi
}%
\providecommand \@ifx [1]{%
 \ifx #1\expandafter \@firstoftwo
 \else \expandafter \@secondoftwo
 \fi
}%
\providecommand \natexlab [1]{#1}%
\providecommand \enquote  [1]{``#1''}%
\providecommand \bibnamefont  [1]{#1}%
\providecommand \bibfnamefont [1]{#1}%
\providecommand \citenamefont [1]{#1}%
\providecommand \href@noop [0]{\@secondoftwo}%
\providecommand \href [0]{\begingroup \@sanitize@url \@href}%
\providecommand \@href[1]{\@@startlink{#1}\@@href}%
\providecommand \@@href[1]{\endgroup#1\@@endlink}%
\providecommand \@sanitize@url [0]{\catcode `\\12\catcode `\$12\catcode
  `\&12\catcode `\#12\catcode `\^12\catcode `\_12\catcode `\%12\relax}%
\providecommand \@@startlink[1]{}%
\providecommand \@@endlink[0]{}%
\providecommand \url  [0]{\begingroup\@sanitize@url \@url }%
\providecommand \@url [1]{\endgroup\@href {#1}{\urlprefix }}%
\providecommand \urlprefix  [0]{URL }%
\providecommand \Eprint [0]{\href }%
\providecommand \doibase [0]{http://dx.doi.org/}%
\providecommand \selectlanguage [0]{\@gobble}%
\providecommand \bibinfo  [0]{\@secondoftwo}%
\providecommand \bibfield  [0]{\@secondoftwo}%
\providecommand \translation [1]{[#1]}%
\providecommand \BibitemOpen [0]{}%
\providecommand \bibitemStop [0]{}%
\providecommand \bibitemNoStop [0]{.\EOS\space}%
\providecommand \EOS [0]{\spacefactor3000\relax}%
\providecommand \BibitemShut  [1]{\csname bibitem#1\endcsname}%
\let\auto@bib@innerbib\@empty
\bibitem [{\citenamefont {Drechsel}\ \emph {et~al.}(2003)\citenamefont
  {Drechsel}, \citenamefont {Pasquini},\ and\ \citenamefont
  {Vanderhaeghen}}]{Drechsel:2002ar}%
  \BibitemOpen
  \bibfield  {author} {\bibinfo {author} {\bibfnamefont {D.}~\bibnamefont
  {Drechsel}}, \bibinfo {author} {\bibfnamefont {B.}~\bibnamefont {Pasquini}},
  \ and\ \bibinfo {author} {\bibfnamefont {M.}~\bibnamefont {Vanderhaeghen}},\
  }\href {\doibase 10.1016/S0370-1573(02)00636-1} {\bibfield  {journal}
  {\bibinfo  {journal} {Phys. Rept.}\ }\textbf {\bibinfo {volume} {378}},\
  \bibinfo {pages} {99} (\bibinfo {year} {2003})}\BibitemShut {NoStop}%
\bibitem [{\citenamefont {Griesshammer}\ \emph {et~al.}(2012)\citenamefont
  {Griesshammer}, \citenamefont {McGovern}, \citenamefont {Phillips},\ and\
  \citenamefont {Feldman}}]{Griesshammer:2012we}%
  \BibitemOpen
  \bibfield  {author} {\bibinfo {author} {\bibfnamefont {H.~W.}\ \bibnamefont
  {Griesshammer}}, \bibinfo {author} {\bibfnamefont {J.~A.}\ \bibnamefont
  {McGovern}}, \bibinfo {author} {\bibfnamefont {D.~R.}\ \bibnamefont
  {Phillips}}, \ and\ \bibinfo {author} {\bibfnamefont {G.}~\bibnamefont
  {Feldman}},\ }\href {\doibase 10.1016/j.ppnp.2012.04.003} {\bibfield
  {journal} {\bibinfo  {journal} {Prog. Part. Nucl. Phys.}\ }\textbf {\bibinfo
  {volume} {67}},\ \bibinfo {pages} {841} (\bibinfo {year} {2012})}\BibitemShut
  {NoStop}%
\bibitem [{\citenamefont {Hagelstein}\ \emph {et~al.}(2016)\citenamefont
  {Hagelstein}, \citenamefont {Miskimen},\ and\ \citenamefont
  {Pascalutsa}}]{Hagelstein:2015egb}%
  \BibitemOpen
  \bibfield  {author} {\bibinfo {author} {\bibfnamefont {F.}~\bibnamefont
  {Hagelstein}}, \bibinfo {author} {\bibfnamefont {R.}~\bibnamefont
  {Miskimen}}, \ and\ \bibinfo {author} {\bibfnamefont {V.}~\bibnamefont
  {Pascalutsa}},\ }\href {\doibase 10.1016/j.ppnp.2015.12.001} {\bibfield
  {journal} {\bibinfo  {journal} {Prog. Part. Nucl. Phys.}\ }\textbf {\bibinfo
  {volume} {88}},\ \bibinfo {pages} {29} (\bibinfo {year} {2016})}\BibitemShut
  {NoStop}%
\bibitem [{\citenamefont {Pasquini}\ and\ \citenamefont
  {Vanderhaeghen}(2018)}]{Pasquini:2018wbl}%
  \BibitemOpen
  \bibfield  {author} {\bibinfo {author} {\bibfnamefont {B.}~\bibnamefont
  {Pasquini}}\ and\ \bibinfo {author} {\bibfnamefont {M.}~\bibnamefont
  {Vanderhaeghen}},\ }\href {\doibase 10.1146/annurev-nucl-101917-020843}
  {\bibfield  {journal} {\bibinfo  {journal} {Ann. Rev. Nucl. Part. Sci.}\
  }\textbf {\bibinfo {volume} {68}},\ \bibinfo {pages} {75} (\bibinfo {year}
  {2018})}\BibitemShut {NoStop}%
\bibitem [{\citenamefont {Guichon}\ \emph {et~al.}(1995)\citenamefont
  {Guichon}, \citenamefont {Liu},\ and\ \citenamefont
  {Thomas}}]{Guichon:1995pu}%
  \BibitemOpen
  \bibfield  {author} {\bibinfo {author} {\bibfnamefont {P.~A.}\ \bibnamefont
  {Guichon}}, \bibinfo {author} {\bibfnamefont {G.}~\bibnamefont {Liu}}, \ and\
  \bibinfo {author} {\bibfnamefont {A.~W.}\ \bibnamefont {Thomas}},\ }\href
  {\doibase 10.1016/0375-9474(95)00217-O} {\bibfield  {journal} {\bibinfo
  {journal} {Nucl. Phys. A}\ }\textbf {\bibinfo {volume} {591}},\ \bibinfo
  {pages} {606} (\bibinfo {year} {1995})}\BibitemShut {NoStop}%
\bibitem [{\citenamefont {Guichon}\ and\ \citenamefont
  {Vanderhaeghen}(1998)}]{Guichon:1998xv}%
  \BibitemOpen
  \bibfield  {author} {\bibinfo {author} {\bibfnamefont {P.~A.~M.}\
  \bibnamefont {Guichon}}\ and\ \bibinfo {author} {\bibfnamefont
  {M.}~\bibnamefont {Vanderhaeghen}},\ }\href {\doibase
  10.1016/S0146-6410(98)00056-8} {\bibfield  {journal} {\bibinfo  {journal}
  {Prog. Part. Nucl. Phys.}\ }\textbf {\bibinfo {volume} {41}},\ \bibinfo
  {pages} {125} (\bibinfo {year} {1998})}\BibitemShut {NoStop}%
\bibitem [{\citenamefont {Fonvieille}\ \emph {et~al.}(2020)\citenamefont
  {Fonvieille}, \citenamefont {Pasquini},\ and\ \citenamefont
  {Sparveris}}]{Fonvieille:2019eyf}%
  \BibitemOpen
  \bibfield  {author} {\bibinfo {author} {\bibfnamefont {H.}~\bibnamefont
  {Fonvieille}}, \bibinfo {author} {\bibfnamefont {B.}~\bibnamefont
  {Pasquini}}, \ and\ \bibinfo {author} {\bibfnamefont {N.}~\bibnamefont
  {Sparveris}},\ }\href {\doibase 10.1016/j.ppnp.2020.103754} {\bibfield
  {journal} {\bibinfo  {journal} {Prog. Part. Nucl. Phys.}\ }\textbf {\bibinfo
  {volume} {113}},\ \bibinfo {pages} {103754} (\bibinfo {year}
  {2020})}\BibitemShut {NoStop}%
\bibitem [{\citenamefont {Gorchtein}\ \emph {et~al.}(2010)\citenamefont
  {Gorchtein}, \citenamefont {Lorce}, \citenamefont {Pasquini},\ and\
  \citenamefont {Vanderhaeghen}}]{Gorchtein:2009qq}%
  \BibitemOpen
  \bibfield  {author} {\bibinfo {author} {\bibfnamefont {M.}~\bibnamefont
  {Gorchtein}}, \bibinfo {author} {\bibfnamefont {C.}~\bibnamefont {Lorce}},
  \bibinfo {author} {\bibfnamefont {B.}~\bibnamefont {Pasquini}}, \ and\
  \bibinfo {author} {\bibfnamefont {M.}~\bibnamefont {Vanderhaeghen}},\ }\href
  {\doibase 10.1103/PhysRevLett.104.112001} {\bibfield  {journal} {\bibinfo
  {journal} {Phys. Rev. Lett.}\ }\textbf {\bibinfo {volume} {104}},\ \bibinfo
  {pages} {112001} (\bibinfo {year} {2010})}\BibitemShut {NoStop}%
\bibitem [{\citenamefont {Pohl}\ \emph {et~al.}(2010)\citenamefont {Pohl} \emph
  {et~al.}}]{Pohl:2010zza}%
  \BibitemOpen
  \bibfield  {author} {\bibinfo {author} {\bibfnamefont {R.}~\bibnamefont
  {Pohl}} \emph {et~al.},\ }\href {\doibase 10.1038/nature09250} {\bibfield
  {journal} {\bibinfo  {journal} {Nature}\ }\textbf {\bibinfo {volume} {466}},\
  \bibinfo {pages} {213} (\bibinfo {year} {2010})}\BibitemShut {NoStop}%
\bibitem [{\citenamefont {Mohr}\ \emph {et~al.}(2012)\citenamefont {Mohr},
  \citenamefont {Taylor},\ and\ \citenamefont {Newell}}]{RevModPhys.84.1527}%
  \BibitemOpen
  \bibfield  {author} {\bibinfo {author} {\bibfnamefont {P.~J.}\ \bibnamefont
  {Mohr}}, \bibinfo {author} {\bibfnamefont {B.~N.}\ \bibnamefont {Taylor}}, \
  and\ \bibinfo {author} {\bibfnamefont {D.~B.}\ \bibnamefont {Newell}},\
  }\href {\doibase 10.1103/RevModPhys.84.1527} {\bibfield  {journal} {\bibinfo
  {journal} {Rev. Mod. Phys.}\ }\textbf {\bibinfo {volume} {84}},\ \bibinfo
  {pages} {1527} (\bibinfo {year} {2012})}\BibitemShut {NoStop}%
\bibitem [{\citenamefont {Pohl}\ \emph {et~al.}(2013)\citenamefont {Pohl},
  \citenamefont {Gilman}, \citenamefont {Miller},\ and\ \citenamefont
  {Pachucki}}]{Pohl:2013yb}%
  \BibitemOpen
  \bibfield  {author} {\bibinfo {author} {\bibfnamefont {R.}~\bibnamefont
  {Pohl}}, \bibinfo {author} {\bibfnamefont {R.}~\bibnamefont {Gilman}},
  \bibinfo {author} {\bibfnamefont {G.~A.}\ \bibnamefont {Miller}}, \ and\
  \bibinfo {author} {\bibfnamefont {K.}~\bibnamefont {Pachucki}},\ }\href
  {\doibase 10.1146/annurev-nucl-102212-170627} {\bibfield  {journal} {\bibinfo
   {journal} {Ann. Rev. Nucl. Part. Sci.}\ }\textbf {\bibinfo {volume} {63}},\
  \bibinfo {pages} {175} (\bibinfo {year} {2013})}\BibitemShut {NoStop}%
\bibitem [{\citenamefont {Carlson}(2015)}]{Carlson:2015jba}%
  \BibitemOpen
  \bibfield  {author} {\bibinfo {author} {\bibfnamefont {C.~E.}\ \bibnamefont
  {Carlson}},\ }\href {\doibase 10.1016/j.ppnp.2015.01.002} {\bibfield
  {journal} {\bibinfo  {journal} {Prog. Part. Nucl. Phys.}\ }\textbf {\bibinfo
  {volume} {82}},\ \bibinfo {pages} {59} (\bibinfo {year} {2015})}\BibitemShut
  {NoStop}%
\bibitem [{\citenamefont {Gao}\ and\ \citenamefont
  {Vanderhaeghen}(2021)}]{Gao:2021sml}%
  \BibitemOpen
  \bibfield  {author} {\bibinfo {author} {\bibfnamefont {H.}~\bibnamefont
  {Gao}}\ and\ \bibinfo {author} {\bibfnamefont {M.}~\bibnamefont
  {Vanderhaeghen}},\ }\href@noop {} {\  (\bibinfo {year} {2021})},\ \Eprint
  {http://arxiv.org/abs/2105.00571} {arXiv:2105.00571 [hep-ph]} \BibitemShut
  {NoStop}%
\bibitem [{\citenamefont {Carlson}\ and\ \citenamefont
  {Vanderhaeghen}(2011)}]{Carlson:2011zd}%
  \BibitemOpen
  \bibfield  {author} {\bibinfo {author} {\bibfnamefont {C.~E.}\ \bibnamefont
  {Carlson}}\ and\ \bibinfo {author} {\bibfnamefont {M.}~\bibnamefont
  {Vanderhaeghen}},\ }\href {\doibase 10.1103/PhysRevA.84.020102} {\bibfield
  {journal} {\bibinfo  {journal} {Phys. Rev. A}\ }\textbf {\bibinfo {volume}
  {84}},\ \bibinfo {pages} {020102} (\bibinfo {year} {2011})}\BibitemShut
  {NoStop}%
\bibitem [{\citenamefont {Birse}\ and\ \citenamefont
  {McGovern}(2012)}]{Birse:2012eb}%
  \BibitemOpen
  \bibfield  {author} {\bibinfo {author} {\bibfnamefont {M.~C.}\ \bibnamefont
  {Birse}}\ and\ \bibinfo {author} {\bibfnamefont {J.~A.}\ \bibnamefont
  {McGovern}},\ }\href {\doibase 10.1140/epja/i2012-12120-8} {\bibfield
  {journal} {\bibinfo  {journal} {Eur. Phys. J. A}\ }\textbf {\bibinfo {volume}
  {48}},\ \bibinfo {pages} {120} (\bibinfo {year} {2012})}\BibitemShut
  {NoStop}%
\bibitem [{\citenamefont {Antognini}\ \emph {et~al.}(2013)\citenamefont
  {Antognini}, \citenamefont {Kottmann}, \citenamefont {Biraben}, \citenamefont
  {Indelicato}, \citenamefont {Nez},\ and\ \citenamefont
  {Pohl}}]{Antognini:2013rsa}%
  \BibitemOpen
  \bibfield  {author} {\bibinfo {author} {\bibfnamefont {A.}~\bibnamefont
  {Antognini}}, \bibinfo {author} {\bibfnamefont {F.}~\bibnamefont {Kottmann}},
  \bibinfo {author} {\bibfnamefont {F.}~\bibnamefont {Biraben}}, \bibinfo
  {author} {\bibfnamefont {P.}~\bibnamefont {Indelicato}}, \bibinfo {author}
  {\bibfnamefont {F.}~\bibnamefont {Nez}}, \ and\ \bibinfo {author}
  {\bibfnamefont {R.}~\bibnamefont {Pohl}},\ }\href {\doibase
  10.1016/j.aop.2012.12.003} {\bibfield  {journal} {\bibinfo  {journal} {Annals
  Phys.}\ }\textbf {\bibinfo {volume} {331}},\ \bibinfo {pages} {127} (\bibinfo
  {year} {2013})}\BibitemShut {NoStop}%
\bibitem [{\citenamefont {Zyla}\ \emph {et~al.}(2020)\citenamefont {Zyla} \emph
  {et~al.}}]{ParticleDataGroup:2020ssz}%
  \BibitemOpen
  \bibfield  {author} {\bibinfo {author} {\bibfnamefont {P.~A.}\ \bibnamefont
  {Zyla}} \emph {et~al.} (\bibinfo {collaboration} {Particle Data Group}),\
  }\href {\doibase 10.1093/ptep/ptaa104} {\bibfield  {journal} {\bibinfo
  {journal} {PTEP}\ }\textbf {\bibinfo {volume} {2020}},\ \bibinfo {pages}
  {083C01} (\bibinfo {year} {2020})}\BibitemShut {NoStop}%
\bibitem [{\citenamefont {Lensky}\ \emph {et~al.}(2018)\citenamefont {Lensky},
  \citenamefont {Hagelstein}, \citenamefont {Pascalutsa},\ and\ \citenamefont
  {Vanderhaeghen}}]{Lensky:2017bwi}%
  \BibitemOpen
  \bibfield  {author} {\bibinfo {author} {\bibfnamefont {V.}~\bibnamefont
  {Lensky}}, \bibinfo {author} {\bibfnamefont {F.}~\bibnamefont {Hagelstein}},
  \bibinfo {author} {\bibfnamefont {V.}~\bibnamefont {Pascalutsa}}, \ and\
  \bibinfo {author} {\bibfnamefont {M.}~\bibnamefont {Vanderhaeghen}},\ }\href
  {\doibase 10.1103/PhysRevD.97.074012} {\bibfield  {journal} {\bibinfo
  {journal} {Phys. Rev. D}\ }\textbf {\bibinfo {volume} {97}},\ \bibinfo
  {pages} {074012} (\bibinfo {year} {2018})}\BibitemShut {NoStop}%
\bibitem [{\citenamefont {Alarcon}\ \emph {et~al.}(2014)\citenamefont
  {Alarcon}, \citenamefont {Lensky},\ and\ \citenamefont
  {Pascalutsa}}]{Alarcon:2013cba}%
  \BibitemOpen
  \bibfield  {author} {\bibinfo {author} {\bibfnamefont {J.~M.}\ \bibnamefont
  {Alarcon}}, \bibinfo {author} {\bibfnamefont {V.}~\bibnamefont {Lensky}}, \
  and\ \bibinfo {author} {\bibfnamefont {V.}~\bibnamefont {Pascalutsa}},\
  }\href {\doibase 10.1140/epjc/s10052-014-2852-0} {\bibfield  {journal}
  {\bibinfo  {journal} {Eur. Phys. J. C}\ }\textbf {\bibinfo {volume} {74}},\
  \bibinfo {pages} {2852} (\bibinfo {year} {2014})}\BibitemShut {NoStop}%
\bibitem [{\citenamefont {Tomalak}\ and\ \citenamefont
  {Vanderhaeghen}(2016)}]{Tomalak:2015hva}%
  \BibitemOpen
  \bibfield  {author} {\bibinfo {author} {\bibfnamefont {O.}~\bibnamefont
  {Tomalak}}\ and\ \bibinfo {author} {\bibfnamefont {M.}~\bibnamefont
  {Vanderhaeghen}},\ }\href {\doibase 10.1140/epjc/s10052-016-3966-3}
  {\bibfield  {journal} {\bibinfo  {journal} {Eur. Phys. J. C}\ }\textbf
  {\bibinfo {volume} {76}},\ \bibinfo {pages} {125} (\bibinfo {year}
  {2016})}\BibitemShut {NoStop}%
\bibitem [{\citenamefont {Pauk}\ \emph {et~al.}(2020)\citenamefont {Pauk},
  \citenamefont {Carlson},\ and\ \citenamefont {Vanderhaeghen}}]{Pauk:2020gjv}%
  \BibitemOpen
  \bibfield  {author} {\bibinfo {author} {\bibfnamefont {V.}~\bibnamefont
  {Pauk}}, \bibinfo {author} {\bibfnamefont {C.~E.}\ \bibnamefont {Carlson}}, \
  and\ \bibinfo {author} {\bibfnamefont {M.}~\bibnamefont {Vanderhaeghen}},\
  }\href {\doibase 10.1103/PhysRevC.102.035201} {\bibfield  {journal} {\bibinfo
   {journal} {Phys. Rev. C}\ }\textbf {\bibinfo {volume} {102}},\ \bibinfo
  {pages} {035201} (\bibinfo {year} {2020})}\BibitemShut {NoStop}%
\bibitem [{\citenamefont {Ji}(1997{\natexlab{a}})}]{Ji:1996ek}%
  \BibitemOpen
  \bibfield  {author} {\bibinfo {author} {\bibfnamefont {X.-D.}\ \bibnamefont
  {Ji}},\ }\href {\doibase 10.1103/PhysRevLett.78.610} {\bibfield  {journal}
  {\bibinfo  {journal} {Phys. Rev. Lett.}\ }\textbf {\bibinfo {volume} {78}},\
  \bibinfo {pages} {610} (\bibinfo {year} {1997}{\natexlab{a}})}\BibitemShut
  {NoStop}%
\bibitem [{\citenamefont {M\"uller}\ \emph {et~al.}(1994)\citenamefont
  {M\"uller}, \citenamefont {Robaschik}, \citenamefont {Geyer}, \citenamefont
  {Dittes},\ and\ \citenamefont {Ho\v{r}ej\v{s}i}}]{Mueller:1998fv}%
  \BibitemOpen
  \bibfield  {author} {\bibinfo {author} {\bibfnamefont {D.}~\bibnamefont
  {M\"uller}}, \bibinfo {author} {\bibfnamefont {D.}~\bibnamefont {Robaschik}},
  \bibinfo {author} {\bibfnamefont {B.}~\bibnamefont {Geyer}}, \bibinfo
  {author} {\bibfnamefont {F.-M.}\ \bibnamefont {Dittes}}, \ and\ \bibinfo
  {author} {\bibfnamefont {J.}~\bibnamefont {Ho\v{r}ej\v{s}i}},\ }\href
  {\doibase 10.1002/prop.2190420202} {\bibfield  {journal} {\bibinfo  {journal}
  {Fortsch. Phys.}\ }\textbf {\bibinfo {volume} {42}},\ \bibinfo {pages} {101}
  (\bibinfo {year} {1994})}\BibitemShut {NoStop}%
\bibitem [{\citenamefont {Radyushkin}(1996)}]{Radyushkin:1996nd}%
  \BibitemOpen
  \bibfield  {author} {\bibinfo {author} {\bibfnamefont {A.}~\bibnamefont
  {Radyushkin}},\ }\href {\doibase 10.1016/0370-2693(96)00528-X} {\bibfield
  {journal} {\bibinfo  {journal} {Phys. Lett. B}\ }\textbf {\bibinfo {volume}
  {380}},\ \bibinfo {pages} {417} (\bibinfo {year} {1996})}\BibitemShut
  {NoStop}%
\bibitem [{\citenamefont {Ji}(1997{\natexlab{b}})}]{Ji:1996nm}%
  \BibitemOpen
  \bibfield  {author} {\bibinfo {author} {\bibfnamefont {X.-D.}\ \bibnamefont
  {Ji}},\ }\href {\doibase 10.1103/PhysRevD.55.7114} {\bibfield  {journal}
  {\bibinfo  {journal} {Phys. Rev. D}\ }\textbf {\bibinfo {volume} {55}},\
  \bibinfo {pages} {7114} (\bibinfo {year} {1997}{\natexlab{b}})}\BibitemShut
  {NoStop}%
\bibitem [{\citenamefont {Goeke}\ \emph {et~al.}(2001)\citenamefont {Goeke},
  \citenamefont {Polyakov},\ and\ \citenamefont
  {Vanderhaeghen}}]{Goeke:2001tz}%
  \BibitemOpen
  \bibfield  {author} {\bibinfo {author} {\bibfnamefont {K.}~\bibnamefont
  {Goeke}}, \bibinfo {author} {\bibfnamefont {M.~V.}\ \bibnamefont {Polyakov}},
  \ and\ \bibinfo {author} {\bibfnamefont {M.}~\bibnamefont {Vanderhaeghen}},\
  }\href {\doibase 10.1016/S0146-6410(01)00158-2} {\bibfield  {journal}
  {\bibinfo  {journal} {Prog. Part. Nucl. Phys.}\ }\textbf {\bibinfo {volume}
  {47}},\ \bibinfo {pages} {401} (\bibinfo {year} {2001})}\BibitemShut
  {NoStop}%
\bibitem [{\citenamefont {Diehl}(2003)}]{Diehl:2003ny}%
  \BibitemOpen
  \bibfield  {author} {\bibinfo {author} {\bibfnamefont {M.}~\bibnamefont
  {Diehl}},\ }\href {\doibase 10.1016/j.physrep.2003.08.002} {\bibfield
  {journal} {\bibinfo  {journal} {Phys. Rept.}\ }\textbf {\bibinfo {volume}
  {388}},\ \bibinfo {pages} {41} (\bibinfo {year} {2003})}\BibitemShut
  {NoStop}%
\bibitem [{\citenamefont {Belitsky}\ and\ \citenamefont
  {Radyushkin}(2005)}]{Belitsky:2005qn}%
  \BibitemOpen
  \bibfield  {author} {\bibinfo {author} {\bibfnamefont {A.}~\bibnamefont
  {Belitsky}}\ and\ \bibinfo {author} {\bibfnamefont {A.}~\bibnamefont
  {Radyushkin}},\ }\href {\doibase 10.1016/j.physrep.2005.06.002} {\bibfield
  {journal} {\bibinfo  {journal} {Phys. Rept.}\ }\textbf {\bibinfo {volume}
  {418}},\ \bibinfo {pages} {1} (\bibinfo {year} {2005})}\BibitemShut {NoStop}%
\bibitem [{\citenamefont {Boffi}\ and\ \citenamefont
  {Pasquini}(2007)}]{Boffi:2007yc}%
  \BibitemOpen
  \bibfield  {author} {\bibinfo {author} {\bibfnamefont {S.}~\bibnamefont
  {Boffi}}\ and\ \bibinfo {author} {\bibfnamefont {B.}~\bibnamefont
  {Pasquini}},\ }\href {\doibase 10.1393/ncr/i2007-10025-7} {\bibfield
  {journal} {\bibinfo  {journal} {Riv. Nuovo Cim.}\ }\textbf {\bibinfo {volume}
  {30}},\ \bibinfo {pages} {387} (\bibinfo {year} {2007})}\BibitemShut
  {NoStop}%
\bibitem [{\citenamefont {Guidal}\ \emph {et~al.}(2013)\citenamefont {Guidal},
  \citenamefont {Moutarde},\ and\ \citenamefont
  {Vanderhaeghen}}]{Guidal:2013rya}%
  \BibitemOpen
  \bibfield  {author} {\bibinfo {author} {\bibfnamefont {M.}~\bibnamefont
  {Guidal}}, \bibinfo {author} {\bibfnamefont {H.}~\bibnamefont {Moutarde}}, \
  and\ \bibinfo {author} {\bibfnamefont {M.}~\bibnamefont {Vanderhaeghen}},\
  }\href {\doibase 10.1088/0034-4885/76/6/066202} {\bibfield  {journal}
  {\bibinfo  {journal} {Rept. Prog. Phys.}\ }\textbf {\bibinfo {volume} {76}},\
  \bibinfo {pages} {066202} (\bibinfo {year} {2013})}\BibitemShut {NoStop}%
\bibitem [{\citenamefont {Kumericki}\ \emph {et~al.}(2016)\citenamefont
  {Kumericki}, \citenamefont {Liuti},\ and\ \citenamefont
  {Moutarde}}]{Kumericki:2016ehc}%
  \BibitemOpen
  \bibfield  {author} {\bibinfo {author} {\bibfnamefont {K.}~\bibnamefont
  {Kumericki}}, \bibinfo {author} {\bibfnamefont {S.}~\bibnamefont {Liuti}}, \
  and\ \bibinfo {author} {\bibfnamefont {H.}~\bibnamefont {Moutarde}},\ }\href
  {\doibase 10.1140/epja/i2016-16157-3} {\bibfield  {journal} {\bibinfo
  {journal} {Eur. Phys. J. A}\ }\textbf {\bibinfo {volume} {52}},\ \bibinfo
  {pages} {157} (\bibinfo {year} {2016})}\BibitemShut {NoStop}%
\bibitem [{\citenamefont {Cardman}\ \emph {et~al.}(2011)\citenamefont
  {Cardman}, \citenamefont {Ent}, \citenamefont {Isgur}, \citenamefont {Laget},
  \citenamefont {Leemann}, \citenamefont {Meyer},\ and\ \citenamefont
  {Meziani}}]{osti_1345054}%
  \BibitemOpen
  \bibfield  {author} {\bibinfo {author} {\bibfnamefont {L.}~\bibnamefont
  {Cardman}}, \bibinfo {author} {\bibfnamefont {R.}~\bibnamefont {Ent}},
  \bibinfo {author} {\bibfnamefont {N.}~\bibnamefont {Isgur}}, \bibinfo
  {author} {\bibfnamefont {J.-M.}\ \bibnamefont {Laget}}, \bibinfo {author}
  {\bibfnamefont {C.}~\bibnamefont {Leemann}}, \bibinfo {author} {\bibfnamefont
  {C.}~\bibnamefont {Meyer}}, \ and\ \bibinfo {author} {\bibfnamefont {Z.-E.}\
  \bibnamefont {Meziani}},\ }\href@noop {} {\  (\bibinfo {year}
  {2011})}\BibitemShut {NoStop}%
\bibitem [{\citenamefont {Accardi}\ \emph {et~al.}(2016)\citenamefont {Accardi}
  \emph {et~al.}}]{Accardi:2012qut}%
  \BibitemOpen
  \bibfield  {author} {\bibinfo {author} {\bibfnamefont {A.}~\bibnamefont
  {Accardi}} \emph {et~al.},\ }\href {\doibase 10.1140/epja/i2016-16268-9}
  {\bibfield  {journal} {\bibinfo  {journal} {Eur. Phys. J. A}\ }\textbf
  {\bibinfo {volume} {52}},\ \bibinfo {pages} {268} (\bibinfo {year}
  {2016})}\BibitemShut {NoStop}%
\bibitem [{\citenamefont {Guidal}\ and\ \citenamefont
  {Vanderhaeghen}(2003)}]{Guidal:2002kt}%
  \BibitemOpen
  \bibfield  {author} {\bibinfo {author} {\bibfnamefont {M.}~\bibnamefont
  {Guidal}}\ and\ \bibinfo {author} {\bibfnamefont {M.}~\bibnamefont
  {Vanderhaeghen}},\ }\href {\doibase 10.1103/PhysRevLett.90.012001} {\bibfield
   {journal} {\bibinfo  {journal} {Phys. Rev. Lett.}\ }\textbf {\bibinfo
  {volume} {90}},\ \bibinfo {pages} {012001} (\bibinfo {year}
  {2003})}\BibitemShut {NoStop}%
\bibitem [{\citenamefont {Belitsky}\ and\ \citenamefont
  {Mueller}(2003)}]{Belitsky:2002tf}%
  \BibitemOpen
  \bibfield  {author} {\bibinfo {author} {\bibfnamefont {A.~V.}\ \bibnamefont
  {Belitsky}}\ and\ \bibinfo {author} {\bibfnamefont {D.}~\bibnamefont
  {Mueller}},\ }\href {\doibase 10.1103/PhysRevLett.90.022001} {\bibfield
  {journal} {\bibinfo  {journal} {Phys. Rev. Lett.}\ }\textbf {\bibinfo
  {volume} {90}},\ \bibinfo {pages} {022001} (\bibinfo {year}
  {2003})}\BibitemShut {NoStop}%
\bibitem [{\citenamefont {Accardi}\ \emph {et~al.}(2020)\citenamefont {Accardi}
  \emph {et~al.}}]{Accardi:2020swt}%
  \BibitemOpen
  \bibfield  {author} {\bibinfo {author} {\bibfnamefont {A.}~\bibnamefont
  {Accardi}} \emph {et~al.},\ }\href@noop {} {\  (\bibinfo {year} {2020})},\
  \Eprint {http://arxiv.org/abs/2007.15081} {arXiv:2007.15081 [nucl-ex]}
  \BibitemShut {NoStop}%
\bibitem [{\citenamefont {Vanderhaeghen}\ \emph {et~al.}(2000)\citenamefont
  {Vanderhaeghen}, \citenamefont {Friedrich}, \citenamefont {Lhuillier},
  \citenamefont {Marchand}, \citenamefont {Van~Hoorebeke},\ and\ \citenamefont
  {Van~de Wiele}}]{Vanderhaeghen:2000ws}%
  \BibitemOpen
  \bibfield  {author} {\bibinfo {author} {\bibfnamefont {M.}~\bibnamefont
  {Vanderhaeghen}}, \bibinfo {author} {\bibfnamefont {J.}~\bibnamefont
  {Friedrich}}, \bibinfo {author} {\bibfnamefont {D.}~\bibnamefont
  {Lhuillier}}, \bibinfo {author} {\bibfnamefont {D.}~\bibnamefont {Marchand}},
  \bibinfo {author} {\bibfnamefont {L.}~\bibnamefont {Van~Hoorebeke}}, \ and\
  \bibinfo {author} {\bibfnamefont {J.}~\bibnamefont {Van~de Wiele}},\ }\href
  {\doibase 10.1103/PhysRevC.62.025501} {\bibfield  {journal} {\bibinfo
  {journal} {Phys. Rev. C}\ }\textbf {\bibinfo {volume} {62}},\ \bibinfo
  {pages} {025501} (\bibinfo {year} {2000})}\BibitemShut {NoStop}%
\bibitem [{\citenamefont {Heller}\ \emph {et~al.}(2018)\citenamefont {Heller},
  \citenamefont {Tomalak},\ and\ \citenamefont
  {Vanderhaeghen}}]{Heller:2018ypa}%
  \BibitemOpen
  \bibfield  {author} {\bibinfo {author} {\bibfnamefont {M.}~\bibnamefont
  {Heller}}, \bibinfo {author} {\bibfnamefont {O.}~\bibnamefont {Tomalak}}, \
  and\ \bibinfo {author} {\bibfnamefont {M.}~\bibnamefont {Vanderhaeghen}},\
  }\href {\doibase 10.1103/PhysRevD.97.076012} {\bibfield  {journal} {\bibinfo
  {journal} {Phys. Rev. D}\ }\textbf {\bibinfo {volume} {97}},\ \bibinfo
  {pages} {076012} (\bibinfo {year} {2018})}\BibitemShut {NoStop}%
\bibitem [{\citenamefont {Heller}\ \emph {et~al.}(2019)\citenamefont {Heller},
  \citenamefont {Tomalak}, \citenamefont {Vanderhaeghen},\ and\ \citenamefont
  {Wu}}]{Heller:2019dyv}%
  \BibitemOpen
  \bibfield  {author} {\bibinfo {author} {\bibfnamefont {M.}~\bibnamefont
  {Heller}}, \bibinfo {author} {\bibfnamefont {O.}~\bibnamefont {Tomalak}},
  \bibinfo {author} {\bibfnamefont {M.}~\bibnamefont {Vanderhaeghen}}, \ and\
  \bibinfo {author} {\bibfnamefont {S.}~\bibnamefont {Wu}},\ }\href {\doibase
  10.1103/PhysRevD.100.076013} {\bibfield  {journal} {\bibinfo  {journal}
  {Phys. Rev. D}\ }\textbf {\bibinfo {volume} {100}},\ \bibinfo {pages}
  {076013} (\bibinfo {year} {2019})}\BibitemShut {NoStop}%
\bibitem [{\citenamefont {Heller}\ \emph {et~al.}(2021)\citenamefont {Heller},
  \citenamefont {Keil},\ and\ \citenamefont {Vanderhaeghen}}]{Heller:2020lnm}%
  \BibitemOpen
  \bibfield  {author} {\bibinfo {author} {\bibfnamefont {M.}~\bibnamefont
  {Heller}}, \bibinfo {author} {\bibfnamefont {N.}~\bibnamefont {Keil}}, \ and\
  \bibinfo {author} {\bibfnamefont {M.}~\bibnamefont {Vanderhaeghen}},\ }\href
  {\doibase 10.1103/PhysRevD.103.036009} {\bibfield  {journal} {\bibinfo
  {journal} {Phys. Rev. D}\ }\textbf {\bibinfo {volume} {103}},\ \bibinfo
  {pages} {036009} (\bibinfo {year} {2021})}\BibitemShut {NoStop}%
\bibitem [{\citenamefont {Tarrach}(1975)}]{Tarrach:1975tu}%
  \BibitemOpen
  \bibfield  {author} {\bibinfo {author} {\bibfnamefont {R.}~\bibnamefont
  {Tarrach}},\ }\href {\doibase 10.1007/BF02894857} {\bibfield  {journal}
  {\bibinfo  {journal} {Nuovo Cim. A}\ }\textbf {\bibinfo {volume} {28}},\
  \bibinfo {pages} {409} (\bibinfo {year} {1975})}\BibitemShut {NoStop}%
\bibitem [{\citenamefont {Drechsel}\ \emph {et~al.}(1998)\citenamefont
  {Drechsel}, \citenamefont {Knochlein}, \citenamefont {Korchin}, \citenamefont
  {Metz},\ and\ \citenamefont {Scherer}}]{Drechsel:1997xv}%
  \BibitemOpen
  \bibfield  {author} {\bibinfo {author} {\bibfnamefont {D.}~\bibnamefont
  {Drechsel}}, \bibinfo {author} {\bibfnamefont {G.}~\bibnamefont {Knochlein}},
  \bibinfo {author} {\bibfnamefont {A.}~\bibnamefont {Korchin}}, \bibinfo
  {author} {\bibfnamefont {A.}~\bibnamefont {Metz}}, \ and\ \bibinfo {author}
  {\bibfnamefont {S.}~\bibnamefont {Scherer}},\ }\href {\doibase
  10.1103/PhysRevC.57.941} {\bibfield  {journal} {\bibinfo  {journal} {Phys.
  Rev. C}\ }\textbf {\bibinfo {volume} {57}},\ \bibinfo {pages} {941} (\bibinfo
  {year} {1998})}\BibitemShut {NoStop}%
\bibitem [{\citenamefont {Lomon}\ and\ \citenamefont
  {Pacetti}(2012)}]{Lomon:2012pn}%
  \BibitemOpen
  \bibfield  {author} {\bibinfo {author} {\bibfnamefont {E.~L.}\ \bibnamefont
  {Lomon}}\ and\ \bibinfo {author} {\bibfnamefont {S.}~\bibnamefont
  {Pacetti}},\ }\href {\doibase 10.1103/PhysRevD.86.039901} {\bibfield
  {journal} {\bibinfo  {journal} {Phys. Rev. D}\ }\textbf {\bibinfo {volume}
  {85}},\ \bibinfo {pages} {113004} (\bibinfo {year} {2012})},\ \bibinfo {note}
  {[Erratum: Phys.Rev.D 86, 039901 (2012)]}\BibitemShut {NoStop}%
\bibitem [{\citenamefont {Drechsel}\ \emph {et~al.}(2007)\citenamefont
  {Drechsel}, \citenamefont {Kamalov},\ and\ \citenamefont
  {Tiator}}]{Drechsel:2007if}%
  \BibitemOpen
  \bibfield  {author} {\bibinfo {author} {\bibfnamefont {D.}~\bibnamefont
  {Drechsel}}, \bibinfo {author} {\bibfnamefont {S.}~\bibnamefont {Kamalov}}, \
  and\ \bibinfo {author} {\bibfnamefont {L.}~\bibnamefont {Tiator}},\ }\href
  {\doibase 10.1140/epja/i2007-10490-6} {\bibfield  {journal} {\bibinfo
  {journal} {Eur. Phys. J. A}\ }\textbf {\bibinfo {volume} {34}},\ \bibinfo
  {pages} {69} (\bibinfo {year} {2007})}\BibitemShut {NoStop}%
\bibitem [{\citenamefont {Tiator}\ \emph {et~al.}(2011)\citenamefont {Tiator},
  \citenamefont {Drechsel}, \citenamefont {Kamalov},\ and\ \citenamefont
  {Vanderhaeghen}}]{Tiator:2011pw}%
  \BibitemOpen
  \bibfield  {author} {\bibinfo {author} {\bibfnamefont {L.}~\bibnamefont
  {Tiator}}, \bibinfo {author} {\bibfnamefont {D.}~\bibnamefont {Drechsel}},
  \bibinfo {author} {\bibfnamefont {S.}~\bibnamefont {Kamalov}}, \ and\
  \bibinfo {author} {\bibfnamefont {M.}~\bibnamefont {Vanderhaeghen}},\ }\href
  {\doibase 10.1140/epjst/e2011-01488-9} {\bibfield  {journal} {\bibinfo
  {journal} {Eur. Phys. J. ST}\ }\textbf {\bibinfo {volume} {198}},\ \bibinfo
  {pages} {141} (\bibinfo {year} {2011})}\BibitemShut {NoStop}%
\bibitem [{\citenamefont {Pascalutsa}\ \emph {et~al.}(2007)\citenamefont
  {Pascalutsa}, \citenamefont {Vanderhaeghen},\ and\ \citenamefont
  {Yang}}]{Pascalutsa:2006up}%
  \BibitemOpen
  \bibfield  {author} {\bibinfo {author} {\bibfnamefont {V.}~\bibnamefont
  {Pascalutsa}}, \bibinfo {author} {\bibfnamefont {M.}~\bibnamefont
  {Vanderhaeghen}}, \ and\ \bibinfo {author} {\bibfnamefont {S.~N.}\
  \bibnamefont {Yang}},\ }\href {\doibase 10.1016/j.physrep.2006.09.006}
  {\bibfield  {journal} {\bibinfo  {journal} {Phys. Rept.}\ }\textbf {\bibinfo
  {volume} {437}},\ \bibinfo {pages} {125} (\bibinfo {year}
  {2007})}\BibitemShut {NoStop}%
\bibitem [{\citenamefont {Holstein}\ \emph {et~al.}(2000)\citenamefont
  {Holstein}, \citenamefont {Drechsel}, \citenamefont {Pasquini},\ and\
  \citenamefont {Vanderhaeghen}}]{Holstein:1999uu}%
  \BibitemOpen
  \bibfield  {author} {\bibinfo {author} {\bibfnamefont {B.~R.}\ \bibnamefont
  {Holstein}}, \bibinfo {author} {\bibfnamefont {D.}~\bibnamefont {Drechsel}},
  \bibinfo {author} {\bibfnamefont {B.}~\bibnamefont {Pasquini}}, \ and\
  \bibinfo {author} {\bibfnamefont {M.}~\bibnamefont {Vanderhaeghen}},\ }\href
  {\doibase 10.1103/PhysRevC.61.034316} {\bibfield  {journal} {\bibinfo
  {journal} {Phys. Rev. C}\ }\textbf {\bibinfo {volume} {61}},\ \bibinfo
  {pages} {034316} (\bibinfo {year} {2000})}\BibitemShut {NoStop}%
\bibitem [{\citenamefont {Vanderhaeghen}\ \emph {et~al.}(1998)\citenamefont
  {Vanderhaeghen}, \citenamefont {Guichon},\ and\ \citenamefont
  {Guidal}}]{Vanderhaeghen:1998uc}%
  \BibitemOpen
  \bibfield  {author} {\bibinfo {author} {\bibfnamefont {M.}~\bibnamefont
  {Vanderhaeghen}}, \bibinfo {author} {\bibfnamefont {P.~A.}\ \bibnamefont
  {Guichon}}, \ and\ \bibinfo {author} {\bibfnamefont {M.}~\bibnamefont
  {Guidal}},\ }\href {\doibase 10.1103/PhysRevLett.80.5064} {\bibfield
  {journal} {\bibinfo  {journal} {Phys. Rev. Lett.}\ }\textbf {\bibinfo
  {volume} {80}},\ \bibinfo {pages} {5064} (\bibinfo {year}
  {1998})}\BibitemShut {NoStop}%
\bibitem [{\citenamefont {Vanderhaeghen}\ \emph {et~al.}(1999)\citenamefont
  {Vanderhaeghen}, \citenamefont {Guichon},\ and\ \citenamefont
  {Guidal}}]{Vanderhaeghen:1999xj}%
  \BibitemOpen
  \bibfield  {author} {\bibinfo {author} {\bibfnamefont {M.}~\bibnamefont
  {Vanderhaeghen}}, \bibinfo {author} {\bibfnamefont {P.~A.}\ \bibnamefont
  {Guichon}}, \ and\ \bibinfo {author} {\bibfnamefont {M.}~\bibnamefont
  {Guidal}},\ }\href {\doibase 10.1103/PhysRevD.60.094017} {\bibfield
  {journal} {\bibinfo  {journal} {Phys. Rev. D}\ }\textbf {\bibinfo {volume}
  {60}},\ \bibinfo {pages} {094017} (\bibinfo {year} {1999})}\BibitemShut
  {NoStop}%
\bibitem [{\citenamefont {Guidal}\ \emph {et~al.}(2005)\citenamefont {Guidal},
  \citenamefont {Polyakov}, \citenamefont {Radyushkin},\ and\ \citenamefont
  {Vanderhaeghen}}]{Guidal:2004nd}%
  \BibitemOpen
  \bibfield  {author} {\bibinfo {author} {\bibfnamefont {M.}~\bibnamefont
  {Guidal}}, \bibinfo {author} {\bibfnamefont {M.}~\bibnamefont {Polyakov}},
  \bibinfo {author} {\bibfnamefont {A.}~\bibnamefont {Radyushkin}}, \ and\
  \bibinfo {author} {\bibfnamefont {M.}~\bibnamefont {Vanderhaeghen}},\ }\href
  {\doibase 10.1103/PhysRevD.72.054013} {\bibfield  {journal} {\bibinfo
  {journal} {Phys. Rev. D}\ }\textbf {\bibinfo {volume} {72}},\ \bibinfo
  {pages} {054013} (\bibinfo {year} {2005})}\BibitemShut {NoStop}%
\bibitem [{\citenamefont {Pasquini}\ \emph {et~al.}(2014)\citenamefont
  {Pasquini}, \citenamefont {Polyakov},\ and\ \citenamefont
  {Vanderhaeghen}}]{Pasquini:2014vua}%
  \BibitemOpen
  \bibfield  {author} {\bibinfo {author} {\bibfnamefont {B.}~\bibnamefont
  {Pasquini}}, \bibinfo {author} {\bibfnamefont {M.}~\bibnamefont {Polyakov}},
  \ and\ \bibinfo {author} {\bibfnamefont {M.}~\bibnamefont {Vanderhaeghen}},\
  }\href {\doibase 10.1016/j.physletb.2014.10.047} {\bibfield  {journal}
  {\bibinfo  {journal} {Phys. Lett. B}\ }\textbf {\bibinfo {volume} {739}},\
  \bibinfo {pages} {133} (\bibinfo {year} {2014})}\BibitemShut {NoStop}%
\bibitem [{\citenamefont {'t~Hooft}\ and\ \citenamefont
  {Veltman}(1972)}]{tHooft:1972tcz}%
  \BibitemOpen
  \bibfield  {author} {\bibinfo {author} {\bibfnamefont {G.}~\bibnamefont
  {'t~Hooft}}\ and\ \bibinfo {author} {\bibfnamefont {M.~J.~G.}\ \bibnamefont
  {Veltman}},\ }\href {\doibase 10.1016/0550-3213(72)90279-9} {\bibfield
  {journal} {\bibinfo  {journal} {Nucl. Phys. B}\ }\textbf {\bibinfo {volume}
  {44}},\ \bibinfo {pages} {189} (\bibinfo {year} {1972})}\BibitemShut
  {NoStop}%
\bibitem [{\citenamefont {'t~Hooft}\ and\ \citenamefont
  {Veltman}(1979)}]{tHooft:1978jhc}%
  \BibitemOpen
  \bibfield  {author} {\bibinfo {author} {\bibfnamefont {G.}~\bibnamefont
  {'t~Hooft}}\ and\ \bibinfo {author} {\bibfnamefont {M.}~\bibnamefont
  {Veltman}},\ }\href {\doibase 10.1016/0550-3213(79)90605-9} {\bibfield
  {journal} {\bibinfo  {journal} {Nucl. Phys. B}\ }\textbf {\bibinfo {volume}
  {153}},\ \bibinfo {pages} {365} (\bibinfo {year} {1979})}\BibitemShut
  {NoStop}%
\bibitem [{\citenamefont {Gryniuk}\ \emph {et~al.}(2015)\citenamefont
  {Gryniuk}, \citenamefont {Hagelstein},\ and\ \citenamefont
  {Pascalutsa}}]{Gryniuk:2015eza}%
  \BibitemOpen
  \bibfield  {author} {\bibinfo {author} {\bibfnamefont {O.}~\bibnamefont
  {Gryniuk}}, \bibinfo {author} {\bibfnamefont {F.}~\bibnamefont {Hagelstein}},
  \ and\ \bibinfo {author} {\bibfnamefont {V.}~\bibnamefont {Pascalutsa}},\
  }\href {\doibase 10.1103/PhysRevD.92.074031} {\bibfield  {journal} {\bibinfo
  {journal} {Phys. Rev. D}\ }\textbf {\bibinfo {volume} {92}},\ \bibinfo
  {pages} {074031} (\bibinfo {year} {2015})}\BibitemShut {NoStop}%
\bibitem [{\citenamefont {Ellis}\ and\ \citenamefont
  {Zanderighi}(2008)}]{Ellis:2007qk}%
  \BibitemOpen
  \bibfield  {author} {\bibinfo {author} {\bibfnamefont {R.~K.}\ \bibnamefont
  {Ellis}}\ and\ \bibinfo {author} {\bibfnamefont {G.}~\bibnamefont
  {Zanderighi}},\ }\href {\doibase 10.1088/1126-6708/2008/02/002} {\bibfield
  {journal} {\bibinfo  {journal} {JHEP}\ }\textbf {\bibinfo {volume} {02}},\
  \bibinfo {pages} {002} (\bibinfo {year} {2008})}\BibitemShut {NoStop}%
\end{thebibliography}%

\end{document}